\documentstyle[epsfig]{ar209}

\def\be{\begin{eqnarray}}
\def\ee{\end{eqnarray}}
\def\xp{x_p}

\begin{document}

\input epsf.tex
\input epsf.def
\input psfig.sty

\jname{}
\jyear{}
\jvol{ }
\ARinfo{ }

\title{Study of the Fundamental Structure of Matter with an 
Electron-Ion Collider}

\markboth{A. Deshpande et al.}{Study of Fundamental Structure of Matter 
\ldots}

\author{Abhay Deshpande\affiliation{Department of Physics \& Astronomy, 
SUNY at Stony Brook,  New York 11794, U.S.A. \& \\
RIKEN-BNL Research Center, Brookhaven National Laboratory, Upton, 
NY 11973, U.S.A.; email: abhay@bnl.gov}
Richard Milner\affiliation{Physics Department and Laboratory for Nuclear 
Science, Massachusetts Institute of Technology, Cambridge, MA 02139.
U.S.A.; email: milner@mit.edu}
Raju Venugopalan\affiliation{Physics Department, 
Brookhaven National Laboratory, Upton, NY 11973, U.S.A.; \\email: 
raju@quark.phy.bnl.gov}
Werner Vogelsang\affiliation{Physics Department and RIKEN-BNL Research 
Center,
Brookhaven National Laboratory, Upton, NY 11973, U.S.A.; email: 
vogelsan@quark.phy.bnl.gov}}

\begin{keywords}
Quantum Chromodynamics, DIS structure functions, Polarized ep Scattering, 
Nucleon Spin, DIS off Nuclei, Saturation, Color Glass Condensate, EIC, eRHIC
\end{keywords}

\begin{abstract}
We present an overview of the scientific opportunities that would be offered 
by a high-energy electron-ion collider. We discuss the relevant physics of 
polarized and unpolarized electron-proton collisions and of electron-nucleus 
collisions. We also describe the current accelerator and detector plans for 
a future electron-ion collider. \\[-16cm]
BNL-NT-05/19, RBRC-526
\vspace*{15.3cm}
\end{abstract}

\maketitle

\section{Introduction}

Understanding the fundamental structure of matter is one of the central goals of scientific research.  In the closing decades of the twentieth century, physicists developed a beautiful theory, Quantum Chromodynamics (QCD), which explains all of strongly interacting matter in terms of point-like quarks interacting by the exchange of gauge bosons, known as gluons. The gluons of QCD, unlike the photons of QED, can interact with each other. The color force which governs the interaction of quarks and gluons is responsible for more than 99\% of the observable mass in the physical universe and explains the structure of nucleons and their composite structures, 
atomic nuclei, as well as astrophysical objects such as neutron stars.  

During the last 30 years, experiments have verified QCD quantitatively in collisions 
involving a very large momentum exchange between the participants. These 
collisions occur over very short distances much smaller than the size of the proton.  In 
these experiments, the confined quarks and gluons act as if they are nearly 
free pointlike 
particles and exhibit many properties that are predicted by perturbative  QCD (pQCD).  This 
experimental phenomenon was first discovered in deeply inelastic scattering (DIS) experiments of electrons off nucleons.  The discovery 
resulted in the 1990  Nobel Prize in Physics being 
awarded to Friedman, Kendall and Taylor. The phenomenon, that quarks and gluons are quasi-free at 
short distances, follows from a fundamental property of QCD known as {\it asymptotic freedom}.  Gross, Politzer and Wilczek, who first identified and understood this unique characteristic of QCD were awarded the 2004 Nobel Prize in Physics.

When the interaction distance between the quarks and gluons becomes comparable to or larger than the typical size of hadrons, the fundamental constituents of the nucleon are no longer free. They are confined by the strong force that does not allow for the observation of any ``colored'' object.  
In this strong coupling QCD regime, where most hadronic matter exists,  the symmetries of the underlying quark-gluon theory are hidden, and QCD 
computations in terms of the dynamical properties of quarks and 
gluons are difficult.  
A major effort is underway worldwide to carry out {\it ab initio} QCD calculations in the strong QCD regime using Monte-Carlo simulations on large scale computers.

The experimental underpinnings for QCD are derived from decades of work at the CERN, DESY, Fermilab and SLAC
accelerator facilities. Some highlights include  the determination of the nucleon quark momentum and spin distributions and the nucleon gluon momentum distribution, the verification of the QCD prediction for the running of the strong coupling constant $\alpha_s$, the discovery of jets, and the discovery that quark and gluon momentum 
distributions in a nucleus differ from those in a free nucleon.

However, thirty years after QCD has been established as the Standard Model of the strong force, and despite impressive progress made in the intervening decades,  understanding how QCD works in detail remains one of the outstanding issues in physics. Some crucial open questions that need to be addressed are listed below.

\pagebreak
\begin{itemize}
\item[--]{\it What is the gluon momentum distribution in the atomic nucleus?}\hfill\break
QCD tells us that the nucleon is primarily made up of specks of matter (quarks) bound by tremendously powerful gluon fields.  Thus atomic nuclei are primarily composed of glue.  
Very little is known about the gluon momentum distribution in a nucleus. 
Determining these gluon distributions is therefore a fundamental measurement of high priority. This quantity is also essential for an understanding of other important questions in hadronic physics.  For example, the interpretation of experiments searching for a deconfined quark-gluon state in relativistic heavy ion collisions is dependent on the knowledge of the initial quark and gluon configuration in a heavy nucleus.  This will 
be especially true for heavy ion experiments at the Large Hadron Collider 
(LHC) at CERN. Further, there are predictions that gluonic matter at high parton 
densities has novel properties that can be probed in 
hard scattering experiments on nuclear targets. Hints of the existence of this state may have been seen in Deuteron-Gold experiments at the Relativistic 
Heavy Ion Collider (RHIC) at Brookhaven.

\item[--] {\it How is the spin structure of the nucleon understood to arise from the quark and gluon constituents?}\hfill\break  
High energy spin-dependent lepton scattering experiments from polarized nucleon targets have produced surprising results.  The spins of the quarks 
account for only about $20\%$ of the spin of the proton.  The contribution of the gluons may be large.  Dramatic effects are predicted for measurements beyond the capability of any existing accelerator.  There are hints from other experiments that the contribution of orbital angular momentum may be large.  

\item[--] {\it Testing QCD}\hfill\break
It is imperative to continue to subject 
QCD to stringent tests because there is so much about the theory that remains a mystery. QCD can be tested in two ways: one is by precision measurements, 
and the other is by looking 
for novel physics which is sensitive to the confining properties 
of the theory. Both 
of these can be achieved at a high luminosity lepton-ion collider with a detector that has a wide rapidity and angular coverage. An example of precision physics is the Bjorken Sum Rule in spin-dependent lepton scattering from a polarized nucleon.  This fundamental sum rule relates inclusive spin-dependent lepton scattering to the ratio of axial to vector coupling constants in neutron $\beta$-decay.  Present experiments test it to about $\pm 10\%$: it would be highly desirable to push these tests to about $\pm$1\%. Further, with lattice QCD expected to make substantial progress in the ability to make {\it ab initio} QCD calculations during the next decade, precise measurements of the calculable observables will be required. An example of a physics measurement 
sensitive to confinement is hard diffraction, where 
large mass final states are formed with large ``color-less" gaps in rapidity separating them from the hadron or nucleus. At the Hadron Electron Ring Accelerator (HERA) at DESY, 
roughly 10\% of events are of this nature. The origins of these rapidity gaps, which must be intimately 
related to the confining properties of the theory, can be better understood with detectors that are able to provide detailed maps of the structure of events in DIS.
\end{itemize}

This article motivates and describes the next generation 
accelerator required by nuclear and particle physicists
to study and test QCD, namely a polarized lepton-ion collider.  The basic characteristics of the collider are motivated as follows:

\begin{itemize}
\item[--] {\it lepton beam} \hfill\break
The lepton probe employs the best understood interaction in nature (QED) to study hadron structure.  Electrons and positrons couple directly to the quarks. The experimental conditions which maximize sensitivity to valence and sea quarks as well as probe gluons are well understood.  Further, the availability of both positron and electron beams will enable experiments that are sensitive to the exchange of the parity violating Z and W-bosons.

\item[--] {\it range of center-of-mass (CM) energies} \hfill\break
To cleanly interact with  quarks, a minimum center-of-mass (CM) energy of about 10 GeV is required.  To explore and utilize the powerful $Q^2$  evolution equations of QCD, CM energies of order 100 GeV are desirable.  This consideration strongly motivates the collider geometry.

\item[--] {\it high luminosity}\hfill\break
The QED interaction between the lepton probe and the hadron target is relatively weak.  Thus precise and definitive measurements demand a high collision luminosity of order $10^{33}$ nucleons cm$^{-2}$ s$^{-1}$.
  
\item[--] {\it polarized beams} \hfill\break
Polarized lepton and nucleon beams are essential to address the central question of the spin structure of the nucleon.  Both polarized proton and neutron (effectively polarized $^2$H or $^3$He) are required for tests of the fundamental Bjorken Sum Rule.  The polarization direction of at least one of the beams must be reversible on a rapid timescale to minimize systematic uncertainties.

\item[--] {\it nuclear beams} \hfill\break
Light nuclear targets are useful for probing the 
spin and flavor content of parton distributions.  Heavy nuclei are essential for experiments probing the behavior of quarks and gluons in the nuclear medium. 

\item[--]{\it detector considerations} \hfill\break
The collider geometry has a significant advantage over fixed-target experiments at high energy because it makes feasible the detection of complete final-states.  A central collider detector with momentum and energy measurements and particle identification for both leptons and hadrons will be essential for many experiments.  Special purpose detectors that provide wide angular and rapidity coverage will be essential for several specific measurements.
\end{itemize}

\begin{figure}[!h]
\begin{center}
\epsfig{file=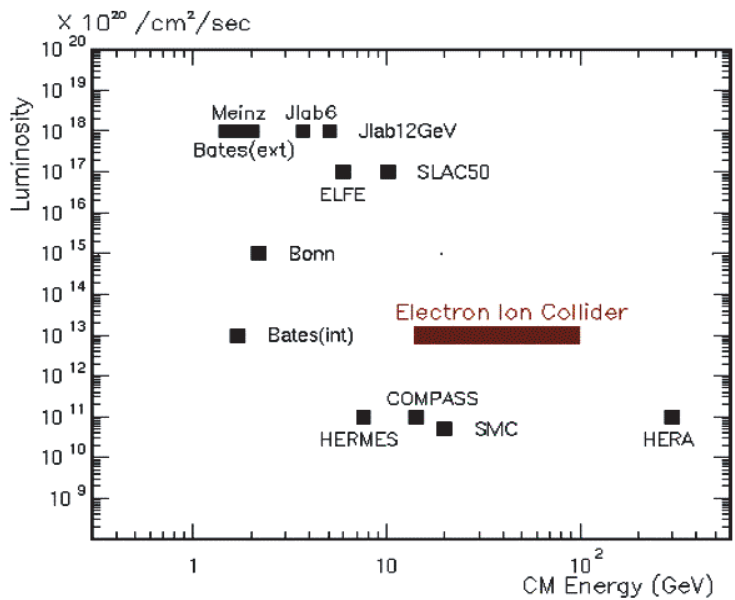,width=10cm,angle=0}
\caption{ The center-of-mass energy vs. 
luminosity of the proposed Electron-Ion Collider eRHIC 
compared to other lepton scattering facilities. \label{fig:lumi}}
\end{center}
\end{figure}

These considerations constrain the design parameters of the collider to be a 5 to 10 GeV energy electron (or positron) beam colliding with a nucleon beam of energy 25 GeV to 250 GeV.  The collider is 
anticipated to deliver nuclear beams of energies ranging from 
$20-100$ GeV/nucleon.  The lepton and nucleon beams must be highly polarized and the collision luminosity must be of order $10^{33}$ nucleons cm$^{-2}$ s$^{-1}$.  The proposed eRHIC design (described in section 4) realizes the required specifications in a cost effective and timely way by using the existing RHIC facility at BNL. The characteristics of eRHIC are well beyond the capability of any existing accelerator, as is clear from Fig.~\ref{fig:lumi}.

\begin{figure}[!h]
\begin{center}
\epsfig{file=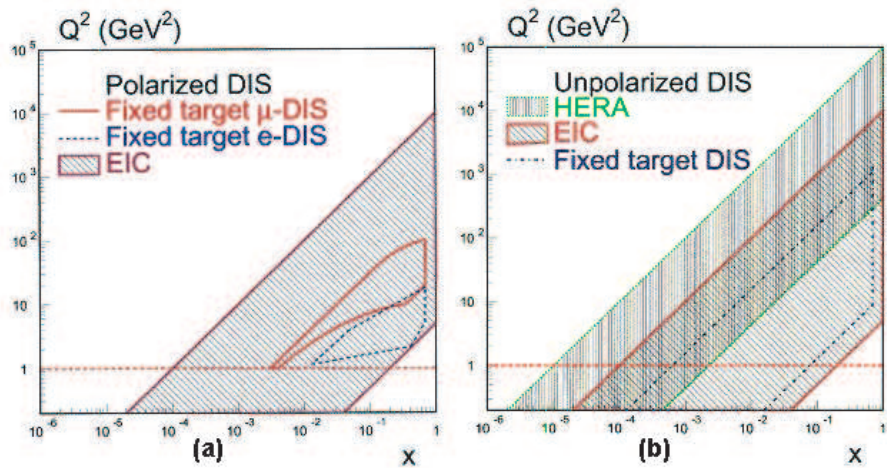,width=12cm,angle=0}
\caption[*]{The $x$-$Q^2$ range of the proposed lepton-ion collider at Brookhaven National Laboratory (eRHIC) in comparison with the past \& present experimental DIS facilities. The left plot is for polarized DIS experiments, and the right  corresponds to the unpolarized DIS experiments. \label{fig:xq2}}
\end{center}
\end{figure}

By  delivering high energies to the collision, the collider provides an increased range for investigating quarks and gluons with small momentum fraction ($x$) and for studying their behavior over a wide range of momentum transfers 
($Q^2$).  In deeply inelastic 
scattering, the accessible values of the Bjorken variable $x$ (defined in section~2) are limited by the available CM energy.  For example, collisions between a 10 GeV lepton beam and nuclear beams of 100 GeV/nucleon provide access to values of $x$ as small as $3 \times 10^{-4}$ for $Q^{2} \sim 1$ GeV$^{2}$.  In a fixed-target configuration, a 2.1 TeV lepton beam would be required to  produce the same CM energy.  Figure~\ref{fig:xq2} 
shows the $x$-$Q^2$ range possible with the proposed eRHIC machine and 
compares that range to the currently explored kinematic region.

In this article, the scientific case and accelerator design for a new facility to study the fundamental quark and gluon structure of strongly interacting matter is presented.  Section 2 describes the current understanding of the quark and gluon structure of hadrons and nuclei.  Section 3 presents highlights of the scientific opportunities available with a 
lepton-ion collider.  Section 4 describes the accelerator design effort and section 5 describes the interaction region and eRHIC detector design.

\section{Status of the Exploration of the 
Partonic Structure of Hadrons and Nuclei \label{status}}

This section will summarize our current understanding of the partonic
structure of hadrons and nuclei in QCD, accumulated during the past
three decades from a variety of deeply inelastic and hadronic
scattering experiments. We will also comment on what new information
may become available  from DIS as well as
from RHIC and other experimental facilities around the world before a future 
electron-ion collider starts taking data. We will outline the status 
of our knowledge on i)
the parton distributions in nucleons, ii) spin and flavor
distributions in the nucleon, iii) nuclear modifications to the
inclusive nucleon distributions such as the European Muon Collaboration (EMC) 
effect and quark and
gluon shadowing, iv) color coherent phenomena in nuclei that probe the
space-time structure of QCD such as color transparency and opacity,
partonic energy loss and the $p_T$ broadening of partons in media.  In
each case, we will outline the most important remaining questions
and challenges. These will be addressed further in Section~\ref{science}.
\subsection{Deeply-inelastic scattering}
\begin{figure}[!h]
\begin{center}
\epsfig{file=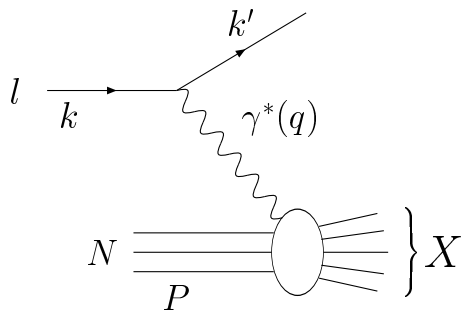,width=6cm,angle=0}
\caption[*]{Deeply-inelastic lepton-nucleon scattering mediated by virtual 
photon exchange. 
\label{fig:dis}}
\end{center}
\end{figure}

The cross-section for the inclusive deeply inelastic scattering (DIS) 
process shown in Fig.~\ref{fig:dis} can be written as a product of the 
leptonic tensor ${\cal L_{\mu\nu}}$ and 
the hadronic tensor ${\cal W^{\mu\nu}}$ as 
\begin{equation} \label{crsec}
\frac{d^2\sigma}{dxdy} \; \propto \; 
{\cal L}_{\mu\nu}^{} (k,q,s) \;{\cal W}^{\mu\nu}_{}(P,q,S) \; ,
\end{equation}
where one defines the Lorentz invariant scalars, the famous Bjorken 
variable $x=-q^2/2P\,  q$, and $y=P\,  q/P\,  k$. Note that as illustrated 
in Fig.~\ref{fig:dis},
$k$ ($k^\prime$) is the 4-momentum of the incoming (outgoing) electron,
$P$ is the 4-momentum of the incoming hadron, and $q=k-k^\prime$ is
the 4-momentum of the virtual photon. The center of mass energy squared is 
$s=(P+k)^2$. From these invariants, one can deduce simply that 
$x\,  y \approx Q^2/s$, where $Q^2 = -q^2 >0$.

The hadronic tensor can be written in full generality as
\begin{eqnarray}
&&{\cal W}^{\mu\nu} (P,q,S) =\frac{1}{4\pi} \int d^4 z \; 
{\rm e}^{i q\,  z}
\langle P,S|\left[ {\cal J}_{\mu}(z),{\cal J}_{\nu}(0) \right] 
|P,S\rangle =-g^{\mu\nu} F_1 (x,Q^2) \nonumber \\
&& + \frac{P^{\mu} P^{\nu}}{P\,  q}
F_2 (x,Q^2)  - i \varepsilon^{\mu\nu\rho\sigma} 
\frac{q_{\rho}P_{\sigma}}{2 \,P\,  q}F_3 (x,Q^2) 
\; +\; i \varepsilon^{\mu\nu\rho\sigma} q_{\rho} \left[ 
\frac{S_{\sigma}}{P\,  q}
\; g_1 (x,Q^2) \right. \nonumber \\
&&\left.+ \frac{S_{\sigma}(P\,  q)-
P_{\sigma} (S\,  q)}{(P\,  q)^2}
\; g_2 (x,Q^2) \right] 
\;+ \; \left[ \frac{P^{\mu} S^{\nu}+
S^{\mu} P^{\nu}}{2P\,  q}  - \frac{S\,  q}{(P\,  q)^2} 
P^{\mu} P^{\nu} \right] \; g_3 (x,Q^2) \nonumber \\
&&\;+ \; \frac{S\,  q}{(P\,  q)^2} 
P^{\mu} P^{\nu}\; g_4 (x,Q^2) \; - \;
\frac{S\,  q}{P\,  q} g^{\mu\nu} 
\; g_5 (x,Q^2)  \; . \label{strcfct}
\end{eqnarray}
The $F_i$ are referred to as the ``unpolarized''
structure functions, whereas the $g_i$ are the ``spin-dependent'' ones,
because their associated tensors depend on the nucleon spin vector
$S^{\mu}$. Note that parity-violating interactions mediated by electroweak boson 
exchange are required for $F_3,g_3,g_4,g_5$ to contribute. 

Inserting Eq.~\ref{strcfct} and the straightforwardly calculated 
leptonic tensor into Eq.~\ref{crsec}, one obtains the DIS
cross section in terms of the structure functions.
If one averages over the hadronic 
spins and restricts oneself to parity conserving (for $Q^2 \ll M_Z^2$)
electron-nucleon scattering alone, one finds the simple expression
\begin{eqnarray}
\frac{d^2\sigma}{dx\,dQ^2} = {2\pi \alpha_{\rm em}^2\over Q^4}
\left[\left(1+ (1-y)^2\right)\,F_2(x,Q^2) -y^2 
F_L(x,Q^2)\right]\, .
\end{eqnarray}
Here, $\alpha_{\rm em}$ is the coupling constant of Quantum
Electrodynamics and  $F_L$ is the ``longitudinal'' structure function, defined 
by the relation 
$F_L = F_2 -2x\,F_1$. 
\begin{figure}[!h]
\begin{center}
\epsfig{file=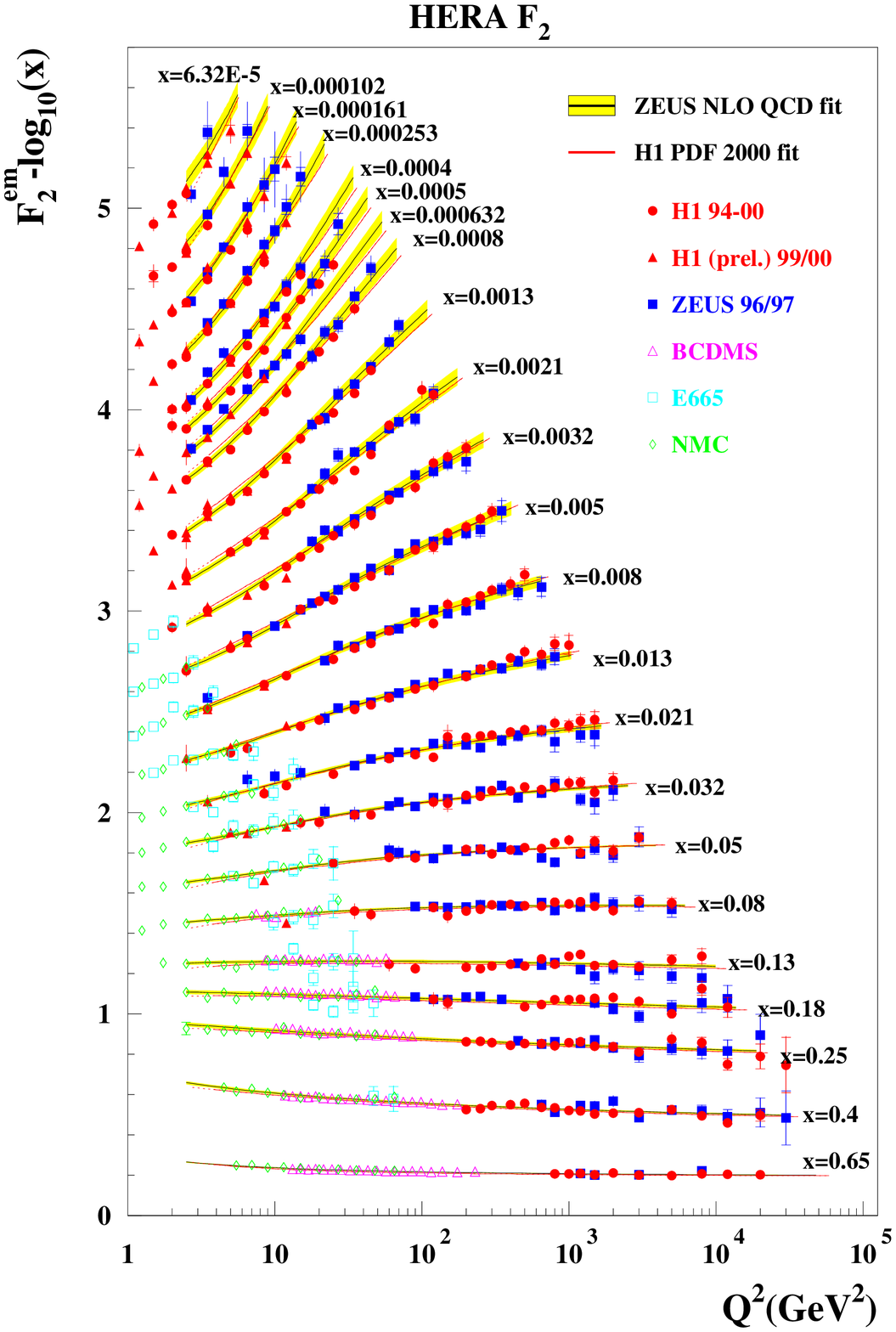,width=6cm,angle=0}
\hspace*{5mm}
\epsfig{file=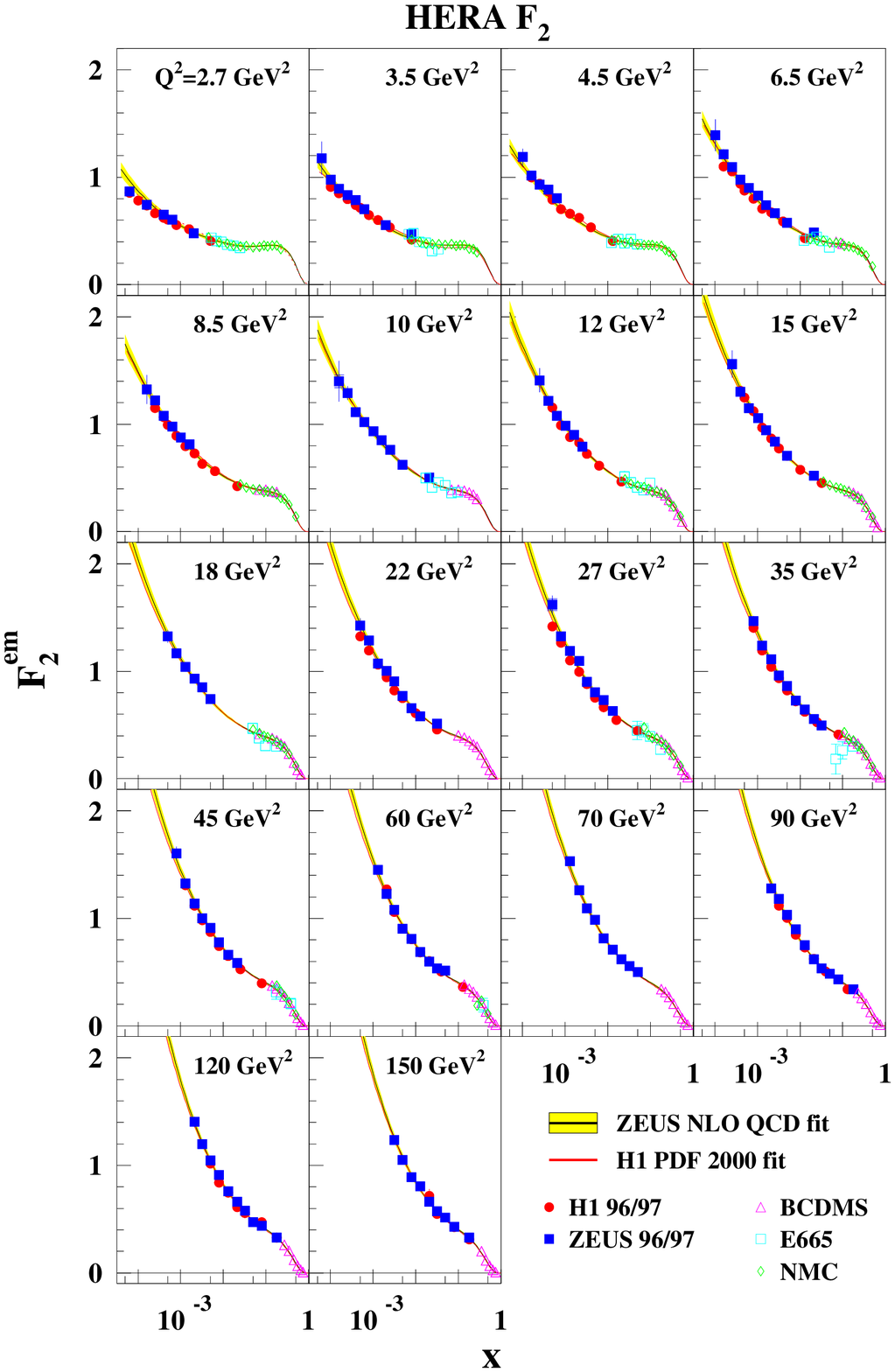,width=6cm,angle=0}
\caption[*]{The plot on the left shows the world data on $F_2$ as a 
function of $Q^2$ for fixed values of $x$. On the right we show the 
converse: $F_2$ as a function of $x$ for fixed values of 
$Q^2$. From~\cite{rizvi}.}  
\label{fig:worldf2}
\end{center}
\end{figure}

In the leading logarithmic approximation of QCD the measured structure 
function $F_2(x,Q^2)$ can be written as 
\begin{eqnarray}
F_2(x,Q^2) = \sum_{\small q=u,d,s,c,b,t} e_q^2 \left(xq(x,Q^2) + 
x{\bar q}(x,Q^2)\right) \, ,
\end{eqnarray}
where $q(x,Q^2)$ (${\bar q}(x,Q^2))$ is the probability density for 
finding a quark (anti-quark) with momentum 
fraction $x$ at a momentum resolution scale $Q^2$; $e_q$ is the
quark charge.

In the simple parton model one has Bjorken scaling, $F_2(x,Q^2)\rightarrow 
F_2(x)$. The ``scaling violations" seen in the $Q^2$-dependence of $F_2(x,Q^2)$ arise from the fact that QCD is not a scale invariant theory and has an 
intrinsic scale $\Lambda_{\rm QCD}\approx 200$ MeV. They are only logarithmic in 
the Bjorken limit of $Q^2\rightarrow \infty$ and $s\rightarrow \infty$ 
with $x\sim Q^2/s$ fixed. 
As one moves away from the asymptotic regime, the scaling violations 
become significant. They can be quantitatively computed in QCD perturbation
theory using for example the machinery of the operator product expansion 
and the renormalization group. The result is most conveniently summarized 
by the Dokshitzer-Gribov-Lipatov-Altarelli-Parisi (DGLAP) evolution 
equations for the parton densities~\cite{dglap,dglap1}:
\begin{equation} \label{dglapeq}
\frac{d}{d \ln Q^2} \left( \begin{array}{c}
\! q \!
\\ \! g \! \end{array} \right)(x,Q^2) 
= \left( \begin{array}{cc}
\! P_{qq}(\alpha_s,x) &  P_{qg}(\alpha_s,x) \! \\
\! P_{gq}(\alpha_s,x) & P_{gg}(\alpha_s,x) \!
\end{array} \right) \;\otimes \;
\left( \begin{array}{c}
\!  q \! \\ \!  g \!
\end{array} \right) \left( x,Q^2 \right) \; ,
\end{equation}
where $\otimes$ denotes a convolution, and the $P_{ij}$ 
are known as ``splitting functions''~\cite{dglap1}
and are evaluated in QCD perturbation theory.
They are now known to three-loop accuracy~\cite{mvv}.
The evolution of the quark densities $q$, $\bar{q}$ involves the 
gluon density $g(x,Q^2)$. The physical picture behind evolution 
is the fact that the virtuality $Q^2$ of the probe sets
a resolution scale for the partons, so that a change in $Q^2$
corresponds to a change in the parton state seen. 
The strategy is then to parameterize the parton distributions at some 
initial scale $Q^2=Q_0^2$, and to determine the parameters by evolving
the parton densities to (usually, larger) $Q^2$ and by comparing to 
experimental data for $F_2(x,Q^2)$.

The pioneering DIS experiments, which first measured Bjorken scaling of $F_2$, 
were performed at SLAC~\cite{section2:1}.
However, because of the (relatively) small energies, these experiments were limited to the region of $x\geq 0.1$. 
With the intense muon beams of CERN and Fermilab, with energies in excess of 100 GeV, the DIS cross-section of the 
proton was measured down to and below $x\sim 10^{-3}$~\cite{section2:2}. In the 1990's, the HERA collider at 
DESY extended the DIS cross-section of the proton to below $x=10^{-4}$~\cite{section2:3,section2:4}. The 
current experimental determination of $F_2^{\rm proton}(x,Q^2)$ extends over 4 orders of magnitude in $x$ and 
$Q^2$. This is shown in Fig.~\ref{fig:worldf2}.
The left panel in Fig.~\ref{fig:worldf2} shows next-to-leading order (NLO) 
QCD global fits by the ZEUS and H1 detector collaborations at HERA
to $F_2$ as a function of $Q^2$ for the 
world DIS data. The data and the QCD fit 
are in excellent agreement over a wide range in $x$ and $Q^2$. In the right panel of 
Fig.~\ref{fig:worldf2}, the $x$ dependence of $F_2$ is shown for different bins in $Q^2$. The rapid rise 
in $F_2$ with decreasing $x$ reflects the sizeable contribution from the sea quark distribution at small $x$. 

\begin{figure}[!h]
\begin{center}
\epsfig{file=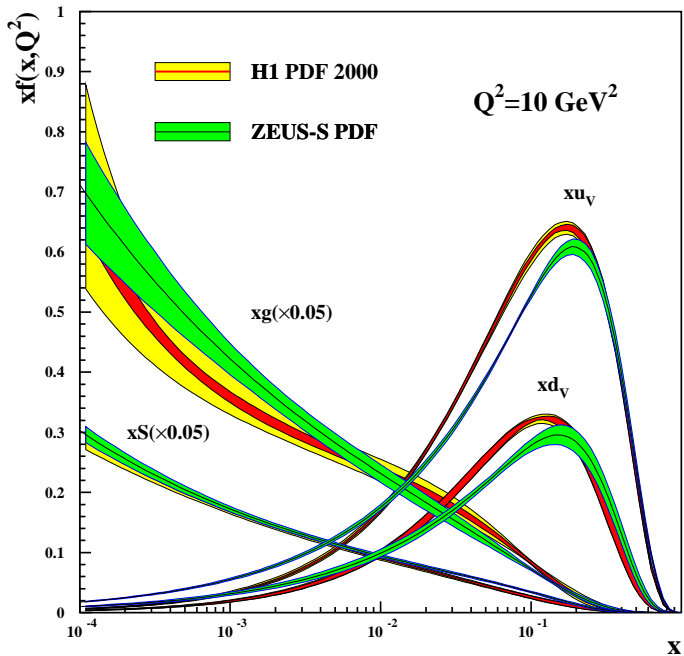,width=10cm,angle=0}
\caption[*]{The valence (up and down) quark, sea quark and gluon distributions plotted as a 
function of $x$ for fixed $Q^2=10$ GeV$^2$. Note that the sea and glue distributions are 
scaled down by a factor of $1/20$. From~\cite{rizvi}.}
\label{fig:flavor}
\end{center}
\end{figure}

In Fig.~\ref{fig:flavor} we show the valence up and down quark distributions as well as the gluon and 
sea quark distributions extracted by the H1 and ZEUS collaborations
as functions of $x$ for fixed $Q^2=10$ GeV$^2$. The valence parton distributions are 
mainly distributed at large $x$ whereas the glue and sea quark distributions dominate hugely at small $x$. 
Indeed, the gluon and sea quark distributions are divided by a factor of 20 to ensure they 
can be shown on the scale of the plot. Already at $x\sim 0.1$, the gluon distribution is nearly a factor of two 
greater than the sum of the up and down quark valence distributions. 

As follows from Eq.~\ref{dglapeq}, 
the gluon distribution in DIS may be extracted from scaling violations of $F_2$: $xg(x,Q^2)\propto 
{\partial F_2(x,Q^2)\over \partial \ln Q^2}$. As one goes to low $Q^2$, 
$xg(x,Q^2)$ becomes small, and some analyses find a preference for a negative
gluon distribution at low $x$, modulo statistical and systematic uncertainties~\cite{section2:8,mrst03}. This is in principle 
not a problem in QCD beyond leading order. However, the resulting longitudinal structure function $F_L$ also comes out close to zero or even negative for $Q^2\sim 2$ GeV$^2$~\footnote{The leading twist expression 
for $F_L$ is simply related to $\alpha_S\,xg(x,Q^2)$.}, which is unphysical
because $F_L$ is a positive-definite quantity. 
A likely explanation for this finding is that contributions to $F_L$ that are suppressed by inverse powers of $Q^2$ 
are playing a significant role at these values of $Q^2$~\cite{BartelsPeters}. These contributions are
commonly referred to as higher twist effects.

\begin{figure}[!h]
\begin{center}
\epsfig{file=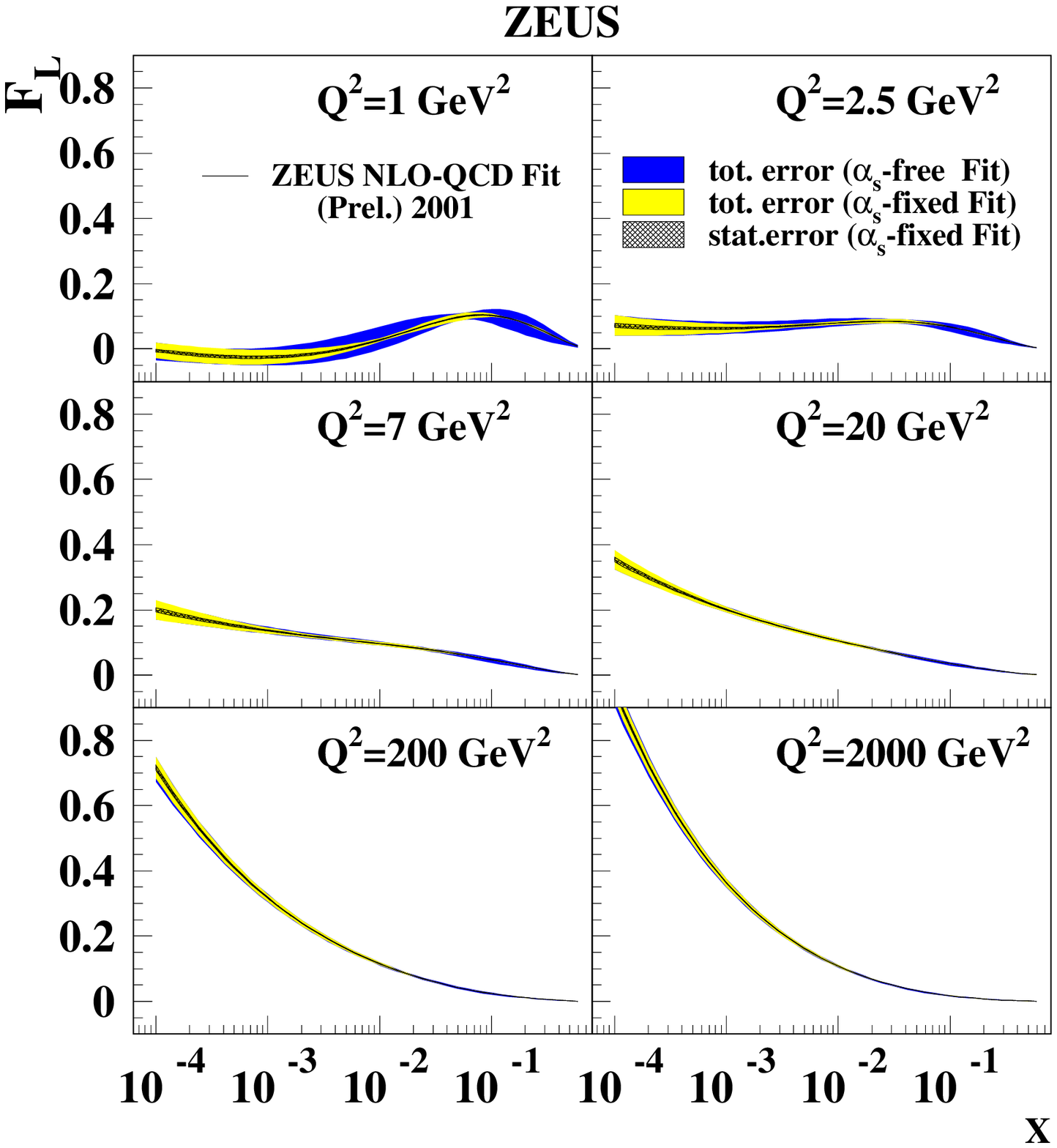,width=6.2cm,angle=0}
\hspace*{0.5cm}
\epsfig{file=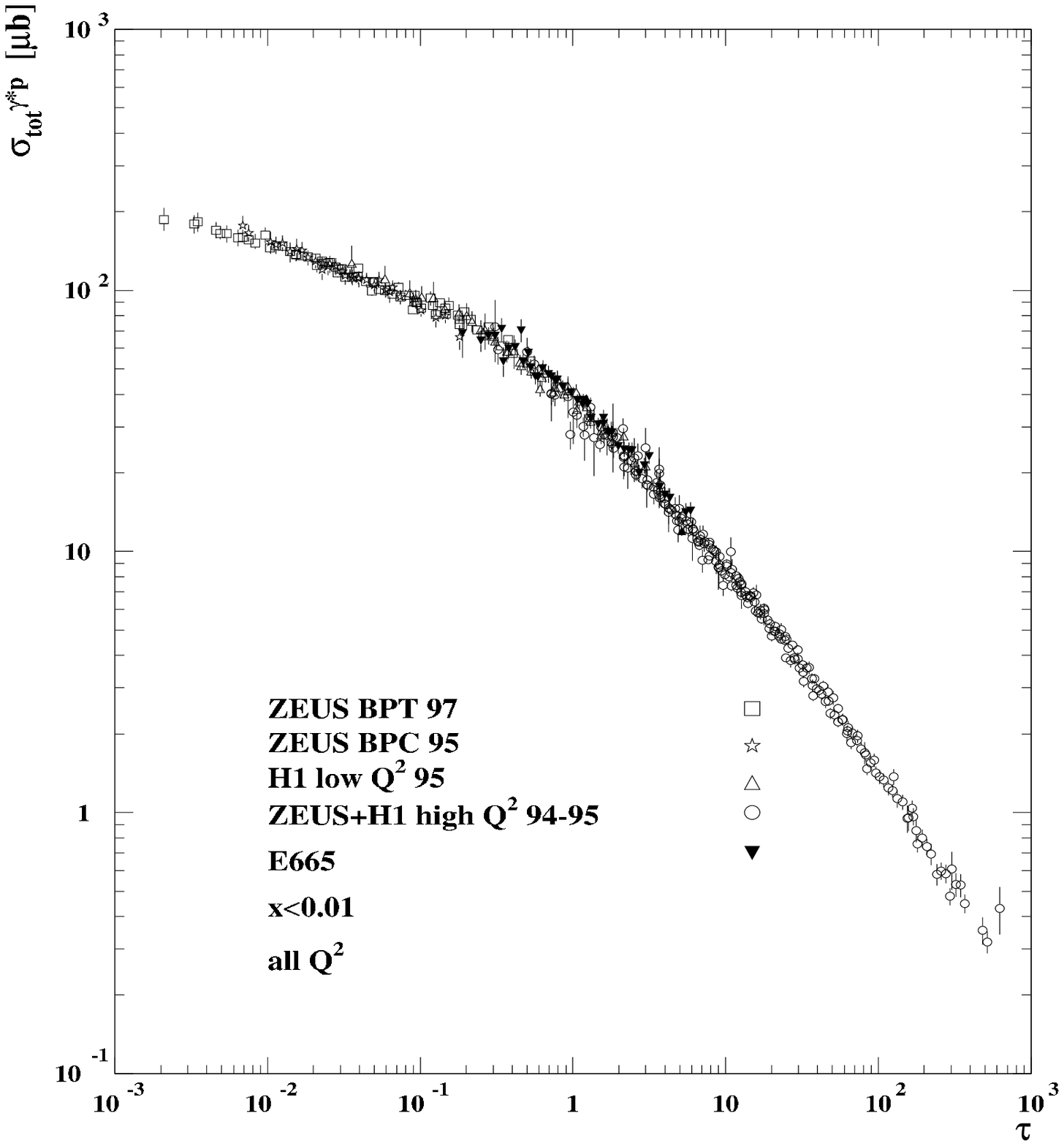,width=6.2cm,angle=0}
\caption[*]{The plot on the left shows the longitudinal structure function $F_L$ as a function of $x$ for different $Q^2$ bins~\cite{section2:8}. On the right (from Ref.~\cite{BSK}) is a plot of the virtual photon-proton cross-section plotted as a function of $\tau =Q^2/Q_s^2$. See text for further explanation.}
\label{fig:geom_scale}
\end{center}
\end{figure}

It has been shown recently~\cite{BSK} that the HERA data on the
virtual photon-proton cross-section ($\sigma^{\gamma^* p} = 4\pi^2
\alpha_{\rm em} F_2(x,Q^2)/Q^2$), for all $x\leq 10^{-2}$ and $0.045
\leq Q^2 < 450$ GeV$^2$, exhibit the phenomenon of ``geometrical scaling''
shown in Fig.~\ref{fig:geom_scale}. The data are shown to scale as a function of $\tau=Q^2/Q_s^2$, where 
$Q_s^2(x)= Q_0^2 (x_0/x)^{-\lambda}$ with $Q_0^2 = 1$ GeV$^2$, $x_0 = 3\cdot 10^{-4}$ and $\lambda\approx  0.3$. The scale $Q_s^2$ is called the saturation 
scale. Geometrical scaling, although very general, is realized in a simple model, the Golec-Biernat-W\"{u}sthoff 
model which includes all twist contributions~\cite{Golec-Biernat}. The model (and variants) provide a phenomenological 
description of the HERA data on diffractive cross-sections and inclusive vector meson production~\cite{Golec1,BartelsGK,KowalskiT,MMS,GoncalvesMachado}. 
The saturation scale and geometrical scaling will be discussed further 
in Section~\ref{science}.

\subsection{Spin structure of the nucleon}
\subsubsection{What we have learned from polarized DIS}\label{g1old}
\begin{figure}[!b]
\hspace*{0mm}
\epsfig{file=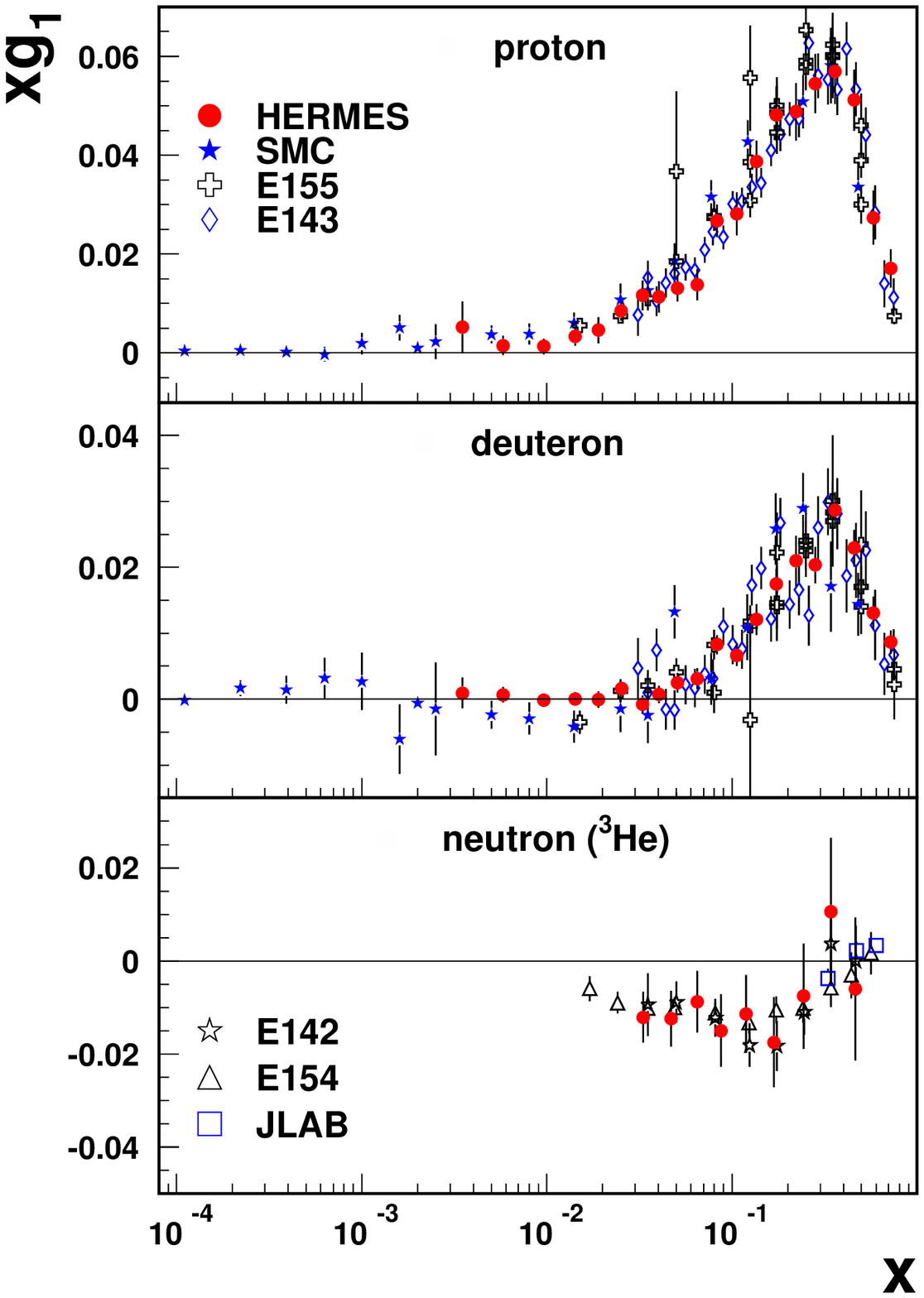,width=5.5cm,angle=0}

\vspace*{-8.cm}
\hspace*{6.5cm}
\epsfig{file=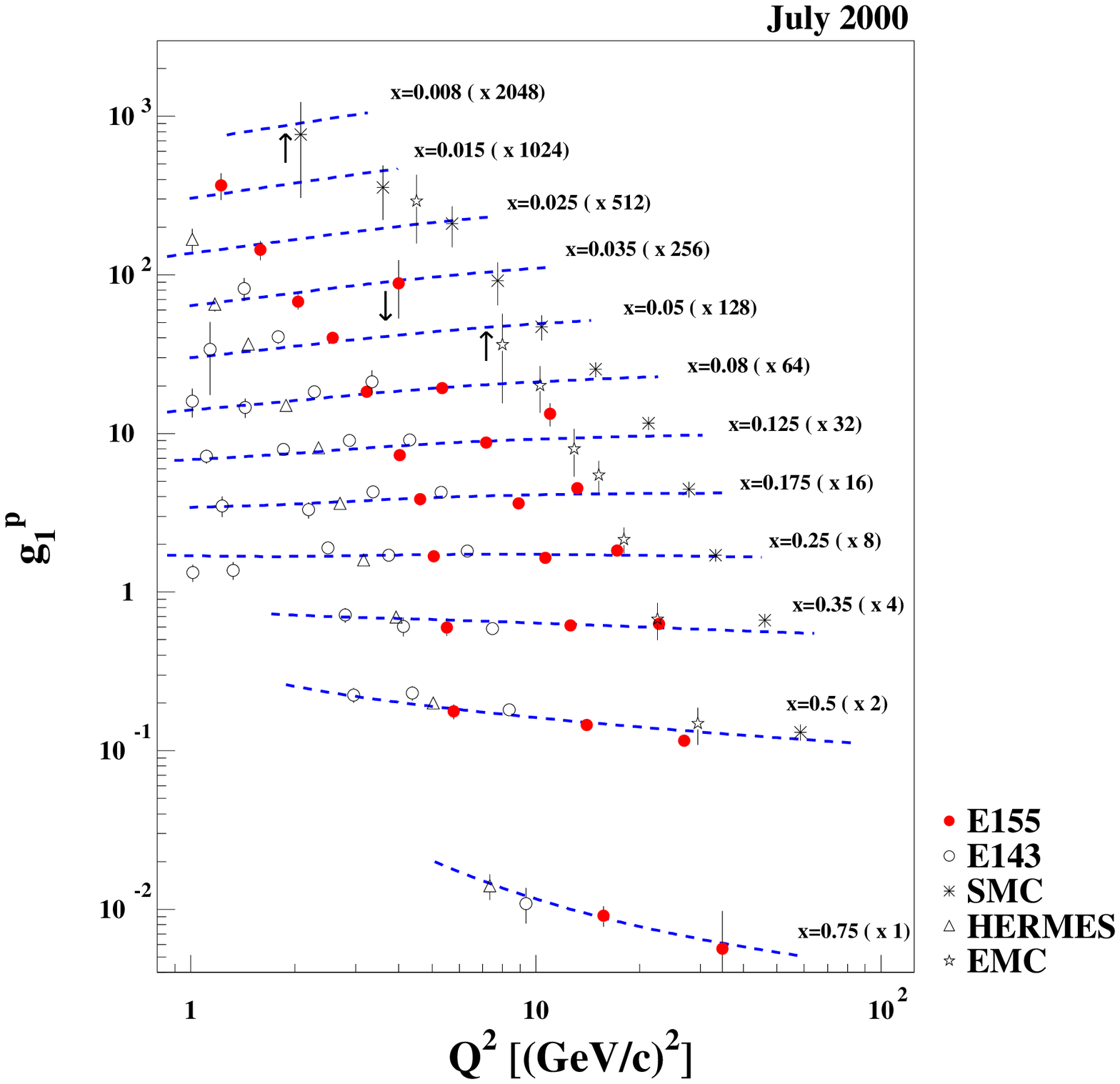,width=6.8cm,angle=0}

\vspace*{12mm}
\caption[*]{Left: world data on the spin structure function
$g_1$ as compiled and shown in \cite{US02}. Right:
$g_1(x,Q^2)$ as a function of $Q^2$ for various $x$. The curves 
are from a phenomenological fit. Taken from~\cite{WATobias}.
\label{fig:g1data}}
\end{figure}

Spin physics has played a prominent role in QCD for several 
decades. The field has been driven by the successful
experimental program of polarized deeply-inelastic lepton-nucleon 
scattering at SLAC, CERN, DESY and the Jefferson Laboratory \cite{hughes}. 
A main focus has been on 
measurements with longitudinally polarized lepton beam and target. 
For leptons with helicity $\lambda$ scattering off nucleons polarized 
parallel or antiparallel to the lepton direction, one has 
\cite{ansel}
\begin{equation}
\frac{d^2\sigma^{\,\lambda\,,\,\Rightarrow}}{dxdQ^2}
-\frac{d^2\sigma^{\,\lambda\,,\,\Leftarrow}}{dxdQ^2} 
\;\propto \;  C\left(G_v,G_a,\lambda\right)\,\left[ \, 
\lambda \, x y (2-y) \, g_1  +
(1-y)  \, g_4  + x y^2  \, g_5 \,  \right] \; ,
\label{longwq}
\end{equation}
where $C(G_v,G_a,\lambda)$ are factors depending on the vector and axial 
couplings of the lepton to the exchanged gauge boson. 
The terms involving $g_4$ and $g_5$ in Eq.~\ref{longwq} are associated with $Z$ and 
$W$ exchange in the DIS process and violate parity. 
In the fixed-target regime, pure-photon exchange strongly dominates,
and scattering off a longitudinally polarized target determines $g_1$. 
Figure~\ref{fig:g1data} (left) shows a recent compilation~\cite{US02} of the 
world data on $g_1(x,Q^2)$, for proton, deuteron, and neutron targets. 
Roughly speaking, $g_1$ is known about as well now as the unpolarized 
$F_2$ was in the mid-eighties, prior to HERA. Figure~\ref{fig:g1data} (right)
shows the measured $Q^2$-dependence of $g_1$; the 
predicted scaling violations are visible in the data.

In leading order of QCD, $g_1$ can be written as 
\begin{equation} \label{g1lo}
g_1 (x,Q^2) = \frac{1}{2} \sum_q e_q^2 \left[ \Delta q(x,Q^2) + \Delta 
\bar{q}(x,Q^2) \right] \; , 
\end{equation}
where
\begin{equation} \label{qdef}
\Delta q  \equiv q^{\rightarrow}_{\Rightarrow} \; - \; 
q^{\leftarrow}_{\Rightarrow}
\,\,\,\,\;\;\;\;\;\, 
(q=u,d,s,\ldots)    \; ,
\end{equation}
$q^{\rightarrow}_{\Rightarrow}$ ($q^{\leftarrow}_{\Rightarrow}$)
denoting the number density of quarks of same (opposite) helicity 
as the nucleon. Clearly, the $\Delta q(x,Q^2),\Delta \bar{q}(x,Q^2)$ contain 
information on the nucleon spin structure. Also in the spin-dependent
case, QCD predicts $Q^2$-dependence of the densities. The associated
evolution equations have the same form as Eq.~\ref{dglapeq}, but
with polarized splitting functions~\cite{dglap1,ahmedross,mvn}. Also,
the spin-dependent gluon density $\Delta g$, defined in analogy 
with Equation~\ref{qdef}, appears. 

\begin{figure}[!h]
\hspace*{20mm}
\epsfig{file=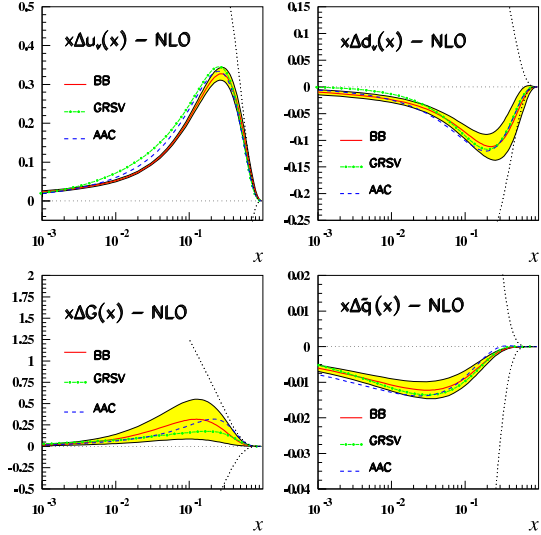,width=8cm,angle=0}
\vspace*{-2.2cm}
\caption[*]{Recent analysis of polarized parton densities of the proton. 
Taken from \cite{bb} 
(``BB''). The additional curves represent the central fits from the analyses 
of~\cite{grsv} (``GRSV'') and~\cite{aac} (``AAC''). 
\label{fig:polpdf}}
\end{figure}

The results of a recent QCD analysis \cite{bb} 
of the data for $g_1(x,Q^2)$ in terms of the polarized
parton densities are shown in Fig.~\ref{fig:polpdf}.
The shaded bands in the figure give estimates of how well
we know the distributions so far. As can be seen, the valence
densities are fairly well known and the sea quark densities to some
lesser extent. This analysis \cite{bb} assumes
flavor-SU(3) symmetry for the sea quarks; the actual uncertainties 
in the individual sea distributions are much larger. Finally, 
Fig.~\ref{fig:polpdf} also shows that we know very little about the 
polarized gluon density. A tendency toward a positive
$\Delta g$ is seen. It is not surprising that the
uncertainty in $\Delta g$ is still large: at LO, $\Delta g$ enters only 
through the $Q^2$-evolution of the structure function $g_1$. 
Because all polarized DIS experiments thus far have been with fixed targets,  
 the lever arm in $Q^2$ has been limited.  This is also 
seen in a comparison of Fig.~\ref{fig:g1data} with Fig.~\ref{fig:worldf2}.

A particular focus in the analysis of $g_1$ has been on the integral 
$\Gamma_1(Q^2)\equiv \int_0^1 g_1(x,Q^2) dx$. 
Ignoring QCD corrections, one has from Eq.~\ref{g1lo}:
\begin{equation} \label{g1nlo1}
\Gamma_1=
\frac{1}{12} \Delta {\cal A}_3 + 
\frac{1}{36}\Delta {\cal A}_8  + \frac{1}{9}\Delta \Sigma   \; ,
\end{equation}
where 
\begin{eqnarray} \label{a3a8s}
\Delta \Sigma &=& \Delta {\cal U}\,+\,\Delta \bar{{\cal U}}\,+\,
\Delta {\cal D}\,+\,\Delta \bar{{\cal D}}\,+\,
\Delta {\cal S}\,+\,\Delta \bar{{\cal S}} \; ,\nonumber \\
\Delta {\cal A}_3&=&\Delta {\cal U}+\,\Delta \bar{{\cal U}}\,-\,
\Delta {\cal D}\,-\,\Delta \bar{{\cal D}}\; ,\nonumber \\
\Delta {\cal A}_8&=&\Delta {\cal U}+\,\Delta \bar{{\cal U}} \,+\,
\Delta {\cal D}\,+\,\Delta \bar{{\cal D}}-2\left(
\Delta {\cal S}\,+\,\Delta \bar{{\cal S}}\right) \; ,
\end{eqnarray}
with $\Delta {\cal Q}=\int_0^1 \Delta q(x,Q^2) dx$, 
which does not evolve with $Q^2$ at lowest order. The flavor 
non-singlet combinations $\Delta {\cal A}_i$ turn out to be proportional
to the nucleon matrix elements of the quark 
non-singlet axial currents, $\langle P,S \,|\, \bar{q} \,
\gamma^{\mu}\, \gamma^5 \,\lambda_i \,q \,|\, P,S \rangle$. 
Such currents typically occur in weak interactions, and by SU(3) rotations
one may relate the matrix elements to the $\beta$-decay 
parameters $F,D$ of the baryon octet~\cite{bjsr,ejsr}. One finds
$\Delta {\cal A}_3=F+D=g_A=1.267$ and $\Delta {\cal A}_8=3F-D\approx
0.58$. The first of these remarkable connections between hadronic
and DIS physics corresponds to the famous Bjorken sum rule~\cite{bjsr}, 
\begin{equation} \label{eqbjsr}
\Gamma_1^p -\Gamma_1^n
 = \frac{1}{6}\Delta {\cal A}_3 \Big[ 1+{\cal O}(\alpha_s)
\Big]= \frac{1}{6}g_A \Big[ 1+{\cal O}(\alpha_s)
\Big] \; ,
\end{equation}
where the superscripts $p$ and $n$ denote the proton and neutron respectively.
The sum rule has been verified experimentally with about $10\%$ 
accuracy \cite{hughes}. The QCD corrections indicated in 
Equation~\ref{eqbjsr} are known \cite{larin} through ${\cal O}(\alpha_s^3)$. 
Assuming the validity of the sum rule, it can be used for a rather 
precise determination of the strong coupling constant \cite{abfr}. 

Determining $\Gamma_1$ from the polarized-DIS data, and using 
the information from $\beta$-decays on $\Delta {\cal A}_3$ and 
$\Delta {\cal A}_8$ as additional input, 
one may determine $\Delta \Sigma$. This quantity is of particular
importance because it measures {\it twice the quark spin contribution to 
the proton spin}. The analysis reveals a small value $\Delta 
\Sigma \approx 0.2$. The experimental finding that the quarks
carry only about 20$\%$ of the proton spin has been one of the 
most remarkable results in the exploration of the structure of
the nucleon. Even though the identification of nucleon with 
parton helicity is not a prediction of QCD (perturbative or 
otherwise) the result came as a major surprise. 
It has sparked tremendous theoretical activity
and has also been the motivation behind a number of dedicated
experiments in QCD spin physics, aimed at further unraveling 
the nucleon spin.

\begin{figure}[!h]
\vspace*{4.6cm}
\begin{center}
\includegraphics{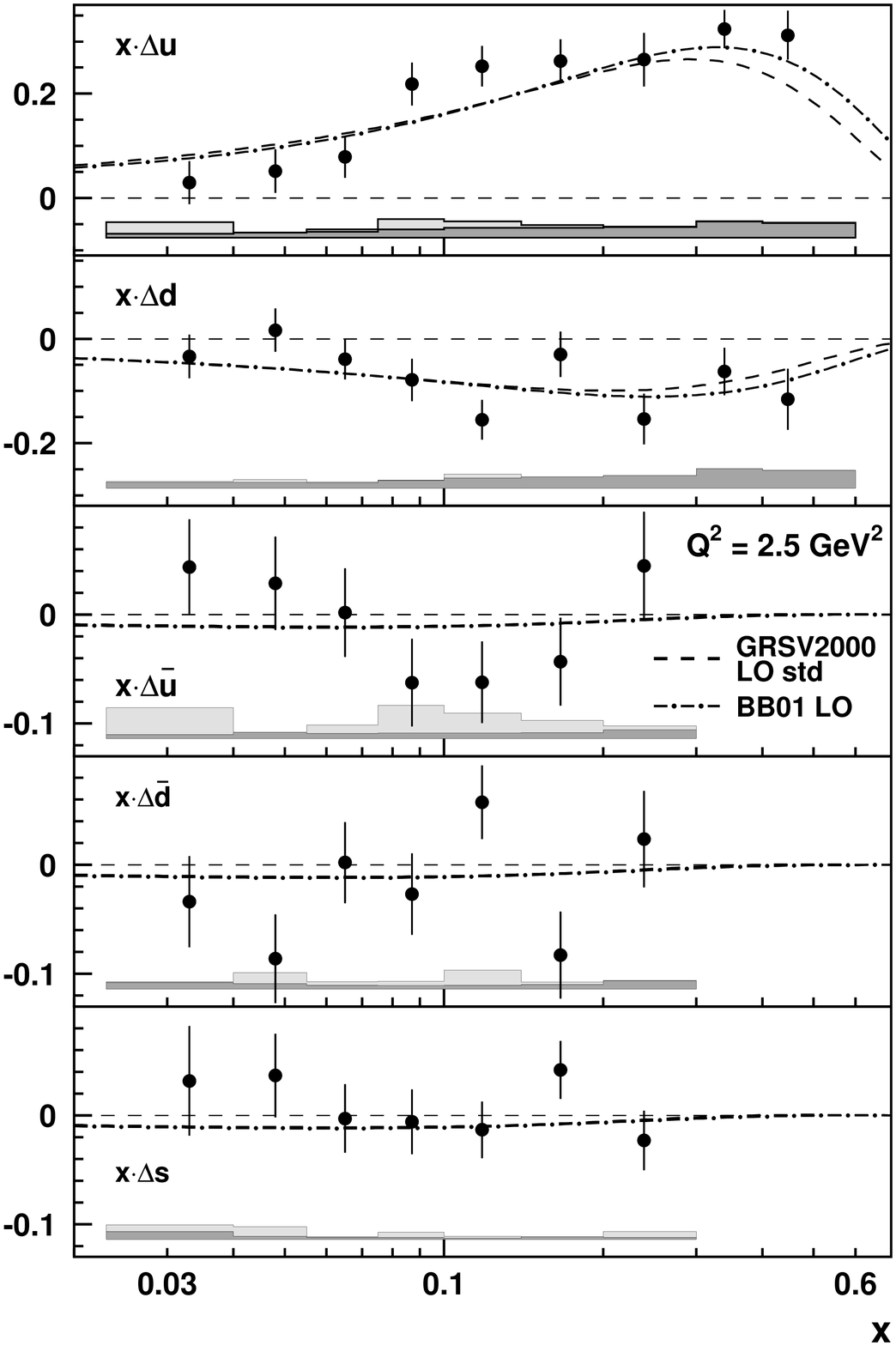}
\vspace*{3.3cm}
\caption[*]{Recent HERMES results \cite{hermesdq} 
for the quark and antiquark
polarizations extracted from semi-inclusive DIS.  \label{fig2.5}}
\vspace*{-1cm}
\end{center}
\end{figure}

A small value for $\Delta \Sigma$ also implies a sizable
negative strange quark polarization in the 
nucleon, $\Delta {\cal S}\,+\,\Delta \bar{{\cal S}}\approx -0.12$.
It would be desirable to have independent experimental
information on this quantity, to eliminate the
uncertainty in the value for $\Delta \Sigma$ due to SU(3) breaking 
effects in the determination of $\Delta {\cal A}_8$ from baryon 
$\beta$ decays \cite{ratcl}. More generally, 
considering Fig.~\ref{fig:polpdf}, 
more information is needed on the polarized sea quark 
distribution functions and their flavor decomposition. 
Such knowledge is also very interesting
for comparisons to model calculations of nucleon structure.
For example, there have been a number of predictions \cite{models} for the 
$\Delta \bar{u}-\Delta \bar{d}$. Progress toward 
achieving a full flavor separation of the nucleon sea
has been made recently, through semi-inclusive measurements in 
DIS (SIDIS) \cite{smc,hermesdq}.  
Inclusive DIS via photon exchange only gives access to the 
combinations $\Delta q+\Delta \bar{q}$, as is evident
from Equation~\ref{g1lo}. If one detects, however, a hadron in the 
final state, the spin-dependent structure function becomes
\begin{equation}
g_1^h (x,z)  \;=\;\frac{1}{2}
\sum_q e_q^2 \left[ \Delta q(x) \;D_q^h(z) + 
\Delta  \bar{q}(x) \;D_{\bar{q}}^h (z) \right] \; .
\end{equation}
Here, the $D_i^h(z)$ are 
fragmentation functions, with $z=E^h/\nu$, where $E^h$ is the energy
of the produced hadron and $\nu$ the energy of the virtual
photon in the Lab frame. Fig.~\ref{fig2.5} shows the 
latest results on the flavor separation by the HERMES
collaboration at HERA \cite{hermesdq}. Uncertainties are still fairly large;  
unfortunately, no further improvements in statistics are expected
from HERMES.  The results are not inconsistent with the large negative 
polarization of $\Delta\bar{u}=\Delta\bar{d}=\Delta \bar{s}$ in 
the sea that has been implemented in many determinations of polarized 
parton distributions from inclusive DIS data (see, e.g., the curves 
in Fig.~\ref{fig:polpdf}). On the other hand, there is no evidence
for a large negative strange quark polarization. 
The results have sparked much renewed theory activity on 
SIDIS~\cite{sidis}. 
We note that at RHIC  $W^{\pm}$ production will be used 
to determine $\Delta u,\Delta \bar{u}, \Delta d,\Delta \bar{d}$ with 
good precision, exploiting the parity-violating couplings of the
$W$ to left-handed quarks and right-handed antiquarks \cite{rhicrev}. 
Comparisons of such data taken at much higher scales with those from 
SIDIS will be extremely interesting. 

A measurement of $\Gamma_1$ 
obviously relies on an estimate of the contribution to the integral 
from $x$ outside the measured region. The extrapolation to small 
$x$ constitutes one main uncertainty in the value of 
$\Delta \Sigma$. As can be seen from Fig.~\ref{fig:g1data}, there
is not much information on $g_1(x,Q^2)$ at $x<0.003$. 
In addition, the data points at the smaller $x$ also have $Q^2$ 
values that are below the DIS regime, making it conceivable that
the ``higher-twist'' contributions to $g_1(x,Q^2)$ are
important and contaminate the extraction of $\Delta \Sigma$.
About half of the data points
shown in Fig.~\ref{fig:g1data} are from the region 
$Q^2 \leq 4$ GeV$^2$, $W^2 =Q^2 (1-x)/x \leq 10$ GeV$^2$,
which in the unpolarized case is usually excluded 
in analyses of parton distribution functions.
Clearly, measurements of polarized DIS and SIDIS at smaller $x$, as well as
at presently available $x$, but higher $Q^2$, will be vital 
for arriving at a definitive understanding of the polarized
quark distributions, and of $\Delta \Sigma$ in particular.

\subsubsection{Contributors to the nucleon spin \label{contrib}}
The partons in the nucleon have to provide the nucleon spin.
When formulating a ``proton spin sum rule'' one has 
in mind the expectation value of the angular momentum 
operator~\cite{orb,orb1},
\begin{equation} \label{spsr}
\frac{1}{2} \;=\; \langle P,1/2\,|\,\hat{J}_3\,|\,P,
1/2\rangle
\;=\;\langle P,1/2\,|\,\int d^3 \,\left[ \vec{x}\times \vec{T}\right]_3 \,|
\,P,1/2\rangle\, ,
\end{equation}
where $T^i\equiv {\cal T}^{0i}$ with ${\cal T}$ the QCD energy-momentum
tensor. Expressing the operator in terms of quark and gluon operators,
one may write:
\begin{equation}
\frac{1}{2}  = \frac{1}{2} \Delta \Sigma +  \Delta G(Q^2) + L_q(Q^2)+ 
L_g(Q^2) \; ,
\end{equation}
where $\Delta G(Q^2)=\int_0^1 \Delta g(x,Q^2)$ is the gluon spin
contribution and the $L_{q,g}$ correspond to orbital angular 
momenta of quarks and gluons. Unlike $\Delta \Sigma$, 
$\Delta G$ and $L_{q,g}$ depend on the resolution scale $Q^2$ already
at lowest order in evolution. The small size of the quark spin 
contribution implies that we must look elsewhere for the proton's
spin: sizable contributions to the nucleon spin should come from 
$\Delta G$ and/or $L_{q,g}$. 

Several current experiments are dedicated to 
a direct determination of $\Delta g(x,Q^2)$. High-transverse momentum 
jet, hadron, and photon final states in polarized $pp$ scattering 
at RHIC offer the best possibilities \cite{rhicrev}.  For example, direct
access to $\Delta g$ is provided by the spin asymmetry for the
reaction $pp\to\gamma X$, owing to the presence of the
QCD Compton process $qg\to \gamma q$. The Spin Muon Collaboration (SMC) and 
COMPASS fixed-target experiments at CERN, and the HERMES 
experiment at DESY, 
access $\Delta g(x,Q^2)$ in charm or high-$p_T$ hadron pair final states
in photon-gluon fusion $\gamma^{\ast}g\to q\bar{q}$ \cite{compass}.
Additional precision measurements with well established
techniques will be needed to determine the integral of the polarized gluon 
distribution, particularly at lower $x$.

Orbital effects are the other candidate for contributions
to the proton spin. Close analysis of the
$\vec{x}\times \vec{T}$ matrix elements in Eq.~\ref{spsr} 
revealed \cite{orb} that they can be measured from a wider class 
of parton distribution functions, the so-called
generalized parton distributions (GPD) \cite{dittes}. These take the general form
$\langle p+\Delta| {\cal O}_{q,g}|p\rangle$, where ${\cal O}_{q,g}$
are suitable quark and gluon operators and $\Delta$ is 
some momentum transfer. The latter is the reason that the GPDs
are also referred to as ``off-forward'' distributions.
The explicit factor $\vec{x}$ in Equation \ref{spsr} forces
one off the forward direction, simply because it requires
a derivative with respect to momentum transfer. This is 
in analogy with the nucleon's Pauli form factor.  In 
fact, matrix elements of the above form interpolate between
DIS structure functions and elastic form factors.

To be more specific \cite{gpdrev}, the {\em total} (spin plus orbital)
angular momentum contribution of a quark to the nucleon spin
is given as \cite{orb}
\begin{equation} \label{Jtot}
J_q\;=\; \frac{1}{2} \;\lim_{\Delta^2\to 0}\;
\int dx \, x\,\left[ H_q (x,\xi,\Delta^2)+E_q (x,\xi,\Delta^2) \right] \; .
\end{equation}
Here, $\xi=\Delta^+/P^+$, where the light-cone momentum 
$\Delta^+\equiv \Delta^0+\Delta^z$, and likewise
for $P^+$. $H_q, E_q$ are defined as form factors of the matrix element \linebreak
$\int dy \, {\rm e}^{iy x}
\; \langle P'\,|\;\overline{\psi}_+(y)
\; \psi_+
\left(0\right)\;|\,P\rangle$.
$H_q$ reduces to the ordinary (forward) quark distribution in 
the limit $\Delta\to 0$, $H_q(x,0,0)=q(x)$, whereas the first moments (in $x$)
of $H_q$ and $E_q$ give the quark's contributions to the nucleon Dirac and 
Pauli form factors, respectively. In addition, Fourier 
transforms of $H_q, E_q$ with respect to the transverse components
of the momentum transfer $\Delta$ give information on the position 
space distributions of partons in the nucleon \cite{ft}, for example:
\begin{equation}
H_q(x,\xi=0,-\vec{\Delta}_{\perp}^2)=\int d^2\vec{b} \,
{\rm e}^{-i \vec{\Delta}_{\perp}\,  \vec{b}} \, q(x,b) \; .
\end{equation}
$q(x,b)$ is the probability density for finding a quark
with momentum fraction $x$ at transverse distance $\vec{b}$
from the center. It thus gives a transverse profile of the
nucleon. GPDs, therefore, may give us remarkably deep new
insight into the nucleon. 

\begin{figure}[h!]
\vspace*{-0.2cm}
\hspace*{0mm}
\epsfig{file=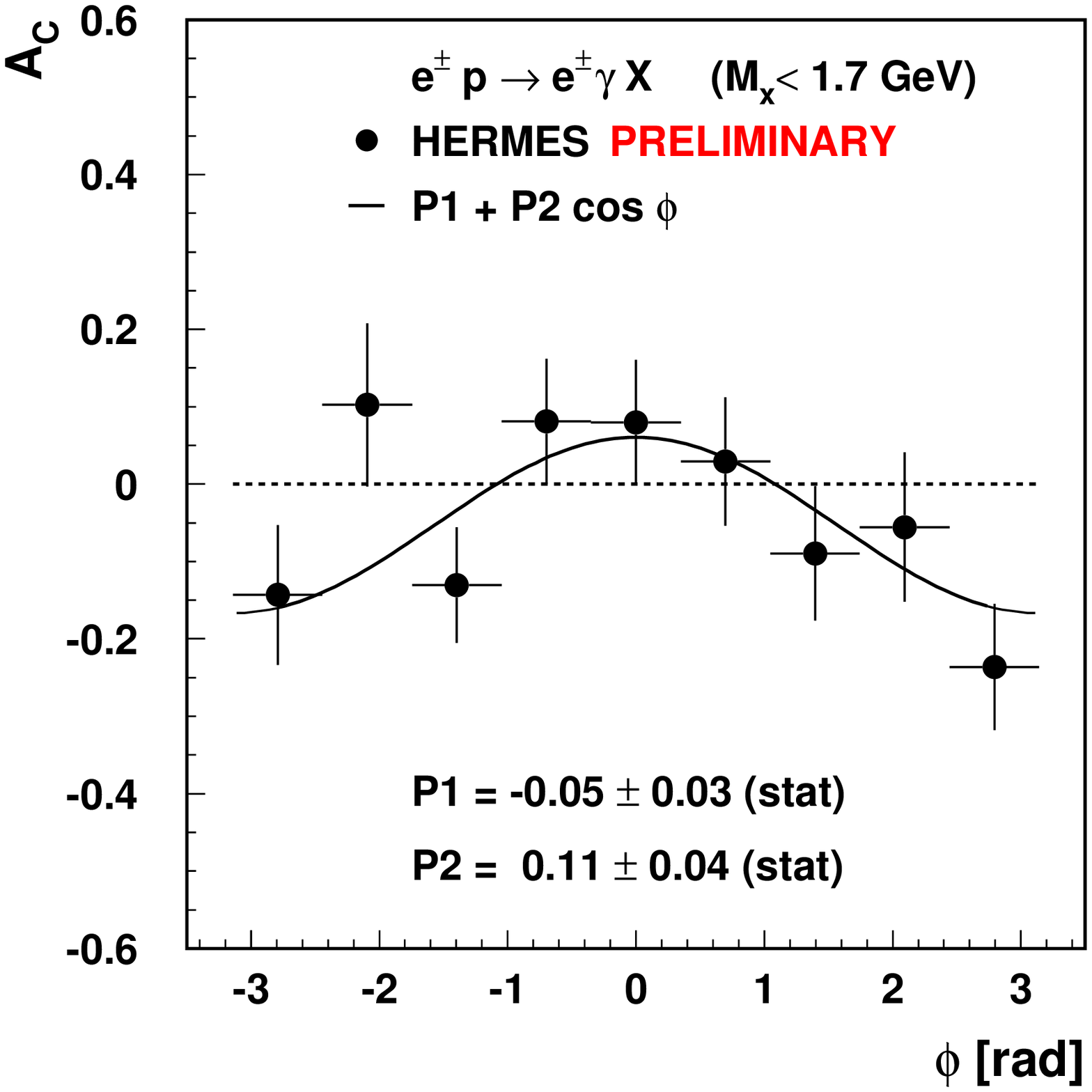,width=6cm,angle=0}

\vspace*{-6.2cm}
\hspace*{7.cm}
\epsfig{file=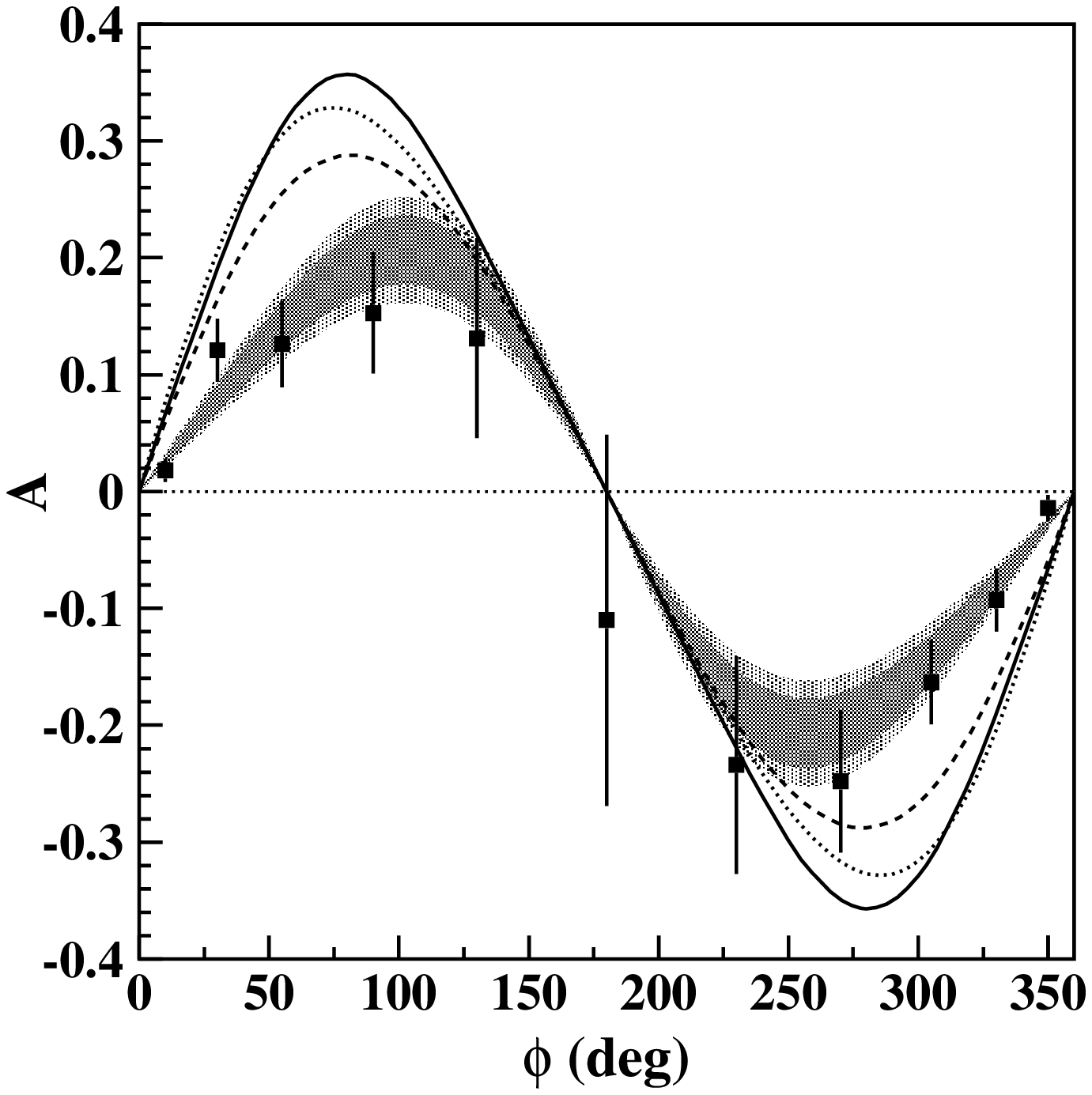,width=6cm,angle=0}
\caption{\label{fig:DVCS-SSA} Data for the beam charge asymmetry
in DVCS from HERMES~\cite{hermdvcs1} 
(left) and for the beam spin asymmetry from 
CLAS~\cite{clasdvcs} (right), as functions of the azimuthal angle $\phi$. 
For the definitions of these asymmetries, see text.}
\label{fig:dvcs}
\end{figure}

The classic reaction for a measurement of the $H_q, E_q$ 
is ``deeply virtual Compton scattering (DVCS)'', $\gamma^* p \rightarrow 
\gamma p$ \cite{orb}. It is the theoretically best explored and 
understood reaction \cite{dvcsth}. Next-to-leading order calculations
are available~\cite{FMcD}.
The GPDs contribute to the reaction at amplitude level. 
The amplitude for DVCS interferes with that for the Bethe-Heitler 
process. The pure Bethe-Heitler part of the differential $ep$ 
cross section is calculable and can in principle be subtracted, 
provided it does not dominate too strongly.  Such a subtraction 
has been performed in DVCS measurements at small $x$ by H1 and ZEUS
\cite{dvcshera}.  A different possibility to eliminate 
the Bethe-Heitler contribution is to take the difference of cross 
sections for opposite beam or target polarization.  In both cases, 
contributions from Compton scattering and the Compton-Bethe-Heitler 
interference survive.  The cleanest separation of these pieces can 
be achieved in experiments with lepton beams of either charge. Because the 
Compton contribution to the $ep$ amplitude is linear and the 
Bethe-Heitler contribution quadratic in the lepton charge, the 
interference term is projected out in the difference $d\sigma(e^+ p) - 
d \sigma(e^- p)$ of cross sections, whereas it is absent in their sum.
Both the ``beam-spin'' asymmetry 
\begin{equation} 
\frac{d\sigma_+(e^- p) -d\sigma_-(e^- p)}{d\sigma_+(e^- p) +d\sigma_-(e^- p)}\; ,
\end{equation}
where $\pm$ denote positive (negative) beam helicities, and the ``beam-charge'' asymmetry
\begin{equation} 
\frac{d\sigma(e^+ p) -d\sigma(e^- p)}{d\sigma(e^+ p) +d\sigma(e^- p)}
\end{equation}
have been observed \cite{hermdvcs1,hermdvcs2,clasdvcs}. Fig.~\ref{fig:dvcs} 
shows some of the results.

Hard exclusive meson production, $\gamma^* p\to M p$, 
is another process that gives access to GPDs, and much activity
has gone into this direction as well \cite{gpdrev,cfs}. Both DVCS and 
exclusive meson production have their practical advantages
and disadvantages. Real photon production is cleaner, but the
price to be paid is an additional power of $\alpha_{\rm em}$. 
Meson production may be easier to detect; however, its amplitude
is suppressed relatively by a power $1/Q$. The importance of using
nucleon polarization in off-forward reactions is well established.
There have also been first studies for DVCS off nuclei \cite{dvcsnucl}.

Practical problems are the fact that GPDs depend
on three variables (plus a scale in which they evolve), and
that they appear in complicated convolutions with the partonic
hard-scattering kernels.
We are still far from the quantitative experimental surveys of 
DVCS and related processes that would allow us to work backwards 
to new insights into off-diagonal matrix elements and angular 
momentum. Nevertheless, a direction for the field has been set.

\subsubsection{Transverse polarization \label{subsectr}}
In addition to the unpolarized and the helicity-dependent
distributions, there is a third set of twist-2 parton 
distributions, namely transversity \cite{rs}. In analogy with Equation~\ref{qdef}
these distributions 
measure the net number (parallel minus antiparallel) 
of partons with transverse polarization in a transversely
polarized nucleon:
\begin{equation}
\delta q(x)=q^{\uparrow}_{\Uparrow}(x) - q^{\downarrow}_{\Uparrow} (x)
\label{qtrdef}\; .
\end{equation}
In a helicity basis \cite{rs}, transversity 
corresponds to an interference of an amplitude in which a
helicity-$+$ quark emerges from a helicity-$+$ nucleon, but is
returned as a quark of negative helicity into a nucleon of
negative helicity. This helicity-flip structure makes transversity 
a probe of chiral symmetry 
breaking in QCD \cite{collins}. Perturbative-QCD interactions 
preserve chirality, and so the helicity flip must
primarily come from soft non-perturbative interactions for which 
chiral symmetry is broken. The required helicity flip
also precludes a gluon transversity distribution at 
leading twist \cite{rs}.

Measurements of transversity are not straightforward. Again
the fact that perturbative interactions in the Standard
Model do not change chirality (or, for massless quarks, helicity) 
means that inclusive DIS is not useful. Collins, however, 
showed \cite{coll93} that properties of fragmentation 
might be exploited to obtain a ``transversity polarimeter'': 
a pion produced in fragmentation will have some transverse 
momentum with respect to the fragmenting parent quark. 
There may then be a correlation of the form $\;i\vec{S}_T \,   
(\vec{P}_{\pi} \times \vec{k}_{\perp})$ among the transverse spin 
$\vec{S}_T$ of the fragmenting quark, the pion momentum $\vec{P}_{\pi}$, and
the transverse momentum $\vec{k}_{\perp}$ of the quark relative to the pion.
The fragmentation function associated with this correlation
is denoted as $H_1^{\perp,q}(z)$, the Collins function. 
If non-vanishing, the Collins function makes a 
{\it leading-power} \cite{coll93,jmy,goeke} contribution to the single-spin 
asymmetry $A_{\perp}$ in the reaction $\;ep^{\uparrow}
\to e\pi X$:
\begin{equation} 
A_{\perp}\propto|\vec{S}_T|
\sin(\phi+\phi_S)\sum_q\,e_q^2
\delta q(x)H_1^{\perp,q}(z) \; ,\label{eq3}
\end{equation}
where $\phi$ ($\phi_S$) is the angle between the lepton plane
and the $(\gamma^* \pi)$ plane (and the transverse target spin).
As shown in Equation~\ref{eq3}, this asymmetry would then
allow access to transversity.

If ``intrinsic'' transverse momentum in the fragmentation
process can play a crucial role in the asymmetry for 
$\;ep^{\uparrow} \to e\pi X$, a natural question is whether 
$k_{\perp}$ in the initial state can be relevant as well. 
Sivers suggested \cite{sivers} that the $k_{\perp}$ distribution 
of a quark in a transversely polarized hadron could have an 
azimuthal asymmetry, $\,\vec{S}_T \,   (\vec{P} \times 
\vec{k}_{\perp})$. It was realized \cite{BHS,gauge} that the Wilson 
lines in the operators defining the Sivers function, required by gauge 
invariance, are crucial for the function 
to be non-vanishing. This intriguing discovery has been one of 
the most important theoretical developments in QCD spin physics 
in the past years. 
Another important aspect of the Sivers 
function is that it arises as an interference of wave functions 
with angular momenta $J_z=\pm 1/2$ and hence contains information
on parton orbital angular momentum \cite{BHS,sivoam},
complementary to that obtainable from DVCS. 

Model calculations and
phenomenological studies of the Sivers functions $f_{1T}^{\perp ,q}$ have 
been presented \cite{models1}. It makes a contribution to
$\;ep^{\uparrow}\to e\pi X$~\cite{jmy},
\begin{equation} 
A_{\perp}\propto|\vec{S}_T|
\sin(\phi-\phi_S)\;\sum_q\,e_q^2\;
f_{1T}^{\perp ,q}(x)\;D_q^{\pi}(z)  \; .\label{eq4}
\end{equation}
This is in competition with the Collins function contribution,
Equation~\ref{eq3}; however, the azimuthal angular dependence
is discernibly different. HERMES has completed a run
with transverse polarization and performed an extraction of
the contributions from the Sivers and Collins 
effects \cite{hermessivcoll}. There are also first results from
COMPASS~\cite{compasstr}. First independent information on
the Collins functions is now coming from Belle measurements in 
$e^+ e^-$ annihilation~\cite{bellecoll}. 
The Collins and Sivers functions are
also likely involved~\cite{sivcollpp} in explanations of experimental
observations of very large single-transverse spin asymmetries
in $pp$ scattering~\cite{singlepp}, where none were expected.
It was pointed out \cite{gauge,bv} that comparisons of DIS results 
and results from $p^{\uparrow}p$ scattering at RHIC 
will be particularly interesting: from the properties of 
the Wilson lines it follows that the Sivers functions violate 
universality of the distribution functions. For example,
the Sivers functions relevant in DIS and in the 
Drell-Yan process should have opposite sign. This is a striking 
prediction awaiting experimental testing.

\subsection{Nuclear modifications}

\begin{figure}[!h]
\hspace*{32mm}
\epsfig{file=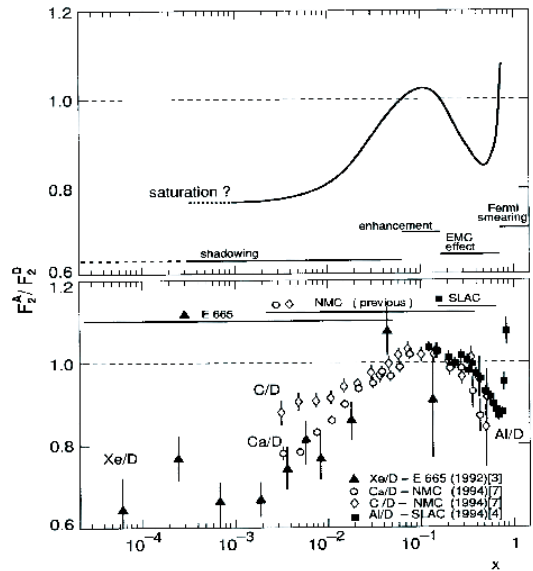,width=8cm,angle=0}
\vspace*{4mm}

\caption[*]{Upper: An idealized depiction of the ratio of the structure function 
of a nucleus, $F_2^A(x,Q^2)$ per nucleon to $F_2^d(x,Q^2)$ of deuterium. 
Lower: Measured $F_2(x,Q^2)$ structure functions for C, 
Ca, and Xe relative to Deuterium. From~\cite{Arneodo}.}
\label{fig:F2shadow}
\end{figure}

The nucleus  is traditionally described as a collection of  weakly bound nucleons 
confined in a potential created by their mutual interaction.  It came as a surprise 
when the EMC experiment~\cite{section2:20} uncovered a systematic nuclear dependence 
to the nuclear structure function $F_2^A (x,Q^2)$ in iron relative to that for 
Deuterium because the effect was as much as  20\% for 
$x \sim 0.5$.  This is significantly larger than the  effect ($< 5$\%) due to the 
natural scale for nuclear effects given by the ratio of the binding energy per 
nucleon to the nucleon mass.  Several dedicated fixed target 
experiments~\cite{section2:21,section2:22,section2:23} confirmed the existence of the 
nuclear dependence observed by the EMC albeit with significant modifications of the 
original EMC results at small $x$.  The upper part of Figure~\ref{fig:F2shadow} 
shows an idealized version of the nuclear modification of the relative structure 
functions per nucleon.  It is $2/A$ times the ratio of a measured nuclear structure 
function of nucleus A to that for Deuterium.  The rise at the largest values of $x$ 
is ascribed to the 
nucleons' Fermi momentum.  The region above $x \geq 0.2$ is referred to as the EMC 
effect region.  When $x\leq 0.05$, the nuclear ratio drops below one and the region 
is referred to as the nuclear shadowing region, 
whereas the region with the slight enhancement in between the shadowing and EMC effect 
regions is called the anti-shadowing region. The lower part 
of Figure~\ref{fig:F2shadow} presents a sample of high precision data of ratios of  
structure functions over a broad range in $A$, $x$ and $Q^2$.  We shall now discuss 
what is known about these regions, focusing in particular on the EMC effect and nuclear 
shadowing regions.
\subsubsection{The EMC effect}

\begin{figure}[!h]
\hspace*{25mm}
\epsfig{file=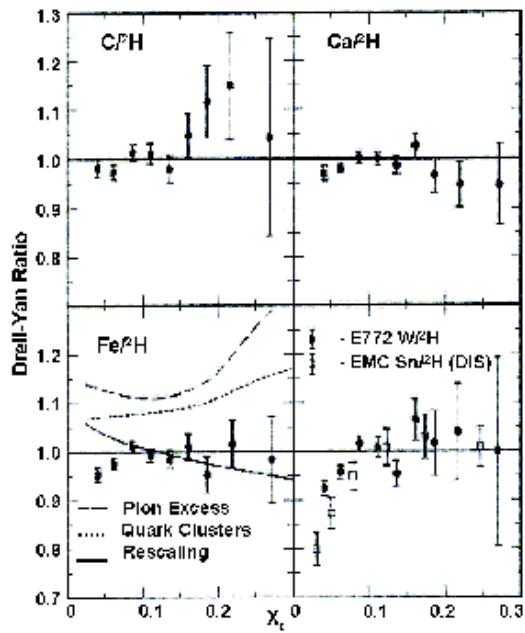,width=7cm,angle=0}
\caption[*]{The ratio of the anti-quark distribution per nucleon in several nuclei relative to  Deuterium. Data are 
shown from a Drell-Yan experiment~\cite{section2:27} 
and compared to theoretical predictions. Also shown is the ratio for Tungsten 
to Deuterium compared to DIS data from the EMC experiment~\cite{section2:20} for the $F_2$ ratio of Tin to Deuterium. 
\label{fig:DYshadow}}
\end{figure}

A  review of the DIS data and various interpretations of the EMC effect since its discovery in the early 1980's can be found in Ref.~\cite{section2:24}.   A common interpretation of the EMC effect is based on models where inter-nucleon interactions at a wide range of inter-nucleon distances are mediated by meson exchanges.  The traditional theory~\cite{section2:25,section2:26} of nuclear interactions predicts a net increase in the distribution of virtual pions with increasing nuclear density relative to that of free nucleons. This is because meson interactions are attractive in nuclei.  In these models, nuclear pions may carry about 5\% of the total momentum to fit the EMC effect at $x \sim 0.3$.  Each pion carries a light-cone fraction of about
 0.2-0.3 of that for a nucleon.  Sea anti-quarks belonging to these nuclear pions may scatter off a hard probe. Hence the predicted enhancement of the nuclear sea of 10\% to 15\% for  $x \sim 0.1$--0.2 and for $A\geq 40$.   The conventional view of nuclear binding is challenged by the constancy with $A$ of the anti-quark distribution extracted from the production of Drell-Yan pairs in proton-nucleus collisions at Fermilab~\cite{section2:27}.  These data are 
shown in Fig.~\ref{fig:DYshadow}. No enhancement was observed at 
the level of 1\%  accuracy in the Drell-Yan experiments. The Drell-Yan data 
was also compared with and showed good agreement with the DIS EMC data for 
the $F_2$ ratio of Tin to Deuterium. 
  
Furthermore, first results~\cite{section2:28} from the
Jefferson Laboratory (TJNAF) experiment E91-003 indicate that 
there is no significant pion excess in the $A (e, e^\prime, B)$ reaction.
(It has however been pointed out ~\cite{section2:29}  that parameters of pion interactions in nuclei can be readily adjusted to reduce the pion excess to conform with the Drell-Yan data.)  In addition, the energy excitation for the residual nuclear system also reduces the contribution of pions to the nuclear parton densities~\cite{section2:30}. Thus all of these observations suggest that pions may not contribute significantly to $F_2^A$ in the EMC region.  
  
The chiral quark-soliton model~\cite{models} is a  phenomenologically successful model that for instance explains the difference in the anti-quark up and down distributions as a function of $x$~\cite{SmithMiller}. 
Interestingly, it has been shown recently~\cite{SmithMiller} to simultaneously provide a good  description of both the EMC effect and the ratio of anti-quark distributions from Drell-Yan pairs. Recently, it has been argued that a key feature of the EMC effect, the factorization 
of the $x$ and $A$ dependence of the EMC ratio, 
can be understood in a model independent way in an 
effective field theory approach~\cite{ChenDetmold}. 
Several joint leading order QCD analyses of the nuclear DIS and Drell-Yan data combined with the application of the baryon charge and momentum sum 
rules~\cite{section2:31,section2:32,section2:33,EKST} provide 
further information on the nuclear effects on  parton densities in this kinematic region.  These analyses indicate that the valence quark distribution in nuclei is enhanced at $x \sim 0.1$--0.2.  Gluons in nuclei carry practically the same fraction of the momentum (within  1\%)  as in a free nucleon.  If one assumes that  gluon shadowing is similar to that for quarks, these analyses predict a significant enhancement of the gluon distribution  in nuclei at $x \sim 0.1$--0.2~\cite{EHKS}. A recent next-to-leading order (NLO)
analysis of nuclear parton distributions~\cite{DS} however finds that 
this gluon ``anti-shadowing" is much smaller than in the LO analysis.  

\subsubsection{Nuclear shadowing}
Nuclear shadowing is the phenomenon, shown in Fig.~\ref{fig:F2shadow}, where the ratio of the nuclear electromagnetic structure function 
$F_2^A$ relative to $A/2$ times the Deuteron electromagnetic structure functions
$F_2^D$ is less than unity for $x\leq 0.05$. Shadowing is greater for decreasing $x$ and with increasing nuclear size. For moderately small $x$, 
 shadowing is observed to decrease slowly with increasing $Q^2$. Unfortunately, because $x$ and $Q^2$ are inversely correlated for fixed energies, much of the very 
small $x$ data ($x\leq 10^{-3}$) is at very low values of 
 $Q^2\leq 1$ GeV$^2$. In addition, as results~\cite{Arneodo} from the fixed target E665 experiment at Fermilab 
and the New Muon Collaboration (NMC) experiment at CERN 
shown in Fig.~\ref{fig:F2shadow}(b) suggest, good quality data exists only for $x > 4\times 10^{-3}$. At high $Q^2$, the shadowing  of $F_2^A$ can be interpreted in terms of shadowing of quark and anti-quark distributions in nuclei at small $x$. Information on quark shadowing can also be obtained from proton-nucleus Drell-Yan 
 experiments~\cite{E772} and from neutrino-nucleus experiments--most recently from NuTeV at Fermilab~\cite{NuTeV}. 
   
\begin{figure}[!t]
\vspace*{15mm}
\epsfig{file=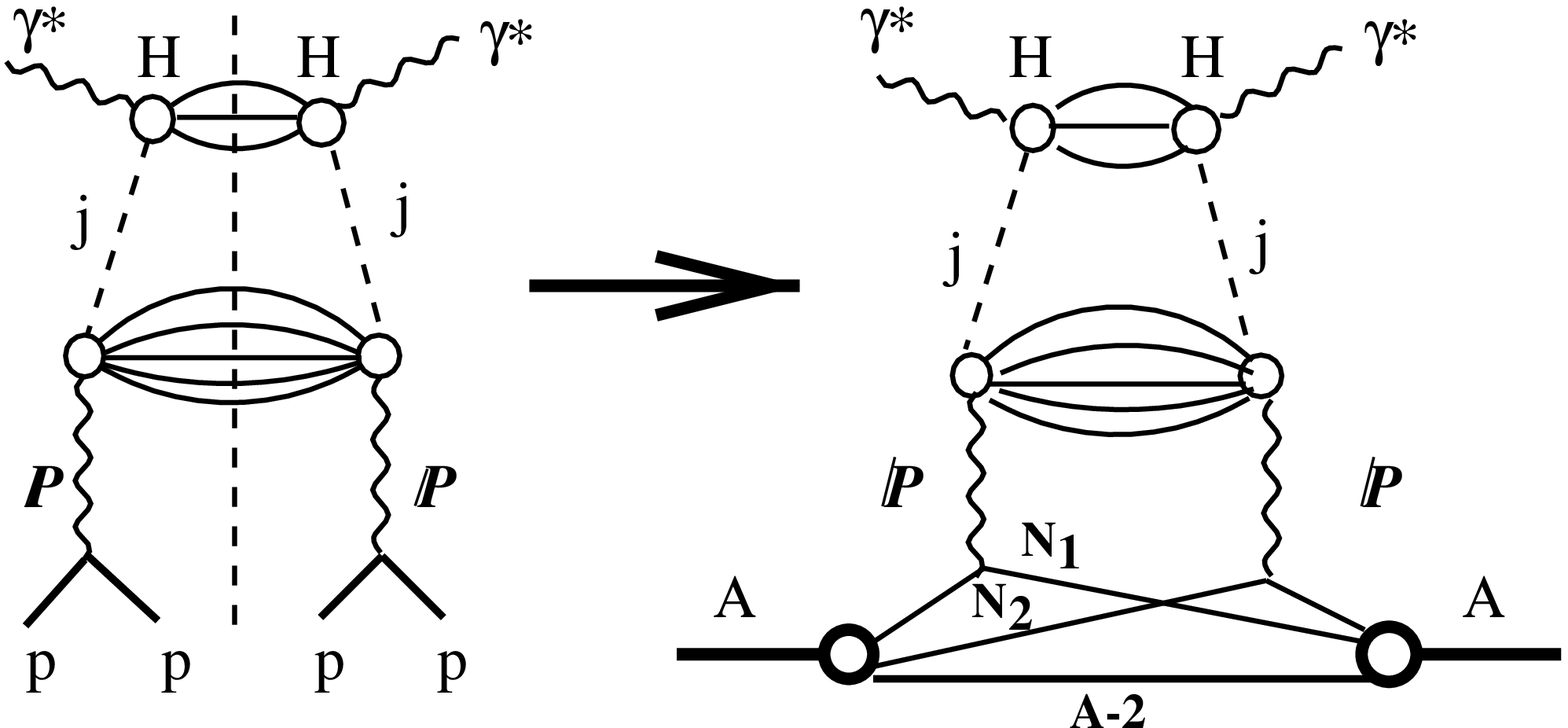,width=6.5cm,angle=0}

\vspace*{-4.5cm}
\hspace*{7.5cm}
\epsfig{file=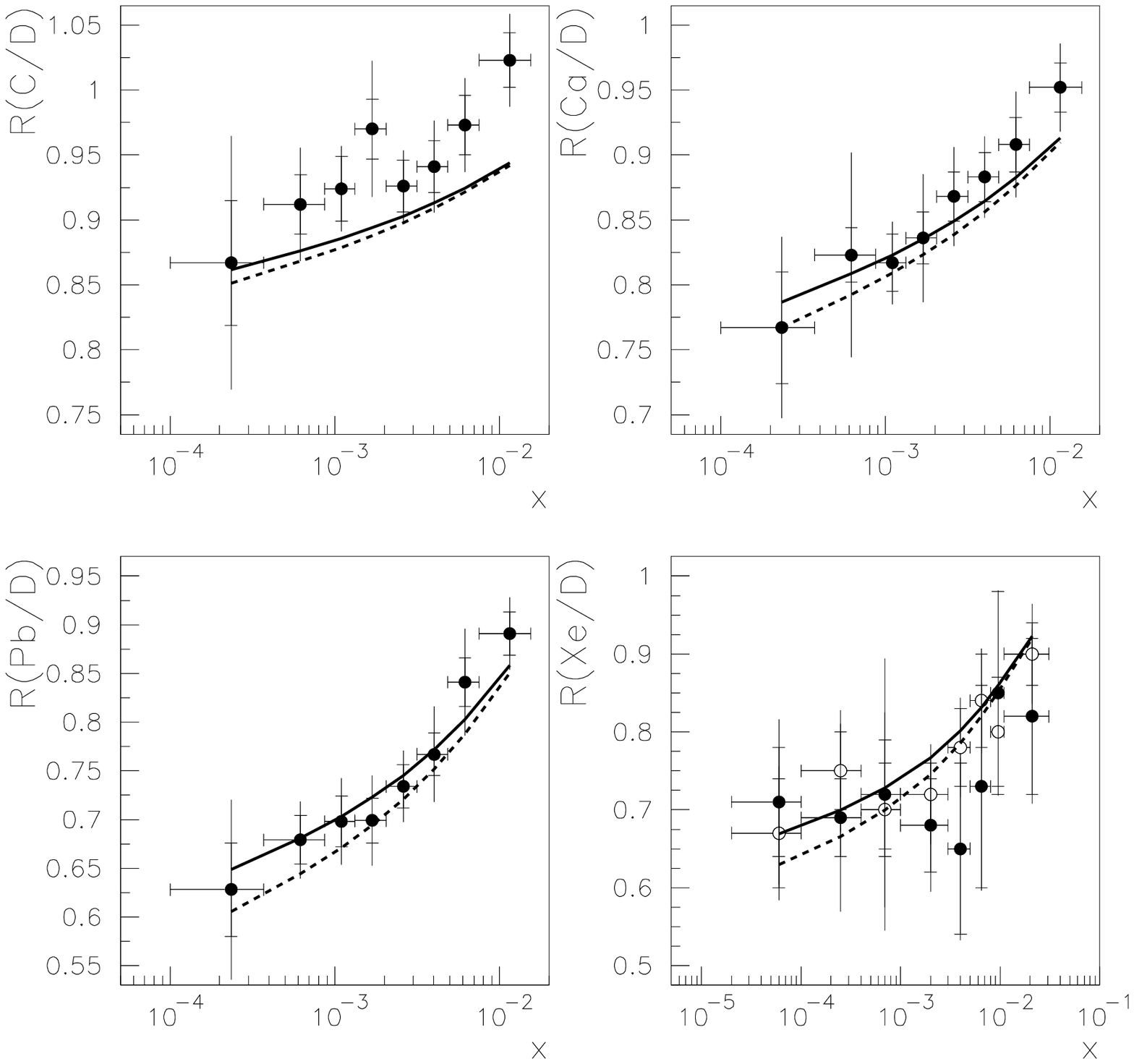,width=6cm,angle=0}
\vspace*{4mm}
\caption[*]{Left: An illustration of the AGK rules relating shadowing 
corrections in nuclei to diffractive scattering on nucleons. Right: A calculation~\cite{CapellaKAS} which uses AGK to fit nuclear $F_2$ 
data using HERA diffractive data. The two curves correspond to two different unitarization 
prescriptions.}
\label{fig:ShadDiff}
\end{figure}

The phenomenon of shadowing has different interpretations depending on the frame in which we consider the space-time evolution of the scattering. Consider for instance the rest frame of a nucleus in $\gamma$-p/A 
 scattering.  The $\gamma$p cross-section is only 0.1mb for energies in excess of 2 GeV, 
corresponding to a mean-free-path of well over 100 fm in nuclear matter.  However, although the high-energy $\gamma$A cross-section might be expected to be proportional to $A$,  the observed increase in the cross-section is smaller than $A$ times the $\gamma$p cross-section.  This is because the photon can fluctuate into a  $q\bar q$-- pair that has a cross-section typical of the strong interactions ($\sim 20$mb) and is absorbed readily (with a mean free path of $\sim 3.5$ fm).  If the fluctuation persists over a length greater than the inter-nucleon separation distance (2 fm), its absorption ÓshadowsÔ it from encountering subsequent nucleons. The coherence length of the virtual photon's 
fluctuation is $l_{\rm coh.}\sim 1/2m_N x$ where $m_N$ is the nucleon mass.  Therefore the onset of shadowing is expected and observed at $x\approx 0.05$.   In this Gribov multiple scattering picture~\cite{Gribov}, there is a 
close relation between shadowing and diffraction. The so-called 
AGK cutting rules~\cite{AGK} relate the first nuclear shadowing correction to  the cross-section for diffractively producing a final state in coherent scattering off a nucleon (integrated over all diffractive final states). See Fig.~\ref{fig:ShadDiff}(a) 
 for an illustration of this correspondence. With these relations (and higher 
order re-scattering generalizations of these) and 
with the HERA diffractive DIS data as input, the NMC nuclear 
shadowing data can be reproduced as shown in the sample 
computation~\cite{CapellaKAS,FrankfurtGMcS} in Fig.~\ref{fig:ShadDiff}(b).

In the infinite momentum frame (IMF), shadowing arises due to gluon recombination and screening in the target. When the density of partons 
in the transverse plane of the nucleus becomes very large, many body recombination and screening effects compete against the growth in the cross-section, leading 
eventually to a saturation of the gluon density~\cite{GLR}.  In the IMF picture, one can again use the AGK rules we discussed previously to relate shadowing and diffraction~\cite{MuellerQiu}, and the result is amenable to a partonic interpretation. 
The saturation regime is characterized by a scale $Q_s(x,A)$, called 
the saturation scale, which grows with decreasing $x$ and increasing $A$. This saturation 
scale arises naturally in the Color Glass Condensate (CGC) framework which is discussed in section 3. 

A natural consequence of saturation physics is the phenomenon of geometrical scaling. 
(See for instance the discussion on geometrical 
scaling of HERA data in section 2.1.) It has been argued that the NMC DIS data also display geometrical scaling~\cite{FreundRSW}--the evidence here albeit interesting is not compelling owing 
to the paucity of nuclear data over a wide range of $x$ and $Q^2$. It is widely 
believed that shadowing is a leading twist effect~\cite{FS,PillerWeise}, 
but some of the IMF discussion in the CGC saturation framework suggests higher twist effects are important for $Q^2 \leq Q_s^2$ because of the large gluon density~\cite{JamalWang}. Constraints from non-linear corrections to the DGLAP framework have also been discussed recently~\cite{EHKQS}. The available data on the $Q^2$  dependence of  shadowing are inconclusive at small $x$.  

\begin{figure}[t!]
\vspace*{5mm}
\hspace*{20mm}
\epsfig{file=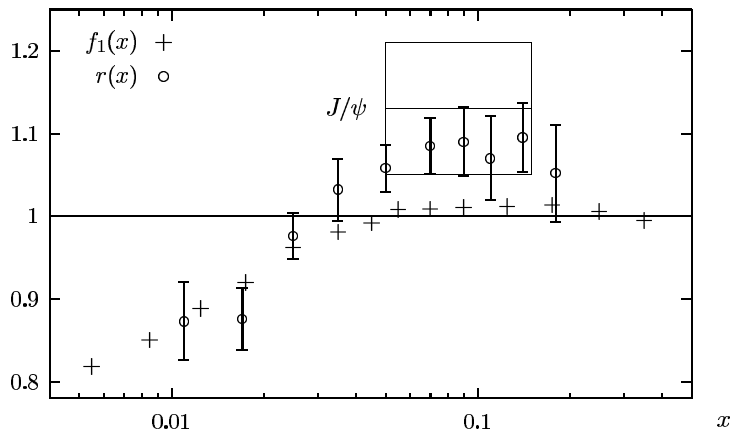,scale=0.8}
\vspace*{3mm}
\caption{The ratio, $r(x)$ of the gluon distributions in Sn relative to 
C and the ratio, $f_1(x)$ of their $F_2(x)$ 
structure functions~\cite{GoussetP}. The box represents the extraction 
of $r(x)$ from $J/\psi$  electro-production in the 
process $\mu + A \rightarrow \mu + J/\psi +X$.
\label{fig:nucleargluon}}
\end{figure}

Our empirical definition of shadowing in DIS refers to quark shadowing. Likewise for quarks and anti--quarks in the Drell--Yan process in hadronic collisions. In DIS gluon distributions are inferred only 
indirectly because 
the virtual photon couples to quarks. The most precise extractions of gluon distributions thus far are from scaling violations of $F_2^A$. To do this properly, one needs a wide window in $x$ and $Q^2$. In contrast to the highly precise data on nucleon gluon distributions from HERA,  our knowledge of nuclear gluon structure functions ($g_A (x,Q^2)$) is {\it nearly non-existent}. This is especially so 
relative to our  knowledge of  quark distributions in nuclei.  The most precise data on  the modification of  gluon distributions in nuclei come from two NMC high precision measurements of the ratio of the scaling violations of
 the structure functions of Tin (Sn) and Carbon (C).  The experiments measure ratios $f_1 =F_2^{\rm Sn}/F_2^{\rm C}$ and $f_2= {\partial \over \partial \ln Q^2}\, f_1$. The ratio 
 $r= g_{\rm Sn}/g_{\rm C}$ can be determined~\cite{GoussetP} from $f_1$ and $f_2$ and the scaling violations of $F_2^{\rm Deuterium}$ (with minimal 
assumptions). The result for $r$ is shown in Fig.~\ref{fig:nucleargluon}. At small $x$, gluon shadowing is observed. The trend suggests that gluon shadowing at small $x$ is greater than that of $F_2$, even though the error bars are too large for a conclusive statement. 
 
 At larger $x$ of $0.1 < x < 0.2$, one observes anti--shadowing of the gluon distributions. This result from scaling violations can be compared to the ratio of gluon distributions  extracted 
 from inclusive $J/\psi$  production in DIS. The latter assumes the gluon fusion model of $J/\psi$-production. The results for $r$ from the latter method are consistent with those from scaling violations. The large experimental uncertainties however leave the extent of anti-shadowing in doubt. Other measurements of scaling violations for the ratio of $F_2^{\rm Sn}/F_2^{\rm C}$ showed an increase of the ratio with the increase of  $Q^2$ consistent with predictions~\cite{section2:31,section2:32}. 
 
The limited data we have may be interpreted to suggest a provocative picture of
nuclear parton densities in the $x\sim 0.1$--$0.2$ region, which corresponds to distances of $\sim 1$--$1.5$ fm, where medium range and short range inter-nucleon forces are expected to be important.  In this region, if the gluon and valence quark fields are enhanced while the sea is somewhat suppressed, as some analyses suggest, gluon-induced interactions between nucleons, as well as valence quark interchanges between nucleons may contribute significantly to nuclear binding~\cite{GoussetP,FS}.
Nuclear gluon distributions can also be further constrained by inclusive hadron distributions recently measured by the 
RHIC experiments~\cite{PHENIXdA,STARdA,BRAHMSdA,PHOBOSdA} in Deuteron-Gold scattering at $\sqrt{s} = 200$ GeV/nucleon. These RHIC results will be discussed in section~3. 
 
\subsection{Space-time correlations in QCD}

The space-time picture of DIS processes strongly depends on 
the value of Bjorken $x$.  An analysis of electromagnetic current correlators in DIS reveals that one
probes the target wave function at space-time points separated by
longitudinal distances $l_{\rm coh.}$ and transverse distances $\sim
1/Q$.  At large $x$, ($x> 0.2$), the virtual photon transforms
into a strongly interacting state very close to the active nucleon,
typically in the middle of the nucleus.  If $Q^2$ is large
enough as well, the produced partonic state interacts weakly with the
medium. At smaller $x$ ($x<0.05$), the longitudinal length scale $l_{\rm
coh.}$ exceeds the nuclear size of the heaviest nuclei.  At
sufficiently small $x$ ($x < 0.005$ for the heaviest nuclei) DIS
processes undergo several spatially separated stages. First, the
virtual photon transforms into a quark-gluon wave packet well before
the nucleus. Time dilation ensures that interactions amongst partons
in the wave packet are frozen over large distances. (These can be
several hundred fermis at EIC energies.) The partons in the wave 
packet interact coherently and instantaneously with the target.  
At high energies, these interactions are eikonal in
nature and do not affect the transverse size of the wave 
packet. Finally, the fast components of the wave packet transform into
a hadronic final state when well past the nucleus. This interval could
be as large as $2\nu/\mu^2$ where $\nu$ is the energy of the virtual
photon and $\mu \leq 1$ GeV is a soft hadronic scale.  

Space-time studies thus far have been limited to semi-exclusive 
experiments that investigate the phenomenon of color transparency, and
more generic inclusive studies of quark propagation through nuclei.
Both these studies involved fixed targets. They are briefly summarized
below.

In pQCD, color singlet objects interact weakly with a 
single nucleon in the target. Additional interactions are suppressed by 
inverse powers of $Q^2$.  This
phenomenon is called ``color transparency'' because the nucleus appears
transparent to the color singlet projectile~\cite{section2:43,section2:44,RalstonJain}.  
At very high energies, even the interaction of small color singlet projectiles
with nuclei can be large.  In this kinematic region,
the phenomenon is termed ``color opacity''~\cite{section2:45,section2:46}.
The earliest study of color transparency in DIS was a study of coherent $J/\psi$
photo-production off nuclei~\cite{section2:47}.  The amplitude of the
process at small $t$ (momentum transfer squared) is approximately
proportional to the nuclear atomic number $A$.  This indicates that the
pair that passes through the nucleus is weakly absorbed.  For hadronic
projectiles, a similar and approximately linear $A$-dependence of the
amplitude was observed recently for coherent diffraction of 500 GeV
pions into two jets~\cite{section2:48}, consistent with
predictions~\cite{section2:45}. 

A number of papers~\cite{section2:49,section2:50,section2:51} predict that the
onset of color transparency at sufficiently large $Q^2$ will give for the
coherent diffractive production of vector mesons
\begin{eqnarray}
{d\sigma(\gamma_L^* +A \rightarrow V +A)\over dt}|_{t=0} \propto A^2 \, .
\end{eqnarray}
For incoherent diffraction at sufficiently large $t$ ($ > 0.1$ GeV$^2$), they predict,
\begin{eqnarray}
R(Q^2) \equiv {{d\sigma(\gamma_L^* +A \rightarrow V +A^\prime)/
dt}\over A\, {d\sigma(\gamma_L^* +N \rightarrow V +N)/ dt}} =1 \, .
\end{eqnarray}
The first measurements of  incoherent diffractive production of
vector mesons were performed by the E665 collaboration at
Fermilab~\cite{section2:52}.  A significant increase of the nuclear
transparency, as reflected in the ratio $R(Q^2)$, was observed.  The
limited luminosity and center-of-mass energy however do not provide a
statistically convincing demonstration of color transparency.  In
addition, the results are complicated by large systematic effects.

Measurements of the inclusive hadron distribution for different final
states as a function of the virtual photon energy $\nu$, its
transverse momentum squared $Q^2$, the fraction $z_h$ of the photon
energy carried by the hadron, and the nuclear size $A$, provide
insight into the propagation of quarks and gluons in nuclear
media. In addition to the time and length scales discussed previously, 
the ``formation time'' $\tau_h$ of a hadron is an additional time scale. 
It is in principle significantly larger than the production time $1/Q$ 
of a color singlet parton.  If the formation time is large, the
``pre-hadron'' can multiple scatter in the nucleus, thereby broadening
its momentum distribution, and also suffer radiative energy loss before
hadronization.
The QCD prediction for transverse momentum broadening resulting 
from multiple scattering is (for quarks) given by the expression~\cite{section2:53} 
\begin{eqnarray} 
\langle\Delta p_\perp^2 \rangle
= {\alpha_S C_F \pi^2\over 2} xg(x,Q^2) \rho L \approx
0.5\, \alpha_S \left({L \over 5\, {\rm fm}}\right)\, {\rm GeV}^2 \, .  
\end{eqnarray}
Here, $C_F=4/3$ is the color Casimir of the quark, $\rho$ is the nuclear matter
density and $L$ is the length of matter traversed.  The Drell-Yan
data in Ref.~\cite{section2:55} agree with this expression and with
the predicted small size of the effect (empirically,
$\langle \Delta
p_\perp^2\rangle\sim 0.12$ GeV$^2$ for heavy nuclei).  
One also observes a large difference in the $A$ dependence of the 
transverse momentum of Drell-Yan di-muons relative 
to those from $J/\psi$ and $\Upsilon$ production and decay~\cite{section2:56}. 
In the former process only the incident quark undergoes strong interactions, 
whereas in the latter, the produced vector mesons interact strongly as well. 
However, the size of the effect and the  comparable broadening of 
the $J/\psi$ and $\Upsilon$ (albeit the latter is appreciably smaller than the former) 
need to be better understood. The $p_\perp$ imbalance of di-jets in nuclear
photo-production suggests a significantly larger $p_\perp$ broadening
effect than in $J/\psi$ production~\cite{section2:57}. This suggests non-universal 
behavior of $p_\perp$ broadening effects but may also occur from a contamination of 
the jets by soft fragments.
Parton $p_\perp$ broadening due to multiple scattering may also be
responsible for the anomalous behavior of inclusive hadron production
in hadron-nucleus scattering at moderate $p_\perp$ of a few GeV.  In this case, the
ratio $R_{\rm pA}$ of inclusive hadron production in hadron-nucleus
scattering to the same process on a nucleon, is suppressed at low $p_\perp$ but exceeds 
unity between $1-2$ GeV.  This ``Cronin effect''~\cite{section2:58} was
discovered in proton-nucleus scattering experiments in the late
70's. The flavor dependence of the Cronin effect provided an early hint that 
scattering of projectile partons off gluons dominates over scattering off quarks~\cite{Krzywicki}. 
The Cronin effect will be discussed further in section 3 in light of the recent RHIC experiments on Deuteron-Gold
scattering~\cite{PHENIXdA,STARdA,PHOBOSdA,BRAHMSdA}. 

The energy loss of partons due to scattering in nuclear matter is complicated 
by vacuum induced energy loss in addition to 
the energy loss due to scattering. One computation suggests that
vacuum energy loss is the dominant effect~\cite{Kopeliovich_review}.  
For a quark jet, the medium induced energy
loss increases quadratically with the length, $L$, and is independent of
the energy for $E\rightarrow \infty$.  For $L = 5$ fm, the asymptotic
energy loss, $\Delta E$ , is estimated to be less than $1$\, GeV in a
cold nuclear medium~\cite{section2:53}.  This makes it difficult to empirically
confirm this remarkable $L$-dependence of the energy
loss. DIS data are qualitatively consistent with small
energy loss~\cite{section2:59,section2:60, section2:61}.  The data
indicate that the multiplicity of the leading hadrons is moderately
reduced (by ~10\%) for virtual photon energies of the order of $10$-$20$
GeV for scattering off Nitrogen-14 nuclei.  At higher energies, the leading
multiplicities gradually become $A$-dependent, indicating 
absorption of the leading partons~\cite{section2:60,section2:61,section2:62,section2:63,section2:64}. 

\begin{figure}[htb!]
\begin{center}
\epsfig{file=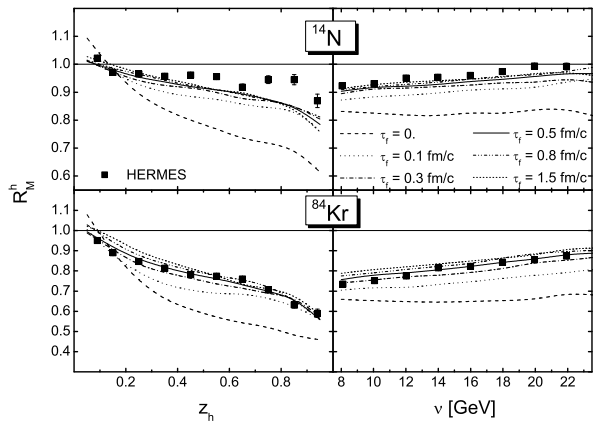,scale=0.8}
\caption{Data from the HERMES experiment ~\cite{section2:59} showing 
the ratio of the inclusive hadron cross section in a nucleus 
relative to that in a nucleon, plotted as a function a) of $z_h$, which is the 
fraction of the quark's momentum carried by a hadron, and b) as a 
function of $\nu$, the photon energy, for two different nuclei. Curves 
denote results of a ``pre-hadron'' 
scattering model~\cite{Falter}, with  differing formation times.}
\label{fig:Falter}
\end{center}
\end{figure} 

A pQCD description of partonic energy loss in 
terms of modified fragmentation functions is claimed to describe 
HERMES data~\cite{Wang}. However, at HERMES energies, 
and perhaps even at EMC energies, descriptions in terms of hadronic re-scattering 
and absorption are at least as successful~\cite{Accardi,Falter}. As previously discussed, however, 
the latter descriptions usually require that the color singlet ``pre-hadrons'' have a 
formation time $\tau_h\sim 0.5$ fm~\cite{Kopeliovich_review,Falter,Greiner}. Fig.~\ref{fig:Falter} 
shows the results from one 
such model as a function of $z_h$ 
(the fraction of the parton momentum carried by a hadron) and $\nu$ (the 
virtual photon energy) compared to the HERMES data.  
At EIC energies, a pQCD approach in terms of 
modified fragmentation functions~\cite{Wang1,Wang2} should be more applicable. 
The results from these analyses will provide an 
important test of jet quenching in hot matter descriptions of the RHIC data.

\section{Scientific Opportunities with an Electron-Ion Collider \label{science}}

This section will discuss  the exciting scientific opportunities 
that will be made possible by the novel features of an electron-ion collider: 
the high luminosity, the possibility to do scans over a wide range in 
energy, polarization of the electron and hadronic beams, a range of 
light and heavy nuclear beams and, not least, the collider geometry of 
the scattering. Scientific firsts  for the electron-ion collider will include a) the first 
high energy polarized electron-polarized proton collider, and b) the 
first high energy electron-nucleus collider. 

\subsection{Unpolarized ep collisions at EIC}

Unpolarized e-p collisions have been studied extensively most recently at the HERA 
collider at DESY. In the eRHIC option for an EIC, the 
center of mass energy in an e-p collision is anticipated to be $\sqrt{s} = 100$ GeV 
compared to $\sqrt{s}$ of over 300 GeV at HERA. 
Although the $x$-$Q^2$ 
reach of an EIC may not be as large as that of HERA, it has significant other 
advantages which we will itemize below.
\begin{itemize}

\item The current design luminosity is approximately 25 times the design 
luminosity of HERA. Inclusive observables will be measured with great 
precision. The additional luminosity will be particularly advantageous for 
studying semi-inclusive and exclusive final states. 

\item The EIC (particularly in the eRHIC version) will be able to vary the 
energies of both the electron and nucleon beams. This will enable a 
first measurement of $F_L$ in the small $x$ regime. The $F_L$ measurement is very important 
in testing QCD fits of structure functions.

\item Electron-Deuteron collisions, with tagging of spectator nucleons, will 
allow high precision studies of the flavor dependence of 
parton distributions. 

\item An eRHIC detector proposed by Caldwell et al.~\cite{Caldwell} would have a 
rapidity coverage nearly twice that of the ZEUS and H1 detectors at HERA. 
This would allow the reconstruction of the event structure of hard forward jets 
with and without rapidity gaps in the final state. With this detector, exclusive 
vector meson and DVCS measurements 
can be performed for a wider range of the photon-proton center of mass energy 
squared $W^2$. It also permits
measurements up to high $|t|$, where $t$ denotes the square of the difference 
in four-momenta of the incoming and outgoing proton.    
These will enable a precise mapping of the energy dependence of final states,
as well as open a window into the spatial distribution of partons down 
to very low impact parameters.

\end{itemize}

We will briefly discuss the physics measurements that can either be done or 
improved upon with the above enumerated capabilities of EIC/eRHIC. 
For inclusive measurements, $F_L$ is clearly a first, ``gold plated" measurement. 
Current QCD fits predict that $F_L$ is very small (and in 
some analyses negative) at small $x$ and small $Q^2 < 2$ GeV$^2$. An independent 
measurement can settle whether this reflects poor extrapolations of 
data (implying leading twist interpretations of data are still adequate
in this regime), or whether higher twist effects are dominant. It will also constrain 
extractions of the gluon distribution because $F_L$ is very sensitive to it. Another 
novel measurement would be that of structure functions in the 
region of large $x\approx 1$. These measurements can be done with 1 fb$^{-1}$ of data 
for up to $x=0.9$ and for $Q^2 < 250$ GeV$^2$. This kinematic window is completely 
unexplored to date. These studies can test perturbative QCD predictions of the helicity 
distribution of the 
valence partons in a proton~\cite{FarrarJackson} as well as the detailed pattern of 
SU(6) symmetry breaking~\cite{Close}. Moments of 
structure functions can be compared to lattice data. These should help quantify the 
influence of higher twist effects. 
Finally ideas such as Bloom-Gilman duality can be further tested in this kinematic 
region~\cite{Ent}.

At small $x$, very little is understood about the quark sea. For instance, the 
origins of the $\bar{u}-\bar{d}$ asymmetry and the suppression of the 
strange sea are not clear. 
High precision measurements of $\pi^\pm$, $K^\pm$, $K_s$ and open charm 
will help separate valence and sea contributions in the small-$x$ 
region. We have already discussed Generalized Parton Distributions and DVCS 
measurements. The high luminosity, wide coverage and measurements at high $|t|$ 
will quantify efforts to extract a 3-D snapshot of the distribution of partons in 
the proton.

\subsection{Polarized $ep$ collisions at EIC}

We expect the EIC to dramatically extend our understanding of the spin 
structure of the proton through measurements of the spin structure 
function $g_1$ over a wide range in $x$ and $Q^2$, of its parity-violating 
counterparts $g_4$ and $g_5$, of gluon polarization $\Delta G$, as well 
as through spin-dependent semi-inclusive measurements,
the study of exclusive reactions, and of polarized photo-production.

\subsubsection{Inclusive spin-dependent structure functions}

We have emphasized in Sec.~\ref{g1old} the need for further 
measurements of $g_{1}(x,Q^{2})$ at lower $x$ and higher $Q^2$. 
A particularly important reason is that one would like to 
reduce the uncertainty in the integral $\Gamma_1(Q^2)$ and hence
in $\Delta \Sigma$. However, the behavior of $g_{1}(x,Q^{2})$ 
at small $x$ is by itself of great interest in QCD. 

\begin{figure}[htbp!]
\hspace*{2.5cm}
\epsfig{file=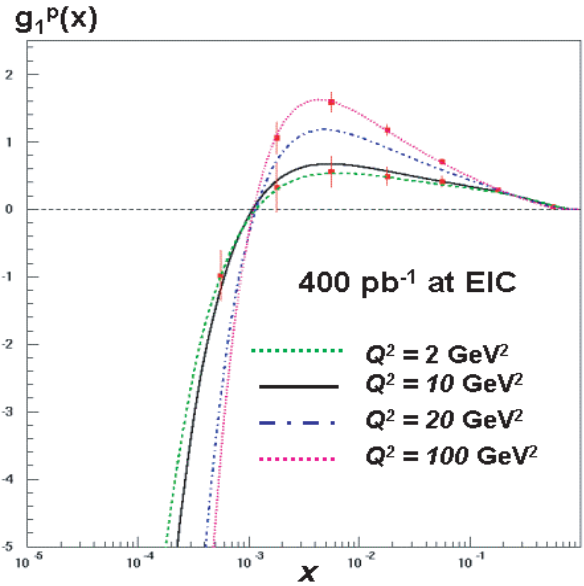,width=8cm,angle=0}
\caption{\label{fig:lowxg1} Possible eRHIC data (statistical accuracy) with $250 
\times 10$ GeV collisions is shown for 400 pb$^{-1}$. Also shown is the 
evolution of $g_1(x,Q^2)$ at low $x$ for different values of $Q^{2}$ for a positive 
gluon polarization~\cite{abfr,smc_QCD}.}
\end{figure}

At very high energies, Regge theory gives guidance to the
expected behavior of $g_{1}(x)$. The prediction \cite{hei} is
that $g_1(x)$ is flat or even slightly vanishing at small $x$,
$g_1(x)\propto x^{-\alpha}$ with $-0.5\leq \alpha \leq 0$.
It is an open question how far one 
can increase $Q^2$ or decrease energy and still trust Regge 
theory. A behavior of the form $g_1 \sim x^{-\alpha}$ 
with $\alpha < 0$ is unstable under DGLAP evolution \cite{bfr,unst}
in the sense that evolution itself will then govern the 
small-$x$ behavior at higher $Q^2$. Under the assumption that
Regge theory expectations are realistic at some (low) scale $Q_0$ one
then obtains ``perturbative predictions'' for $g_{1}(x,Q^{2})$
at $Q\gg Q_0$. The fixed-target polarized DIS data indicate that 
although the non-singlet combination $g_{1}^p-g_{1}^n$ is quite
singular at small $x$, the singlet piece does appear to be
rather flat \cite{bfr,bass}, so that the above reasoning applies
here. It turns out that the leading eigenvector of
small-$x$ evolution is such that the polarized quark singlet 
distribution and gluon density become of opposite sign. For a sizeable
positive gluon polarization, this leads to the striking feature that the 
singlet part of $g_{1}(x,Q^{2})$ is negative at small $x$ 
and large $Q^2$ \cite{bfr}, driven by $\Delta g$. 
Figure~\ref{fig:lowxg1} shows this dramatic behavior
for different values of $Q^2=2,10,20,100$ GeV$^2$ \cite{abhayyale}.  
The projected statistical uncertainties at eRHIC, 
corresponding to 400 pb$^{-1}$ integrated luminosity with an almost $4\pi$
acceptance detector, are also shown. Note that 400 pb$^{-1}$ can 
probably be collected within about one week of e-p running of eRHIC. 
Thus, at eRHIC one will be well positioned to
explore the evolution of the spin structure function $g_1(x,Q^2)$ 
at small $x$. We note that at small $x$ and 
toward $Q^2\to 0$, one could also study the transition region between 
the Regge and pQCD regimes \cite{bass}. 

Predictions for the small-$x$ behavior of $g_1$ have
also been obtained from a perturbative resummation of 
double-logarithms $\alpha_s^k \,\ln^{2k}\left(1/x\right)$  
appearing in the splitting functions \cite{bart,blum,kiyo,ziaja} 
at small $x$ in perturbative QCD. Some of these calculations
indicate a very singular asymptotic behavior of $g_1(x)$.
It has been shown \cite{blum}, however, that subleading terms may still be
very important even far below $x=10^{-3}$. 

The neutron $g_1$ structure function could be measured at eRHIC 
by colliding the electrons with polarized Deuterons or with Helium. 
If additionally the hadronic proton fragments are tagged, a very
clean and direct measurement 
could be performed. As can be seen from Fig.~\ref{fig:g1data},
information on $g_1^n$ at small $x$ is scarce. 
The small-$x$ behavior of the isotriplet
$g_1^p-g_1^n$ is particularly interesting for the Bjorken sum rule
and because of the steep behavior seen in the fixed-target data 
\cite{bfr,bass}. It is estimated \cite{igo} that an accuracy of the 
order of $1\%$ could be achieved for the Bjorken sum rule
in a running time of about one month. 
One can also turn this argument around and use the accurate
measurement of the non-singlet spin structure function and its 
evolution to determine the value of the strong coupling
constant $\alpha_{s}(Q^{2})$\cite{abfr,smc_QCD}. This has been tried, and the 
value one gets from this exercise is comparable to the world average for the strong
coupling constant. It is expected that if precision low-$x$ data from the EIC is 
available and the above mentioned non-singlet structure functions are
measured along with their evolutions, this may result in the most accurate
value of the strong coupling constant $\alpha_{s}(Q^{2})$.

Because of eRHIC's high energy, very large $Q^2$ can be reached. 
Here, the DIS process proceeds not only via photon exchange; 
also the $W$ and $Z$ contribute significantly. 
Equation~\ref{strcfct} shows that in this case new structure functions 
arising from parity violation contribute to the DIS cross section. 
These structure functions contain very rich additional information on parton 
distributions \cite{ansel,svw,eweak}.
As an example, let us consider charged-current (CC) interactions.
Events in the case of $W$ exchange are characterized by a large 
transverse momentum imbalance caused by the inability to detect 
neutrinos from the event.  The charge of the $W$ boson is dictated 
by that of the lepton beam used in the collision. For $W^-$ exchange
one then has for the structure functions $g_1$ and $g_5$ in
Equation~\ref{longwq}:
\begin{equation}\label{g5lo}
g_1^{W^-} (x) = \Delta u (x) + \Delta\bar{d}(x)+
\Delta\bar{s} (x) \; , \; \; \; 
g_5^{W^-} (x) = \Delta u (x) - \Delta\bar{d}(x)- 
\Delta\bar{s} (x) \; .
\end{equation}
These appear in the double-spin asymmetry as defined in Ref.~\cite{ansel},
where the asymmetry can be expressed in terms of structure functions
as
\begin{equation}\label{a5}
A^{W^-} = \frac{2b g_1^{W^-}+a g_5^{W^-}}
{a F_1^{W^-}+b F_3^{W^-}} \; .
\end{equation}
Here $a=2(y^2-2y+2)$, $b=y (2-y)$ and $F_3$ is the unpolarized
parity-violating structure function of Equation~\ref{strcfct}.
Note that the typical scale in the parton densities is $M_W$ here. 
Availability of polarized neutrons and positrons is particularly 
desirable. For example, one finds at lowest order:
\begin{eqnarray} \label{reln}
g_1^{W^-,p} - g_1^{W^+,p} 
&=& \Delta u_v  - \Delta d_v \\
g_5^{W^+,p} + g_5^{W^-,p} &=& \Delta u_v  +
\Delta d_v \\
g_5^{W^+,p} - g_5^{W^-,n} &=& -\left[\Delta u+\Delta \bar{u} -
\Delta d-\Delta \bar{d}\right] \; .
\end{eqnarray}
The last of these relations gives, after integration over all
$x$ and taking into account the first-order QCD correction \cite{svw},
\begin{equation}
\int_0^1 dx \left[ g_5^{W^+,p}-g_5^{W^-,n} \right] 
= - \left( 1-\frac{2 \alpha_s}{3 \pi} \right) g_A \; ,
\end{equation}
equally fundamental as the Bjorken sum rule.

\begin{figure}[htbp!]
\vspace{7mm}
\hspace*{2.5cm}
\epsfig{file=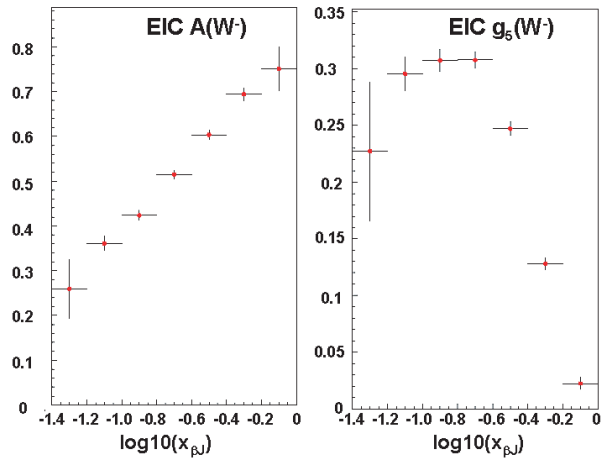,width=8cm,angle=0}
\caption{\label{fig:g5weak} Simulations \cite{contr} for the spin asymmetry
$A^{W^-}$ of Equation~\ref{a5} and the structure function
$g_5^{W^-}$ as functions of $\log_{10}(x)$.}
\end{figure}

A Monte Carlo study, including the detector effects, has shown
that the measurement of the asymmetry in Equation~\ref{a5} and the parity 
violating spin structure functions is feasible at eRHIC.  
Figure \ref{fig:g5weak} shows simulations \cite{contr} for the 
asymmetry and 
the structure function $g_5$ for CC events with an electron beam.
The luminosity was assumed to be 2 fb$^{-1}$. The simulated data shown 
are for $Q^2 > 225$~GeV$^2$. 
Similar estimates exist for $W^+$. Measuring this asymmetry 
would require a positron beam.  
The curves in the figure use the polarized parton distributions of
\cite{gs}. It was assumed that the unpolarized structure functions
will have been measured well by HERA by the time this 
measurement would be performed at eRHIC.  Standard assumptions used 
by the H1 collaboration about the scattered electrons for good detection 
were applied.  The results shown could be obtained (taking into
account machine and detector inefficiencies) in a little 
over one month with the eRHIC luminosity. It is possible that only 
one or both of the electron-proton and positron-proton collisions 
could be performed, depending on which design of the accelerator 
is finally chosen (see Section~4).

\subsubsection{Semi-inclusive measurements}
As we discussed in Sec.~\ref{g1old}, significant
insights into the nucleon's spin and flavor 
structure can be gained from semi-inclusive scattering
$ep\to ehX$. Knowledge of the identity of the produced hadrons 
$h$ allows separation of the contributions from the different quark 
flavors. In fixed target 
experiments, the so-called current hadrons are at forward 
angles in the laboratory frame.  This region is difficult to 
instrument adequately, especially if the luminosity is increased 
to gain significant statistical accuracy. A 
polarized ep collider has the ideal geometry to 
overcome these shortfalls.  The collider kinematics open up the 
final state into a large solid angle in the laboratory which, 
using an appropriately designed detector, allows complete 
identification of the hadronic final state both in the current 
and target kinematic regions of fragmentation phase space.  At 
eRHIC energies the current and target kinematics are well 
separated and may be individually studied. 
At eRHIC higher $Q^2$ will be available than in the 
fixed-target experiments, making the observed spin asymmetries
less prone to higher-twist effects, and the interpretation
cleaner.

\begin{figure}[htbp!]
\hspace*{0.4cm}
\epsfig{file=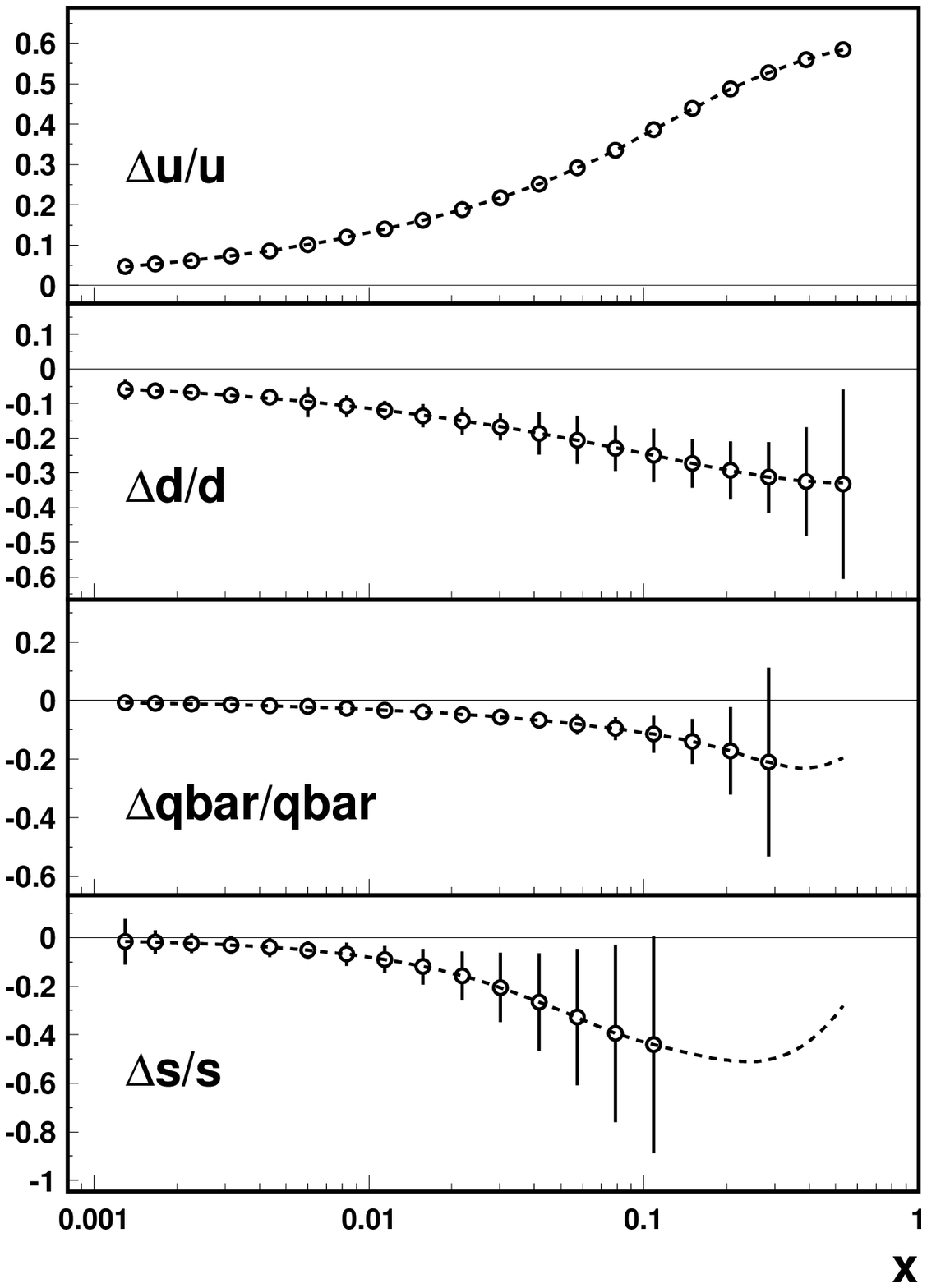,width=5cm,angle=0,bbllx=84,bblly=178,bburx=499,bbury=694}

\hspace*{7cm}
\vspace*{-8cm}
\epsfig{file=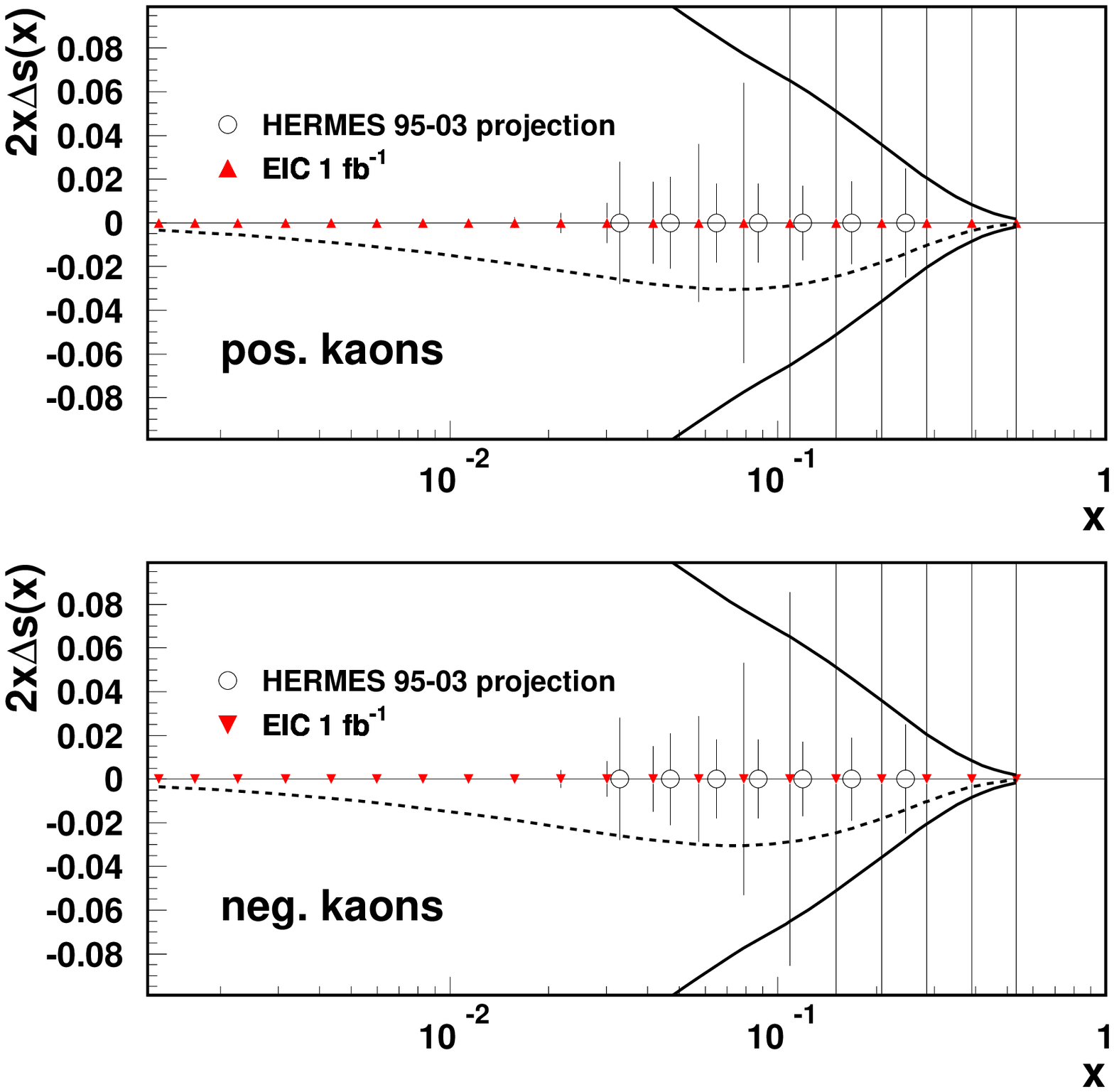,width=5cm,angle=0,bbllx=26,bblly=-433,bburx=537,bbury=57}

\vspace*{3.4cm}
\caption{\label{fig:sidis} Left: projected precision of eRHIC measurements 
of the polarized quark and antiquark distributions \cite{ks}. Right:
expected statistical accuracy of $\Delta s(x)$ from 
spin asymmetries for semi-inclusive 
$K^{\pm}$ measurements for 1~fb$^{-1}$ luminosity operation of eRHIC,
and comparison with the statistical accuracy of the corresponding
{\sc Hermes} measurements.}
\end{figure}

Figure~\ref{fig:sidis} shows simulations \cite{ks} 
of the precision with which one could measure the polarized quark
and antiquark distributions at the EIC. The events were produced 
using the DIS generator LEPTO.  
The plotted uncertainties are statistical only. The simulation 
was based on an integrated luminosity of 1 fb$^{-1}$ for 5 GeV 
electrons on 50 GeV protons, with both beams polarized to 70$\%$.
Inclusive and semi-inclusive asymmetries were analyzed using the 
leading order ``purity'' method developed by the SMC \cite{smc} and 
{\sc Hermes} \cite{hermesdq} collaborations. Excellent 
precision for $\Delta q/q$ can be obtained down to $x \approx 0.001$.  
The measured average $Q^2$ values vary as usual per
$x$ bin;  they are in the range $Q^2=1.1$~GeV$^2$ at the 
lowest $x$ to $Q^2\sim 40$~GeV$^2$ at high $x$. With 
proton beams, one has greater sensitivity to up quarks than to 
down quarks.  Excellent precision for the down quark 
polarizations could be obtained by using deuteron or helium beams.

With identified kaons, and if the up and down quark distributions 
are known sufficiently well, one will have a very good possibility
to determine the strange quark polarization. As we discussed 
in Sec.~\ref{g1old}, $\Delta s(x)$ is one of the most interesting 
quantities in nucleon spin structure. On the right-hand-side
of Figure~\ref{fig:sidis}, we show results expected for $\Delta s(x)$
as extracted from the spin asymmetries for $K^{\pm}$ production. 
As in the previous figure, only statistical 
uncertainties are indicated. The results are compared with the 
precision available in the {\sc Hermes} experiment. 

There is also much interest in QCD in more 
refined semi-inclusive measurements. For example, the transverse
momentum of the observed hadron may be observed. Here, interesting
azimuthal-angle dependences arise at leading twist 
\cite{goeke,koike}, as we discussed in Subsec.~\ref{subsectr}. 
At small transverse momenta, resummations of large
Sudakov logarithms are required \cite{nadolsky}. Measurements at eRHIC 
would extend previous results from HERA \cite{zeussidis} and be 
a testing ground for detailed studies in perturbative QCD.

\subsubsection{Measurements of the polarized gluon 
distribution  $\Delta g(x,Q^{2})$}
One may extract $\Delta g$ from scaling violations of the 
structure function $g_1(x,Q^2)$.
Figure~\ref{fig:polpdf} shows that indeed some
initial information on $\Delta g(x,Q^2)$ has been obtained in this 
way, albeit with very poor accuracy. The uncertainty of the
integral of $\Delta g$ is probably about $100\%$ at the moment 
\cite{abfr}. Measurements at RHIC will vastly improve on this.
eRHIC will offer independent and complementary information. 
Thanks to the large lever arm in $Q^2$, and to the low $x$
that can be reached, scaling violations
alone will constrain $\Delta g(x,Q^2)$ and its integral much better.
Studies \cite{licht} indicate for example that the total uncertainty on
the integral of $\Delta G$ could be reduced to about $5-10\%$ 
by measurements at eRHIC with integrated 
luminosity of  $12$~fb$^{-1}$ ($\sim 2-3$ years of eRHIC operation). 

\begin{figure}[htbp!]
\vspace{-14mm}
\hspace*{2cm}
\epsfig{file=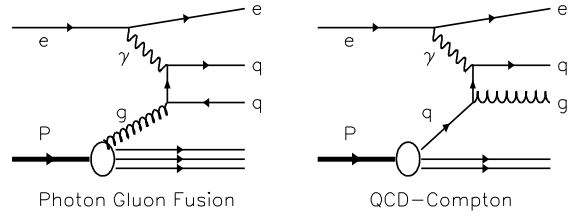,width=10cm,angle=0}

\vspace*{-6mm}
\caption{\label{fig:dijet} Feynman diagrams for the photon-gluon
fusion and the QCD Compton processes.}
\end{figure}

Lepton-nucleon scattering also offers direct ways of accessing
gluon polarization. Here one makes use of the photon-gluon 
fusion (PGF) process, for which the gluon appears at leading 
order. Charm production is one particularly interesting
channel \cite{gr,as,grvo}. It was also proposed 
\cite{as,grvo,oldjets} to use jet pairs,
produced in the reaction $\gamma^{\ast}g\to q\bar{q}$, 
for a determination of $\Delta g$. This process competes
with the QCD Compton process, $\gamma^{\ast}q\to qg$.
Feynman diagrams for these processes are shown 
in Figure~\ref{fig:dijet}. 

\begin{figure}[t!]
\vspace*{5mm}
\hspace*{1.5cm}
\epsfig{file=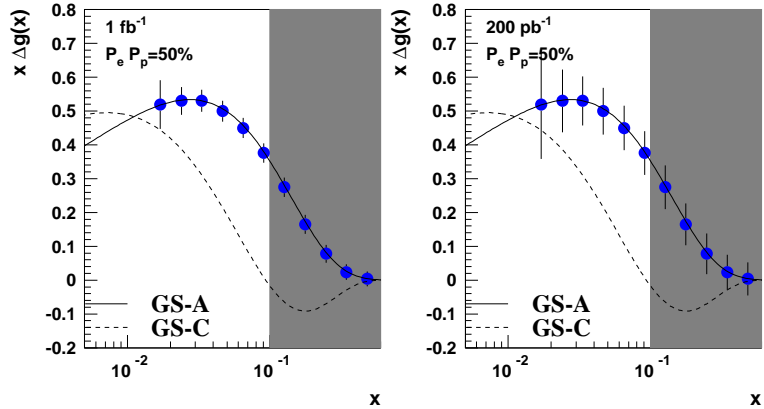,width=10cm,angle=0}
\caption{The statistical precision of $x\Delta g$ from di-jets in LO
for eRHIC, for two different luminosities, with predictions for sets A and C
of the polarized parton densities of Ref.~\cite{gs}. \label{fig:dijglufig2}}
\end{figure}

In the unpolarized case, dijet production has successfully been 
used at HERA to constrain the gluon density \cite{h1jet}. 
Dedicated studies have been performed for dijet production
in polarized collisions
at eRHIC \cite{radel}, using the MEPJET \cite{mepjet} generator. 
The two jets were required to have transverse momenta $>3$~GeV, 
pseudorapidities $-3.5\leq \eta \leq 4$, and invariant mass
$s_{JJ}>100$~GeV$^2$. A $4\pi$ detector coverage was assumed.
The results for the reconstructed
$\Delta g(x)$ are shown in Figure~\ref{fig:dijglufig2}, assuming
luminosities of 1~fb$^{-1}$ (left) and 200~pb$^{-1}$ (right).
The best probe would be 
in the region $0.02\leq x\leq 0.1$; at higher $x$, the 
QCD Compton process becomes dominant. This region is indicated
by the shaded areas in the figure. The region
$0.02\leq x\leq 0.1$ is similar to that probed at RHIC. 
Measurements at eRHIC would thus allow an independent determination
of $\Delta g$ in a complementary physics environment. 


Eventually data for the scaling violations
in $g_1(x,Q^2)$ and for dijet production in DIS will be 
analyzed jointly. Such a combined analysis would determine
the gluon distribution with yet smaller uncertainties. A first
preliminary study for eRHIC \cite{licht1}, following the lines
of \cite{herajet}, indeed confirms this.

\subsubsection{Exploring the partonic structure of polarized photons}
In the photoproduction limit, when the virtuality of the intermediate 
photon is small, the $ep$ cross-section can be approximated 
by a product of a photon flux and an interaction cross section of the 
real photon with the proton.  Measurements at HERA in the photoproduction 
limit have led to a significant improvement in our knowledge
of the {\it hadronic structure} of the photon. 

\begin{figure}[htbp!]

\vspace*{-2mm}
\epsfig{figure=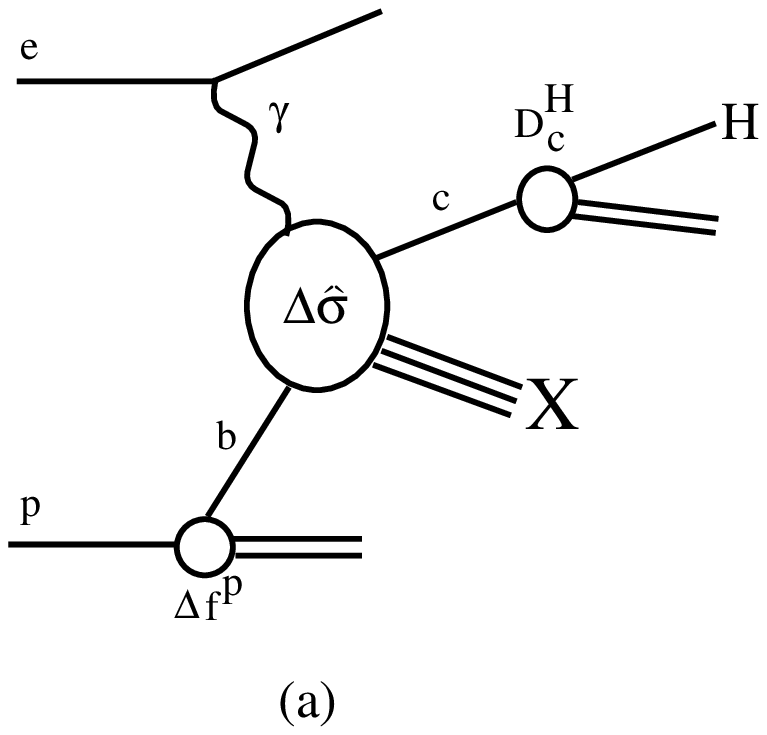,width=0.35\textwidth}

\vspace*{-4.2cm}
\hspace*{6.6cm}
\epsfig{figure=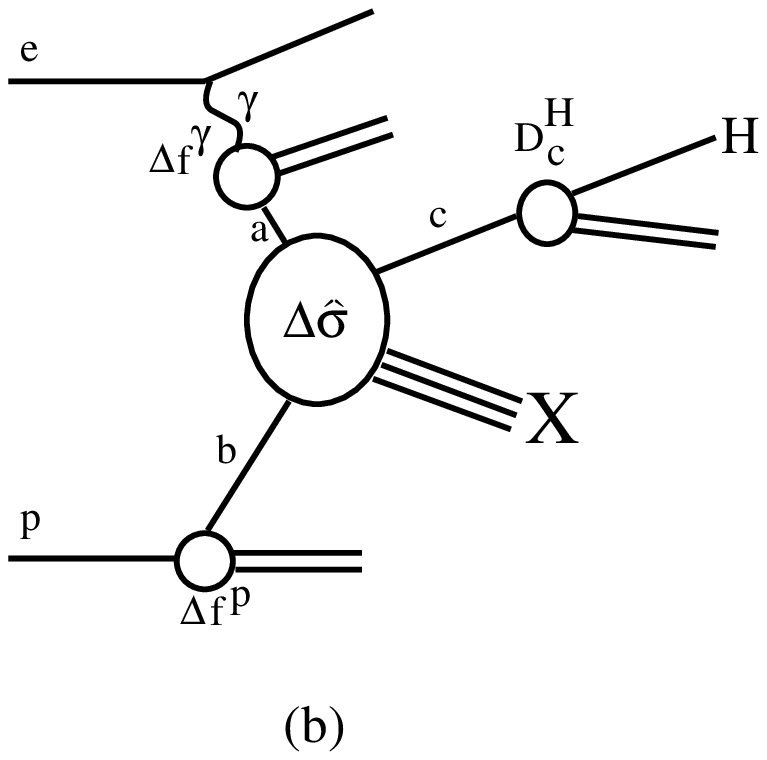,width=0.35\textwidth}

\vspace*{-0.5cm}
\caption{\sf Generic direct (a) and resolved (b) photon contributions to 
the process $lp\rightarrow l^{\prime}HX$. \label{fig:gamma}}
\end{figure}

The structure of the photon manifests itself in so-called ``resolved''
contributions to cross sections. 
We show this in Figure~\ref{fig:gamma}
for the case of photoproduction of hadrons. On the left, the photon
participates itself in the hard scattering, through ``direct'' 
contributions. On the right, the photon behaves like a hadron.
This possibility occurs because of (perturbative) short-time
fluctuations of the photon into $q\bar{q}$ pairs and gluons, and
because of (non-perturbative) fluctuations into vector mesons $\rho,\phi,
\omega$ with the same quantum numbers \cite{klasen}. 
The resolved contributions have been firmly established 
by experiments in $e^+ e^-$ annihilation and $ep$ scattering 
\cite{nisius}. 

A unique application of eRHIC would be to study the parton 
distributions of {\it polarized} quasi-real photons, defined as 
\cite{polga,ref:lomodels}
\begin{equation} \label{qdefgam}
\Delta f^{\gamma}(x)   \equiv f_+^{\gamma_{+}}(x) -  
f_-^{\gamma_{+}}(x) \; ,
\end{equation}
where $f_+^{\gamma_{+}}$ $(f_-^{\gamma_{+}})$ 
denotes the density of a parton $f=u,d,s,\ldots,g$ with positive (negative)
helicity in a photon with positive helicity. The $\Delta f^{\gamma}(x)$
give information on the spin structure of the photon; they
are completely unmeasured so far. 

\begin{figure}[htbp!]
\hspace*{0cm}
\epsfig{figure=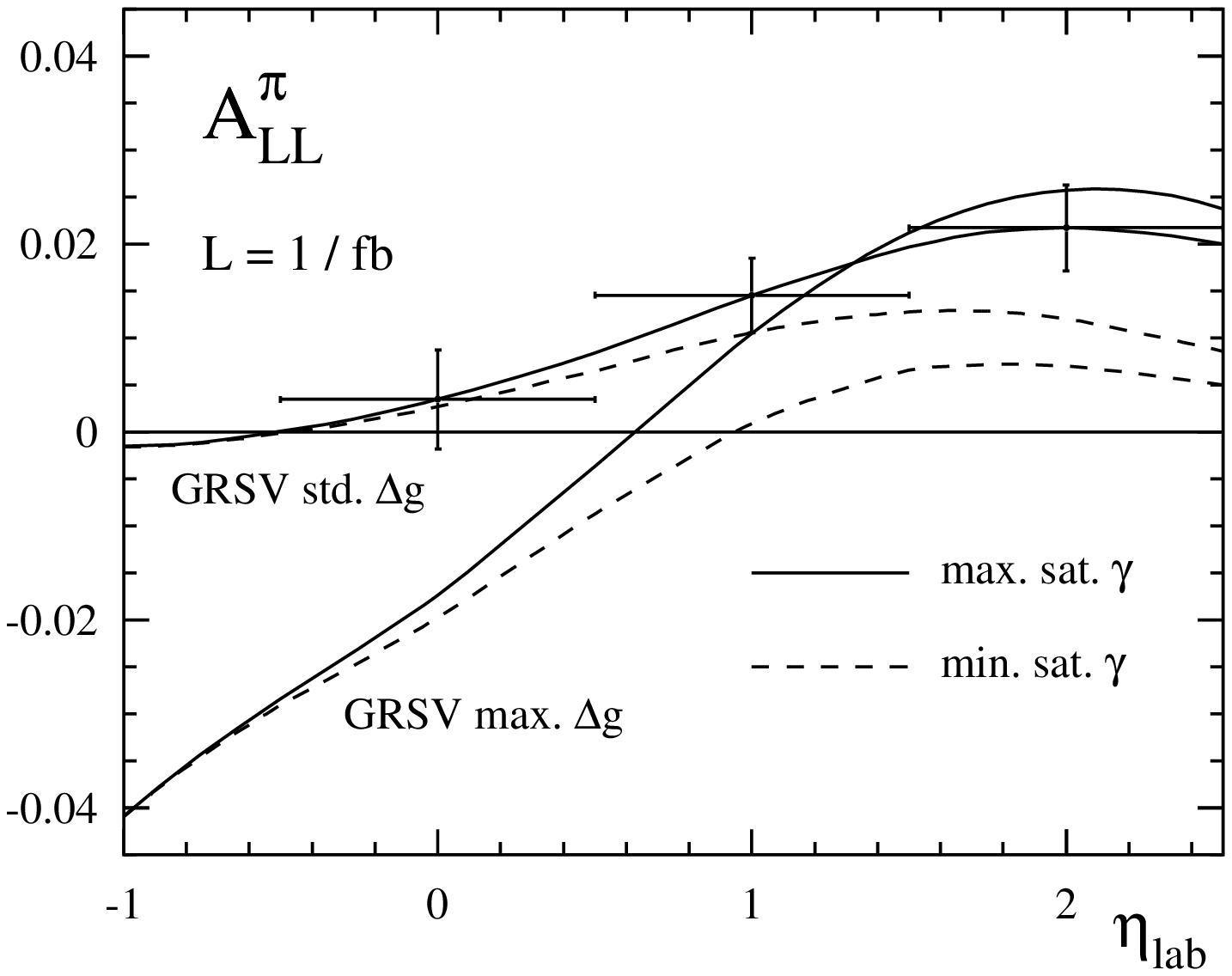,width=0.52\textwidth}

\vspace*{-6cm}
\hspace*{7.4cm}
\epsfig{figure=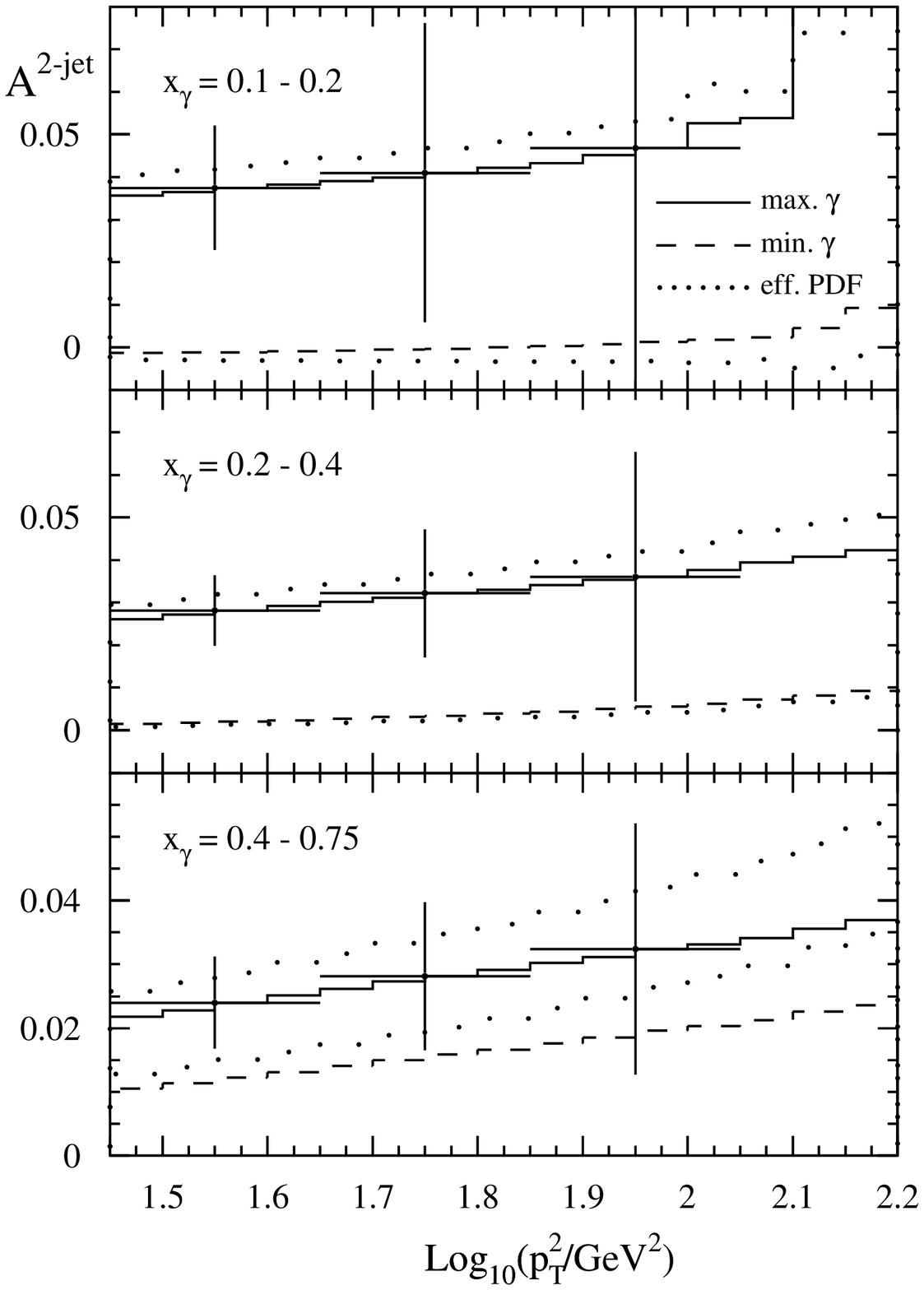,width=0.46\textwidth}

\caption{\label{gammastr} Left: spin asymmetry for $\pi^0$ photoproduction 
in NLO QCD for two sets of polarized photon densities and two different
choices of spin-dependent proton distributions. The error bars
indicate the statistical accuracy anticipated for eRHIC assuming
an integrated luminosity of $1\,\mathrm{fb}^{-1}$. Right: spin asymmetry
for dijet production as a function of the jet transverse momentum,
in three bins of the photon momentum fraction $x_{\gamma}$.}
\end{figure}

Figure~\ref{gammastr} shows samples from studies~\cite{strvo,nlohad} 
for observables at eRHIC that would give information on the 
$\Delta f^{\gamma}(x)$. Two models for the
$\Delta f^{\gamma}(x)$ were considered \cite{ref:lomodels}, one 
with a strong polarization of partons in the photon (``maximal'' set), 
the other with practically unpolarized partons (``minimal'' set).
On the left, we show the double-spin asymmetry for photoproduction
of high-$p_T$ pions, as a function of the pion's pseudorapidity $\eta_{\mathrm{lab}}$
in the eRHIC laboratory frame. The advantage of this observable
is that for negative $\eta_{\mathrm{lab}}$, in the proton 
backward region, the photon mostly interacts ``directly'', via
the process $\gamma g\to q\bar{q}$, 
whereas its partonic content becomes visible at positive 
$\eta_{\mathrm{lab}}$. This may be seen from the figure,
for which we have also used two different sets of polarized 
parton distributions of the proton \cite{grsv}, mainly 
differing in $\Delta g(x)$. 

The right part of Figure~\ref{gammastr} shows predictions for the
spin asymmetry in dijet photoproduction at eRHIC. If one assumes
the jets to be produced by a $2\to 2$ partonic hard scattering,
the jet observables determine the momentum fractions $x_{p,\gamma}$
of the partons in the proton and the photon. Selecting events 
with $x_{\gamma}<1$, one therefore directly extracts the 
``resolved''-photon contribution. At higher orders, this
picture is somewhat diluted, but remains qualitatively intact. Such
measurements of dijet photoproduction cross sections at HERA 
\cite{h1dijet} have been particularly successful in providing information 
on photon structure. This makes the spin asymmetry a good
candidate for learning about the $\Delta f^{\gamma}$ at eRHIC.
In the figure we show results for the asymmetry in three
different bins of $x_{\gamma}$. One can see that with 
1~fb$^{-1}$ luminosity one should be able at eRHIC to 
establish the existence of polarized resolved-photon
contributions, and distinguish between our ``maximal''
and ``minimal'' photon scenarios. For a first exploration
one could also use the approach of ``effective''
parton densities considered in \cite{strvo,h1dijet,ref:effpdf,ref:h3}.

We finally note that measurements of the polarized total photoproduction 
cross section at high energies would also give new valuable information 
on the high-energy contribution to the Drell-Hearn-Gerasimov
sum rule \cite{bass}. The latter relates the total cross
sections with photon-proton angular momentum $3/2$ and $1/2$ 
to the anomalous magnetic moments of the nucleon \cite{dhg}:
\begin{equation}
\int_0^{\infty} \;\frac{d\nu}{\nu}\;\left[
\sigma_{3/2}(\nu) \;-\; \sigma_{1/2}(\nu) \; \right]\;=\;
\frac{2\pi^2 \alpha}{M^2}\kappa^2\;
=\; \left\{ \begin{array}{ll} 204.5\; \mbox{$\mu$b} & \;p\\ 
 232.8\; \mbox{$\mu$b} & \;n  \end{array}\right. \; ,
\end{equation}
where on the right we have given the numerical values
of the sum rule. Currently, the experimental result
for the proton is a few percent high, and the one for
the neutron about $20\%$ low \cite{helbing}. There is practically no
information on the contribution to the sum rule from
photon energies $\nu\geq 3$ GeV; estimates based on 
Regge theory indicate that it is possible that 
a substantial part comes from this region. Measurements
at eRHIC could give definitive answers here. 
The H1 and ZEUS detectors at DESY routinely take data using 
electron taggers situated in the beam pipe 6 - 44 meters 
away from the end of the detectors.  They detect the scattered 
electrons from events at very low $Q^2$ and scattering angles.  
If electron taggers were included in eRHIC, similar 
measurements could be performed.  The $Q^2$ range of such 
measurements at eRHIC is estimated to be $10^{-8}- 10^{-2}$ 
GeV$^2$.

\subsubsection{Hard exclusive processes}

As we have discussed in Subsection~\ref{contrib}, generalized parton
distributions (GPDs) are fundamental elements of nucleon structure.
They contain both the parton distributions and the nucleon form factors 
as limiting cases, and they provide information on the spatial distribution
of partons in the transverse plane. GPDs allow the description of
exclusive processes at large $Q^{2}$, among them DVCS. It is hoped
that eventually these reactions will provide information on the
total angular momenta carried by partons in the proton. See for example
Eq.~(\ref{Jtot}).

The experimental 
requirements for a complete investigation of GPDs are for\-mi\-da\-ble. Many different 
processes need to be investigated at very high luminosities, at large enough $Q^{2}$, 
with polarization, and with suitable resolution to determine reliably the
hadronic final state.
The main difficulty, however, for experimental measurements of exclusive reactions is 
detecting the scattered proton. If the proton is not detected, a ``missing-mass'' 
analysis has to be performed. In case of the DVCS reaction, there may be a significant 
contribution from the Bethe-Heitler process. The amplitude for the Bethe-Heitler process is
known and, as we discussed in Subsection~\ref{contrib}, one may construct beam-spin and 
charge asymmetries to partly eliminate the Bethe-Heitler contribution. Early detector 
design studies have been performed for the EIC~\cite{Roman_pots}. 
These studies indicate that the acceptance can be significantly increased 
by adding stations of Silicon-strip-based Roman Pot Detectors away from the central detector
in a HERA-like configuration.
The detector recently proposed for low-$x$ and low-$Q^{2}$ studies at the 
EIC~\cite{Caldwell} (for details, see Section~\ref{detecsec})
may also be of significant use to measure the scattered proton. Further 
studies are underway and will proceed along with iterations of the design of 
the interaction region and of the beam line. 

Although more detailed studies need yet to be performed, we anticipate that
the EIC would provide excellent possibilities for studying GPDs. 
Measurements at the collider will complement those now underway at 
fixed target experiments and planned with the 12-GeV upgrade at the Jefferson 
Laboratory~\cite{jlab12gev}.

\subsection{Exploring the nucleus with an electron-ion collider}

In this section, we discuss the scientific opportunities available with the 
EIC in DIS off nuclei.  At very high energies, the correct degrees of freedom 
to describe the structure of nuclei are quarks and gluons.  The current 
understanding of partonic structure is just sufficient to suggest that their 
behavior is non-trivial.  The situation is reminiscent of Quantum  
Electrodynamics.  The rich science of condensed matter physics took a 
long time to develop even though the nature of the interaction was well 
understood.  Very little is known about the Ócondensed matterÔ, many-body 
properties of QCD, particularly at high energies.  There are sound reasons 
based in QCD to believe that partons exhibit remarkable collective phenomena 
at high energies. Because the EIC will be the first electron-ion collider, 
we will be entering a {\it terra incognita} in our understanding of 
the properties of quarks and gluons in nuclei. The range in $x$ and $Q^2$ and the 
luminosity will be greater than at any 
previous fixed target DIS experiment. Further, the collider environment is ideal 
for studying semi-inclusive and exclusive 
processes. Finally, it is expected that a wide range of particle species and beam 
energies will be available to study 
carefully the systematic  variation of a wide range of observables with target size 
and energy. 

We will begin our discussion in this section by discussing inclusive "bread and butter" observables such as the 
inclusive nuclear quark and gluon structure functions. As we observed previously, very little is known about nuclear 
structure functions at small $x$ and $Q^2 \gg \Lambda_{\rm QCD}^2\sim 0.04$ GeV$^2$. This is especially true of 
the nuclear gluon distribution.  We will discuss the very significant contributions that the EIC can make in rectifying 
this situation. A first will be a reliable extraction of the longitudinal structure function at small $x$. 
Much progress has been made recently in defining universal diffractive structure functions~\cite{cfs,Collins,VenezianoTrentadue}. 
These structure functions can be measured in nuclei for the first time. Generalized parton 
distributions will help provide a three-dimensional snapshot of the distribution of partons in the nucleus~\cite{ft}. 

We will discuss the properties of partons in a nuclear medium and the experimental observables that will enable us 
to tease out their properties. These include nuclear fragmentation functions that contain valuable information on hadronization in a nuclear environment. The momentum distributions of hadronic final states as functions of $x$, $Q^2$, 
and the fraction of the parton energy carried by a hadron also provide insight into dynamical effects such as parton energy 
loss in the nuclear medium. 

A consequence of small $x$ evolution in QCD is the phenomenon of parton saturation~\cite{GLR}. This arises 
from the competition between attractive Brems\-strahlung~\cite{BFKL} and repulsive screening and recombination 
(many body) effects \cite{MuellerQiu}, which results in a phase space density of partons of order $1/\alpha_S$. At 
such high parton densities, the partons in the wavefunction form a Color Glass Condensate (CGC) for 
reasons we will discuss later~\cite{IV}. The CGC is an effective theory describing the remarkable universal properties of partons 
at high energies. It provides an organizing principle for thinking about high energy scattering and has important ramifications 
for colliders. The evolution of multi-parton correlations 
predicted by the CGC can be studied with high precision in lepton-nucleus collisions.

Experimental observables measured at the EIC can be compared and contrasted with 
observables 
extracted in proton/Deuteron-nucleus and nucleus-nuc\-le\-us 
scattering experiments at RHIC and LHC. The kinematic 
reach of the EIC will significantly overlap with these experiments. Measurements of parton structure functions and 
multi-parton correlations in the nuclear wave function will provide a deeper understanding of the initial conditions 
for the formation of a quark gluon plasma (QGP). Final state interactions in heavy ion collisions such as the energy loss of leading hadrons in hot matter (often termed "jet quenching") are considered strong indicators of the formation of the 
QGP. The EIC will provide benchmark results for cold nuclear matter which will help quantify  energy loss in hot matter. 
Finally, recent results on inclusive hadron production in RHIC D-Au collisions at 200 GeV/nucleon show hints of the high parton density effects predicted by the CGC. We will discuss these and consider the similarities and differences between a p/D-A and an e-A collider.

\subsubsection{Nuclear parton distributions}
\begin{figure}[!h]
\vspace*{5mm}
\hspace*{25mm}
\epsfig{file=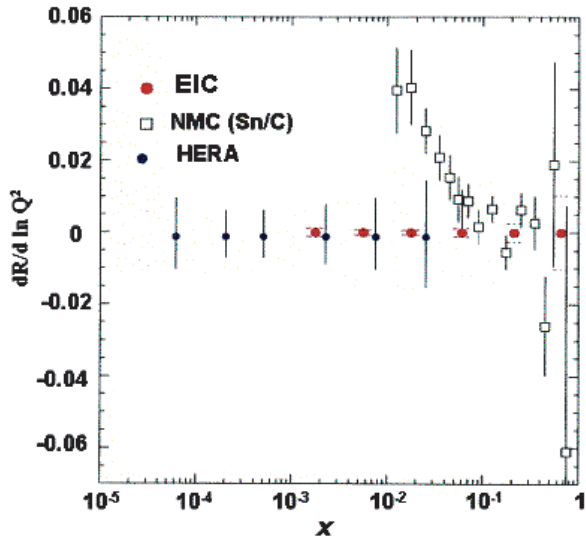,width=8cm,angle=0}
\caption{ The projected statistical accuracy of ${\partial {F_2^A/F_2^N} \over \partial \ln Q^2}$ as a function of $x$ 
for an integrated luminosity of 1 pb$^{-1}$ at the EIC~\cite{Sloan}. The simulated data are compared to previous data from the NMC 
and to data from a hypothetical e-A collider at HERA energies.}
\label{fig:nuclear-simul}
\end{figure}

The range of the EIC in $x$ and $Q^2$ was discussed previously
(see Fig.~\ref{fig:xq2}). It is 
significantly larger than for the previous fixed target 
experiments. The projected statistical accuracy, per inverse picobarn of data, of a measurement of the ratio 
${\partial R \over \partial \ln Q^2}$ versus $x$ at the EIC relative to data from 
previous NMC measurements and a hypothetical future e-A collider at HERA energies is 
shown in Fig.~\ref{fig:nuclear-simul}. Here $R$ denotes the ratio 
of nuclear structure functions, $R=F_2^A/F_2^N$. As discussed previously, 
the logarithmic derivative with $Q^2$ of this ratio can be used 
to extract the nuclear gluon distribution. The EIC is projected to have an integrated luminosity of several hundred pb$^{-1}$ for large 
nuclei, so one can anticipate high precision measurements of nuclear structure functions at small $x$. In particular, because the 
energy of the colliding beams can be varied, the nuclear longitudinal structure function can be measured for the first time 
at small $x$. At small $x$ and large $Q^2$, it is directly proportional to the gluon distribution.  At smaller values of $Q^2$, 
it may be more sensitive to higher twist effects than $F_2$~\cite{BartelsPeters}. 

\begin{figure}[!h]
\hspace*{10mm}
\epsfig{file=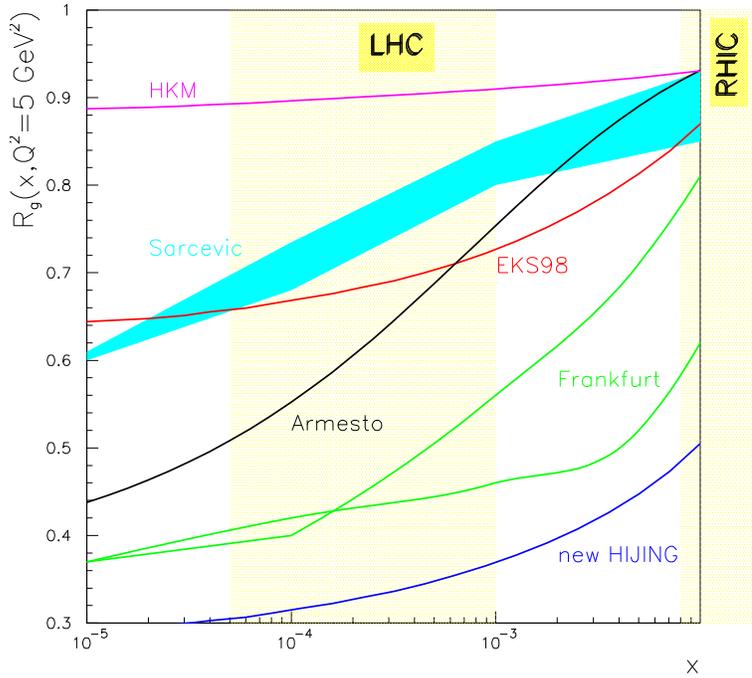,width=10cm,angle=0}
\caption[*]{Ratio of the gluon distribution in Lead to that in a proton, normalized by the number of nucleons, plotted as a function of $x$ for a fixed $Q^2=5$ GeV$^2$. From Ref.~\cite{Nestor}. 
Captions denote models-HKM~\cite{Kumano}, EKS98~\cite{EKS98},  Sarcevic~\cite{Sarcevic}, Armesto~\cite{Armesto1a}, 
Frankfurt~\cite{FrankfurtGMcS}, Hijing~\cite{Hijing}. The vertical bands denote the accessible $x$ regions
at central rapidities at RHIC and LHC.}
\label{fig:Nestor}
\end{figure}

Measurements of nuclear structure functions in the low $x$ kinematic region will test the predictions of the QCD evolution 
equations in this kinematic region. The results of QCD evolution with $Q^2$
depend on input from the structure functions at smaller values of $Q^2$ 
for a range of $x$ values. The data on these is scarce for nuclei. 
These results 
are therefore very sensitive to models of the small $x$ behavior of structure functions at low $Q^2$. A nice plot 
from Ref.~\cite{Nestor} reproduced in Fig.~\ref{fig:Nestor} clearly illustrates the problem. Figure~\ref{fig:Nestor} 
shows results from theoretical models  for the ratio of the gluon distribution in Lead to that in a proton as a function of $x$. Though 
all the models employ the same QCD evolution equations, the range in uncertainty is rather large at small $x$ -- about 
a factor of 3 at $x\sim 10^{-4}$. Although one can try to 
construct better models, the definitive constraint can only come from experiment. 

The shadowing of gluon distributions shown in Fig.~\ref{fig:Nestor} is not understood in a fundamental way. 
We list here some relevant questions which can be addressed 
by a future electron-ion collider.

\begin{itemize}

\item
 Is shadowing a leading twist effect, namely, is it unsuppressed by a power of  $Q^2$ ?  Most models of 
 nuclear structure functions at small $x$ assume this is the case. (For a review,  see Ref.~\cite{PillerWeise}.) Is  
 there a regime of $x$ and $Q^2$, where power corrections due to high parton density effects can be seen?~\cite{GLR,MuellerQiu,MV}
 
 \item 
 What is the relation of shadowing to parton saturation? As we will discuss, parton saturation dynamically gives rise to a semi-hard scale in nuclei. 
 This suggests that shadowing at small $x$ can be understood in a weak coupling analysis. 
 
 \item 
 Is there a minimum to the shadowing ratio for fixed  $Q^2$ and $A$ with decreasing $x$?  If so, is it reached faster for gluons or for quarks?  
 
 \item 
The Gribov relation between shadowing and diffraction that we discussed previously is well established at low 
parton densities. How is it modified at high parton densities? The EIC can test this relation directly 
by measuring diffractive structure functions in ep (and e-A) and shadowing in e-A collisions.

\item  
Is shadowing universal?  For instance, would gluon parton distribution functions extracted from 
p-A collisions at RHIC be identical to those extracted from e-A in the same kinematic regime?  The naive assumption that this is the case may be 
false if higher twist effects are important. Later in this review, we will discuss 
the implications of the possible lack of universality 
for p-A and A-A collisions at the LHC.

\end{itemize}

We now turn to a discussion of diffractive structure functions. At HERA,  hard diffractive events were observed 
where the proton remained intact and the 
virtual photon fragmented into a hard final state producing a large rapidity gap between the projectile and target.  
A rapidity gap is a region in rapidity essentially
devoid of particles. In pQCD, the probability of a
gap is exponentially suppressed as a function of the gap size. At HERA
though, gaps of several units in rapidity are relatively unsuppressed; one finds
that roughly 10\% of the cross--section corresponds to hard
diffractive events with invariant masses $M_X > 3$ GeV. The remarkable
nature of this result is transparent in the proton rest frame: a $50$ TeV
electron slams into the proton and, 10\% of the time, the proton is
unaffected, even though the interaction causes the virtual photon to
fragment into a hard final state.

\begin{figure}[!h]
\hspace*{32mm}
\epsfig{file=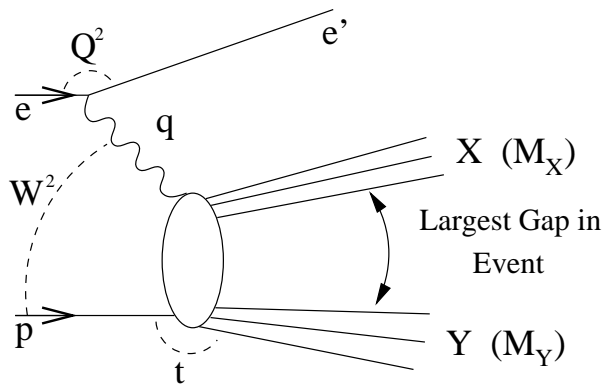,width=8cm,angle=0}
\caption[*]{The diagram of a process with a rapidity gap between the systems 
$X$ and $Y$. The projectile nucleus is denoted here as p. Figure from 
Ref.~\cite{Arneodoetal}.}
\label{fig:madif}
\end{figure}

The interesting question in diffraction is the nature of the
color singlet object (the ``Pomeron'') within the proton that
interacts with the virtual photon. This interaction probes, in a 
novel fashion, the nature of confining
interactions within hadrons. (We will discuss later the possibility that one can study in diffractive events the interplay between 
strong fields produced by confining interactions and those generated by high parton densities.) In hard diffraction, 
because the invariant mass of
the final state is large, one can reasonably ask questions about the
quark and gluon content of the Pomeron. A diffractive structure function 
$F_{2,A}^{D(4)}$ can be defined~\cite{BereraSoper,VenezianoTrentadue,Collins}, in a fashion analogous to $F_2$, as
\be
{d^4 \sigma_{eA\rightarrow eXA}\over {dx_{Bj} dQ^2 d\xp dt}} & &=
A\cdot {4\pi \alpha_{\rm{em}}^2\over x Q^4}
 \left\{ 1-y + 
{y^2\over 2[1+ R_A^{D(4)}(\beta,Q^2,\xp,t)]}\right\}\nonumber \\
&\times & F_{2,A}^{D(4)}(\beta,Q^2,\xp,t) \, ,
\ee
where, $y=Q^2/s x_{Bj}$, and analogously to $F_2$, one has  
$R_A^{D(4)}=F_L^{D(4)}/F_T^{D(4)}$. Further, $Q^2= -q^2 > 0$, $x_{Bj} = Q^2/ 2Pq$, 
$\xp = q(P-P^\prime)/ q P$, $t = (P-P^\prime)^2$
and $\beta = x_{Bj}/\xp$. Here $P$ is the initial nuclear momentum, and 
$P^\prime$ is the net momentum of the fragments $Y$ in the proton 
fragmentation region. Similarly, $M_X$ is the invariant mass of the 
fragments $X$ in the electron fragmentation region. An illustration of 
the hard diffractive event is shown in Fig.~\ref{fig:madif}. 

It is more convenient in practice to measure the structure 
function $F_{2,A}^{D(3)}=\int F_{2,A}^{D(4)} dt$, where 
$|t_{min}|<|t|<|t_{max}|$, where $|t_{min}|$ is the minimal momentum 
transfer to the nucleus, and $|t_{max}|$ is the maximal momentum transfer 
to the nucleus that still 
ensures that the particles in the nuclear fragmentation region $Y$ are 
undetected. An interesting quantity to measure is the ratio 
$R_{A1,A2}(\beta,Q^2,\xp) = {F_{2,A1}^{D(3)}(\beta,Q^2,\xp)\over 
F_{2,A2}^{D(3)}(\beta,Q^2,\xp)}$. The $A$-dependence of this quantity will contain very 
useful information about the universality of the structure of the Pomeron. In a
study for e-A collisions at HERA, it was argued that this ratio could be measured with high 
systematic and statistical accuracy~\cite{Arneodoetal}--the situation for eRHIC should be 
at least comparable, if not better. Unlike $F_2$ however, $F_2^{D}$  is not truly universal--it cannot be applied, for instance, 
to predict diffractive cross sections in p-A scattering; it can 
be applied only in other lepton--nucleus scattering studies~\cite{Collins,cfs}. 
This has been confirmed by a study where 
diffractive structure functions measured at HERA were used as an input in computations 
for hard diffraction at Fermilab. 
The computations vastly overpredicted the Fermilab data on hard diffraction~\cite{Whitmore}.
Some of the topics discussed here will be revisited in our discussion of high 
parton densities. 

\subsubsection{Space-time evolution of partons in a nuclear environment}

The nuclear structure functions  are inclusive observables and 
are a measure of the properties of the nuclear wavefunction. 
Less inclusive observables, which measure these properties in greater detail, will be discussed in the section on the Color Glass 
Condensate. In addition to studying the wave function, we are interested in the properties of partons as they interact with the 
nuclear medium. These are often called final state interactions to distinguish them from the initial state interactions in the wavefunction. 
Separating which effects arise from the wavefunction is not easy because our interpretation of initial state and some final state interactions 
may depend on the gauge in which the computations are 
performed~\cite{KovMueller}. Isolating the two effects in experiments is difficult. 
A case in point is the study of energy loss effects on final states in p-A collisions~\cite{PengGarvey}. These effects are 
not easy to distinguish from shadowing effects in the wavefunction. Nevertheless, in the right kinematics this can be done. 

In section 2, we discussed various final state in-medium QCD processes such as color transparency, parton energy loss and the 
medium modification of fragmentation functions. The EIC will enable qualitative progress in studies of the space-time picture of strong interactions relative to previous fixed target DIS experiments. The reasons for this are as follows.

\begin{itemize}

\item 
The high luminosity of the EIC will increase by many orders of magnitude the current data sample of final states in  DIS scattering off nuclei at high energies.

\item
The EIC will provide a much broader range of $Q^2$ and $x$, making it possible to compare dynamics for approximately the same space-time coherence lengths  as a function of $Q^2$. Fixing the coherence length of partons will allow one to distinguish events wherein a photon is transformed into a strongly interacting system either outside or inside the nucleus. This will help isolate initial state interactions from those in the final state.

\item The collider geometry will enable measurements of final states currently impossible in fixed target kinematics.  In particular, a hermetic detector would clearly isolate coherent processes as well as quasi-elastic processes in DIS off nuclei. In addition, one can study the sizes and distributions of 
rapidity gaps as a function of nuclear size and energy. These will provide a sensitive probe of the interplay between space-time correlations in the final state 
and in the nuclear wavefunction.

\item The detection of nucleons produced in the nuclear fragmentation region would make it feasible to study DIS as a function of the number of the nucleons involved in the interaction.  
In particular, it may be possible to study impact parameter dependence of final states, which will be important to 
understand in detail the nuclear amplification of final state effects. 
In addition, the impact parameter dependence will help distinguish geometrical effects from 
dynamical effects  in event-by-event studies of final states. 

\end{itemize}

In section 2, we discussed Generalized Parton Distributions (GPDs) in the context of 
DIS scattering off  nucleons. These GPDs can 
also be measured in DIS scattering off nuclei~\cite{gpdrev}. The simplest system 
in which to study GPDs is the deuteron. The transition 
from $D\rightarrow p+n$ in the kinematics where the neutron absorbs the momentum 
transfer in the scattering is sensitive to the 
GPD in the neutron with the proton playing the role of a spectator~\cite{CanoPire}. 
Some preliminary studies have been done for 
heavier nuclei~\cite{dvcsnucl}. Certain higher twist correlations in nuclei which 
scale as $A^{4/3}$ 
are sensitive to nucleon GPDs~\cite{Wang1}. This leads us to a discussion of GPDs 
in nuclei at small $x$. 
As we will discuss, high parton density effects are enhanced in large nuclei.
$k_\perp$ dependent GPDs might provide the right 
approach to study this novel regime~\cite{MartinRyskin}. The study of nuclear GPDs 
at moderate and small $x$ is a very promising, albeit 
nascent, direction for further research to uncover the detailed structure of hard 
space-time processes in nuclear media. These nuclear 
distributions can be studied for the first time with the EIC.

\subsubsection{The Color Glass Condensate}

The Color Glass Condensate (CGC) is an effective field theory describing the properties of the dominant parton configurations in 
hadrons and nuclei at high energies~\cite{IV}. The degrees of freedom are partons--which 
carry color charge--hence the ``Color" in CGC. The 
matter behaves like a glass for the following reason. The kinematics of high energy scattering dictates a natural separation 
between large $x$ and small $x$ modes~\cite{Susskind}. The large $x$ partons at high energies behave like frozen random light cone 
sources over time scales that are large compared to the dynamical time scales associated with the small-$x$ partons. 
One can therefore describe an effective theory where the small $x$ partons are dynamical fields and the 
large-$x$ partons are frozen sources~\cite{MV}.
Under quantum evolution~\cite{JIMWLK}, this induces a stochastic coupling between the wee partons via their interaction with the 
sources. This stochastic behavior is very similar to that of  a spin glass. Finally, the Condensate in CGC arises because each of these colored configurations is very similar to 
a Bose-Einstein Condensate. The occupation number of the gluons can be computed to be of order $1/\alpha_S$, and the 
typical momentum of the partons in the configuration is peaked about a typical 
momentum--the saturation momentum $Q_s$.  
These properties are further enhanced by quantum evolution in $x$. Because the occupation number is so large, by the correspondence principle of quantum mechanics, the small $x$ modes can be treated as classical fields. The classical 
field retains its structure while the saturation scale, generated dynamically in the theory, grows with energy: $Q_s(x^\prime) > Q_s(x)$ for $x^\prime < x$. The CGC is sometimes used interchangeably with ``saturation''~\cite{reviews} -- 
both refer to the same phenomenon, the behavior of partons at large occupation numbers.

The Jalilian-Marian-Iancu-McLerran-Weigert-Leonidov-Kovner (JIMWLK) 
re\-nor\-malization group equations describe the properties of partons in the high density regime~\cite{JIMWLK}. They 
form an infinite hierarchy (analogous to the Bogoliubov-Born-Green-Kirkwood-Young (BBGKY) 
hierarchy in statistical mechanics) of ordinary differential equations for the gluon correlators $\langle
A_1 A_2 \cdots A_n\rangle_Y$, where $Y= \ln(1/x)$ is the rapidity. Thus the 
evolution, with $x$, of multi-gluon (semi-inclusive) final states provides 
precise tests of these equations. The full hierarchy of 
equations are difficult to solve\footnote{For a preliminary numerical attempt, 
see Ref.~\cite{RW}.} though there have been major theoretical 
developments in that direction recently~\cite{Ploops}. 

A mean field version of the JIMWLK equation, called the 
Balitsky-Kovchegov (BK) equation~\cite{BK}, describes the inclusive scattering of the 
quark-anti-quark dipole off the hadron in deeply inelastic scattering. In particular, 
the virtual photon-proton cross-section at small $x$ can be written as~\cite{Mueller94,NZ} 
\begin{equation}
\sigma_{T,L}^{\gamma^* p}=\int d^2 r_\perp \int dz |\psi_{T,L}(r_\perp,z,Q^2)|^2 
\sigma_{q{\bar q}N}(r_\perp,x) \, ,
\label{eq:CGC1} 
\end{equation}
where $|\psi_{T,L}|^2$ is the probability for a longitudinally (L) or transversely (T) 
polarized virtual photon to split into a quark with momentum fraction $z$  and 
an anti-quark with momentum fraction $1-z$ of the longitudinal momentum of the virtual 
photon.  For the quark and anti-quark located at ${\vec x_\perp}$ and ${\vec y_\perp}$ 
respectively from the target, their transverse size is ${\vec r_\perp}={\vec x_\perp}-
{\vec y_\perp}$, and the impact parameter of the collision is ${\vec b} =({\vec x_\perp} + 
{\vec y_\perp})/2$. The probability for this splitting is known exactly from 
QED and it is convoluted with the cross-section for the $q\bar q$-pair to 
scatter off the proton. This cross-section for a dipole 
scattering off a target can be expressed as
\begin{equation}
\sigma_{q\bar q N} (x, r_\perp) = 2\, \int d^2 b\, \, {\cal N}_Y (x,r_\perp,b) \, ,
\label{eq:CGC2}
\end{equation}
where ${\cal N}_Y$ is the imaginary part of the forward scattering amplitude. 
The BK equation~\cite{BK} for this amplitude has the operator form
\begin{equation}
{{\partial {\cal N}_Y}\over \partial Y} = {\bar \alpha_S}\, {\cal K}_{\rm BFKL} 
\otimes \left\{ {\cal N}_Y - {\cal N}_Y^2\right\} \, .
\label{eq:CGC3}
\end{equation}
Here ${\cal K}_{\rm BFKL}$ is the well known Balitsky-Fadin-Kuraev-Lipatov (BFKL) 
kernel~\cite{BFKL}. 
When ${\cal N} << 1$, the quadratic term is negligible and one has BFKL growth of 
the number of dipoles; when ${\cal N}$ is close to unity, 
the growth saturates. The approach to unity can be computed 
analytically~\cite{LevinTuchin}. The BK equation is the simplest 
equation including both the Bremsstrahlung responsible for the 
rapid growth of amplitudes at small $x$ as well as the repulsive many 
body effects that lead to a saturation of this growth. 

Saturation models, which incorporate key features of the CGC, explain several features of the HERA data. 
In section 2.1, we discussed 
the property of geometrical scaling observed at HERA which is satisfied by the LHS of Eq.~\ref{eq:CGC1},  where it scaled as a function of the ratio of $Q^2$ to the saturation scale $Q_s^2$. We also mentioned briefly a simple saturation model, the Golec-Biernat model~\cite{Golec-Biernat}, which captured essential features of this phenomenon in both inclusive and diffractive cross sections at HERA. Geometric scaling arises naturally in the Color Glass Condensate~\cite{IIM,MuellerDionysis}, and it has been studied 
extensively both analytically~\cite{MunierPeschanski} and numerically~\cite{Armesto1b,Albacete,GSM} for the BK equation. 
The success of saturation models, as discussed in section 2.1, in explaining less inclusive features of the HERA data is also 
encouraging since their essential features can be understood to follow from the BK equation. Below we will 
discuss the implications of mean field studies with the BK equation, as well as effects beyond BK. 

As mentioned previously, a very important feature of saturation is the dynamical generation of a dimensionful scale $Q_s^2 \gg \Lambda_{\rm QCD}^2$, which 
controls the running of the coupling at high energies: $\alpha_S(Q_s^2) 
\ll 1$. From the BK equation, or more generally, from 
solutions of BFKL in the presence of an absorptive boundary (corresponding to a CGC-like regime of high parton densities), one 
can deduce that, for fixed coupling, $Q_s^2$ has the asymptotic form $Q_s^2 = Q_0^2 \exp\left( c Y\right)$, where $c=4.8 \alpha_S$ and $Y=\ln(x_0/x)$. Here, $Q_0^2$ and $x_0$ are parameters from the initial conditions. Pre-asymptotic $Y$ dependent corrections 
can also be computed and are large. The behavior of $Q_s^2$ changes qualitatively 
when running coupling effects are taken into account. The state of the art is a 
computation of the saturation scale to next-to-leading order in BFKL with additional 
resummation of collinear terms that stabilize the predictions of NLO BFKL~\cite{Dionysis}. One recovers the 
form $Q_s^2 = Q_0^2 \exp\left(\lambda Y\right)$, now with small pre-asymptotic corrections, with $\lambda\approx 0.25$. Remarkably, 
this value is very close to the value extracted in the Golec-Biernat model from fits to the HERA data. 

\begin{figure}[!h]
\hspace*{10mm}
\epsfig{file=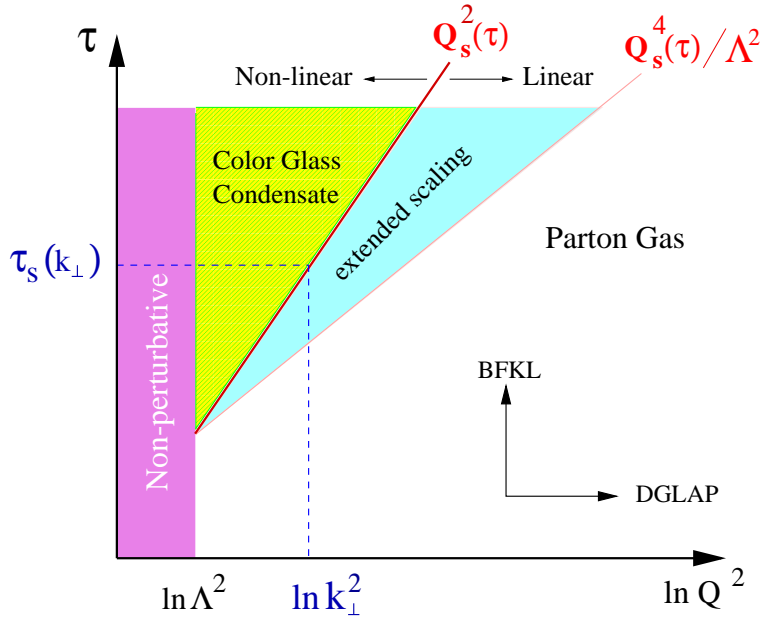,width=10cm,angle=0}
\caption[*]{A schematic plot of the Color Glass Condensate and extended scaling regimes in the $x$-$Q^2$ plane. Here $\tau =\ln(1/x)$ denotes 
the rapidity. From Ref.~\cite{IV}.}
\label{fig:glass}
\end{figure}

Fig.~\ref{fig:glass} shows a schematic plot of the CGC and extended scaling regions 
in the $x$-$Q^2$ plane. 
Clearly, with the wide kinematic range of the EIC, and the large number of available 
measurements--to be discussed later--one has the opportunity to make 
this plot quantitative. One can further 
add an additional axis for the atomic number to see how the kinematic reach of the CGC scales with $A$. In principle, one can 
also study the impact parameter dependence of the saturation scale in addition to the $A$-dependence. 

\subsubsection{Signatures of the CGC}
\vskip 0.1in
{\it Inclusive signatures.}
Inclusive measurements include $F_2$ and $F_L$ for a wide range of nuclei, the latter measurements being done 
independently for the first time. The data will be precise enough to extract derivatives of these with respect to $\ln Q^2$ and 
$\ln x$ in a wide kinematic range in $x$ and $Q^2$. Logarithmic derivatives of 
$F_2$ and $F_L$ will enable the extraction of 
the coefficient $\lambda$ of the saturation scale, which as discussed previously, is defined to be $Q_s^2 = Q_0^2 e^{\lambda Y}$, where $Y=\ln(x_0/x)$, and where $x_0$ and $Q_0^2$
are reference values corresponding to the initial conditions for small-$x$ evolution.
Simulations suggest that a precise extraction of this quantity may be feasible~\cite{Caldwell}. Except at asymptotic energies, $\lambda\equiv \lambda(Y)$. 
Predictions exist for ``universal" pre-asymptotic $Y$-dependent corrections to 
$\lambda$~\cite{MunierPeschanski}. Second derivatives of $F_2$ and $F_L$ with respect to $\ln(x_0/x)$ will be sensitive to these corrections. 
The logarithmic derivatives of $F_2$ and $F_L$ with $Q^2$, especially the latter, will be sensitive to higher twist effects for $Q^2\approx Q_s^2(x,A)$. 
The saturation scale is larger for smaller $x$ and larger $A$--thus deviations of predictions of CGC fits from DGLAP fits should systematically increase 
as a function of both. CGC fits have been shown to fit HERA data at small 
$x$~\cite{IIMunier,Forshaw}. These fits can be extended to nuclei and 
compared to scaling violation data relative to DGLAP fits. The $A$ 
dependence of the saturation scale can also be extracted from nuclear structure 
functions at small $x$. Again, predictions exist for the pre-asymptotic scaling 
of the saturation scale with rapidity (or $x$), for different $A$~\cite{MuellerQsA},
that can be tested against the data.  

In the BK equation (mean field approximation of the CGC renormalization group 
equations), we now have a simple way to make predictions for the effects of high parton 
densities on both inclusive and diffractive~\cite{KovTuchin} structure functions. There 
are now a few  preliminary computations for e-A DIS in this 
framework~\cite{ArmestoSalgado,LevinLublinsky}. Much more remains to be 
done--in particular, comparisons with DGLAP for EIC kinematics and detector cuts.  

\noindent
{\it Semi-inclusive and exclusive signatures.}
The collider geometry of the EIC will greatly enhance the semi-inclusive final states in e-A relative to previous fixed target experiments. 
Inclusive hadron production at $p_\perp\sim Q_s$ 
should be sensitive to higher twist effects for $Q^2\approx Q_s^2(x,A)$. For the largest 
nuclei, these effects should be clearly distinguishable from DGLAP based models. 
Important semi-inclusive observables are coherent (or diffractive) and inclusive vector meson production, which are sensitive measures of the 
nuclear gluon density~\cite{section2:46,section2:51}.
Exclusive vector meson production was suggested by Mueller, Munier and Stasto~\cite{MMS} as a way to extract 
the S-matrix (and therefore the saturation scale in the Golec-Biernat--W\"{u}sthoff 
parameterization) from the $t$-dependence of exclusive $\rho$-meson production.  A similar analysis of $J/\psi$ production was performed by Guzey et al.~\cite{GRSZ}.  These studies for e-A 
collisions will provide an 
independent measure of the energy dependence of the saturation scale in nuclei.
An extensive recent theoretical review of vector meson production of HERA (relevant for EIC studies as well) can be found in Ref.~\cite{INS}.

In hard diffraction, for instance, one should be able to distinguish predictions based on the strong field effects of BK (or hard pomeron based approaches in general) from the soft pomeron physics associated with confinement. As we discussed previously, some saturation models predict that 
hard diffractive events will constitute 30-40\% of the cross-section~\cite{Levin,Strikman}. These computations can be compared with DGLAP 
predictions which match soft Pomeron physics with hard perturbative physics. One anticipates that the latter would result in a much smaller fraction 
of the cross-section and should therefore be easily distinguishable from CGC based ``strong field" diffraction.

The BK renormalization group 
equation is not sensitive to multi-particle correlations. These are sensitive to effects such as Pomeron loops~\cite{Ploops}, although 
phenomenological consequences of these remain to be explored. These effects are reflected in multiplicity fluctuations and 
rapidity correlations over several units in rapidity~\cite{AGK,KovchegovLM}. One anticipates 
quantitative studies of these will be developed in the near future. A wide detector coverage able to resolve the detailed structure of events 
will be optimal for extracting signatures of the novel physics of high parton densities. 

\subsubsection{Exploring the CGC in proton/Deuteron-nucleus collisions}

Although high parton density hot spots may be studied in pp collisions, they are notoriously hard to observe. The proton is a dilute object, except at small impact parameters, and one 
needs to tag on final states over a wide $4\pi$ coverage. Deuteron-nucleus experiments 
are more promising in this regard. They have been performed at RHIC and may be performed 
at LHC in the future. The Cronin effect discovered in the late 70's~\cite{section2:58} 
predicts a hardening of the transverse momentum spectrum in proton-nucleus collisions, 
relative to proton-proton collisions at transverse momenta of order 
$p_\perp \sim 1-2$~GeV. It disappears at much larger $p_\perp$. 
A corresponding depletion is seen at low transverse 
momenta. The effect was interpreted as arising from the multiple scatterings of 
partons from 
the proton off partons from the nucleus \cite{Krzywicki}.  

\begin{figure}[htb!]
\begin{center}
\resizebox*{!}{4.0cm}{\includegraphics{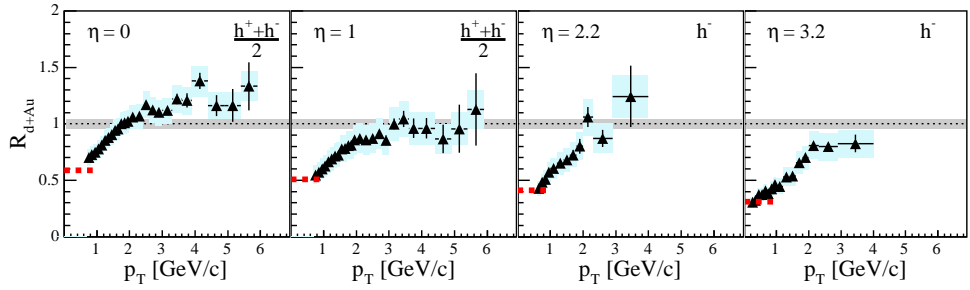}}
\end{center}
\caption{\label{fig:eighteen} Depletion of the Cronin peak from $\eta=0$ to $\eta=3$ for minimum bias events. From Ref.~\cite{BRAHMSdA}.}
\end{figure}

\begin{figure}[htb!]
\begin{center}
\resizebox*{!}{4.0cm}{\includegraphics{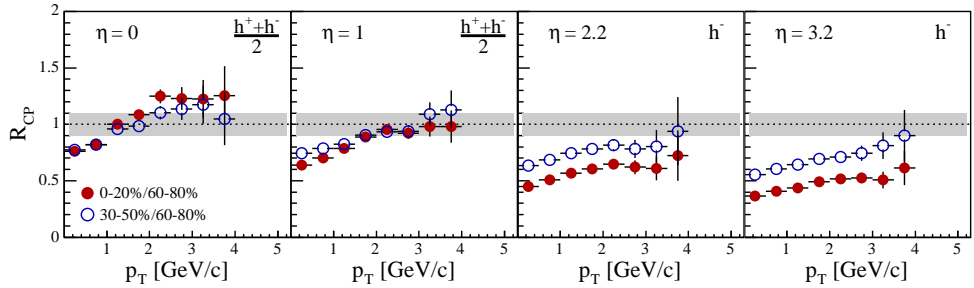}}
\end{center}
\caption{\label{fig:nineteen} Centrality dependence of the Cronin ratio as a function of rapidity. From Ref.~\cite{BRAHMSdA}.}
\end{figure}

First data from RHIC on forward D-Au scattering at $\sqrt{s} = 200$ GeV/nucleon demonstrate how the Cronin effect is
modified with energy or, equivalently, with the rapidity. The $x$
values in nuclei probed in these experiments, at $p_\perp \sim 2$~GeV, range
from $10^{-2}$ in the central rapidity region down to $10^{-4}$ at
very forward rapidities\footnote{It has been argued~\cite{GSV}, however, 
that the forward D-A cross section in the BRAHMS kinematic regime receives
sizable contributions also from rather large $x$ values.}
At central rapidities, one clearly sees a Cronin peak at $p_\perp \sim 1-2$ GeV. A dramatic result obtained by the BRAHMS~\cite{BRAHMSdA} experiment at RHIC\footnote{The 
trends seen by BRAHMS are
also well corroborated by the PHOBOS, PHENIX and STAR experiments at RHIC 
in different kinematic
ranges~\cite{PHENIXdA,STARdA,PHOBOSdA}.} is the rapid shrinking of the Cronin peak with rapidity shown in Fig.~\ref{fig:eighteen}.  In Fig.~\ref{fig:nineteen},  the centrality dependence of the effect is shown. At central rapidities, the Cronin peak is enhanced in more central collisions. For forward rapidities, the trend is reversed: more central
collisions at forward rapidities show a greater suppression than less central collisions! 

Parton distributions in the classical theory of the CGC exhibit the Cronin effect~\cite{DumitruJamal,GelisJamal,BGV1}. However, unlike this classical Glauber picture~\cite{Accardi1}, quantum evolution in the CGC  shows that it breaks down completely when the $x_2$ in the target is such that $\ln(1/x_2)\sim 1/\alpha_S$.  This is precisely the trend observed in the RHIC D-Au experiments~\cite{BRAHMSdA}. The rapid depletion of the Cronin effect 
in the CGC picture is due  to the onset of BFKL evolution, whereas 
the subsequent saturation of this trend reflects the onset of saturation 
effects~\cite{Cronin-theory}. The inversion of the centrality 
dependence can be explained as arising from the onset of BFKL 
anomalous dimensions, that is, 
the nuclear Bremsstrahlung spectrum changes from $Q_s^2/p_\perp^2 
\longrightarrow Q_s/p_\perp$. Finally, an additional piece of evidence in support of the CGC picture is the broadening of azimuthal correlations~\cite{KLM2} for which preliminary data now exists from the STAR collaboration~\cite{STAR}. 
We note that alternative explanations have been given to explain the BRAHMS 
data~\cite{HwaFries-BZK}. 
These ideas can be tested conclusively in photon and di-lepton production 
in D-A collisions at RHIC~\cite{BMS-Jamal-Betemps} as 
well as by more detailed correlation studies. 

Hadronic collisions in pQCD are often interpreted within the framework of collinear 
factorization. At high energies, $k_\perp$ factorization may be applicable~\cite{CCH} 
where the relevant quantities are ``unintegrated'' $k_\perp$ dependent 
parton densities. Strict $k_\perp$-factorization which holds for gluon production in p-A 
collisions~\cite{KovMueller,KKT} is broken for quark production~\cite{BGV2,NSZ}, for 
azimuthal correlations~\cite{JamalYuri} and diffractive 
final states~\cite{KovWied}. For a review, see Ref.~\cite{JK2}.
These cross-sections can still be written in terms of $k_\perp$-dependent multi-parton 
correlation functions~\cite{BGV2} and will also appear in DIS final states~\cite{KovTuchin}. 
DIS will allow us to test the universality of these correlations, that is, whether such 
correlations extracted from p-A collisions can be used to compute e-A final 
states~\cite{GelisJamal,Kopeliovich}.

\subsubsection{The Color Glass Condensate and the Quark Gluon Plasma}

The CGC provides the initial 
conditions for nuclear collisions at high energies. The number and energy of gluons released in a heavy ion collision of identical nuclei can be simply expressed in terms of the saturation scale as~\cite{AR,AYR,Lappi}
\begin{eqnarray}
{1\over \pi R^2}\,{dE\over d\eta} = {c_E\over g^2}\,Q_s^3\,\,\, , \,\,\,
{1\over \pi R^2}\,{dN\over d\eta} = {c_N\over g^2}\, Q_s^2 \, ,
\label{eq:41}
\end{eqnarray}
where $c_E\approx 0.25$ and $c_N\approx 0.3$. Here $\eta$ is the space-time 
rapidity. These simple predictions led to correct predictions for 
the hadron multiplicity at central rapidities in 
Au-Au collisions at RHIC~\cite{AR,ArmestoPajares} 
and for the centrality and rapidity dependence of hadron distributions~\cite{KLN}. 
However, the failure of more detailed comparisons to the RHIC jet quenching data~\cite{KLM} and elliptic flow data~\cite{AYR2}  
suggested that final state effects are important and significantly modify predictions based on the CGC alone. The success of 
hydrodynamic predictions suggests that matter may have thermalized to form a quark gluon plasma~\cite{Huovinen}. Indeed, 
bulk features of multiplicity distributions may be described by the CGC precisely as a consequence of early thermalization--leading 
to entropy conservation~\cite{HiranoNara}. Initial-state effects will be more important in heavy ion collisions at the LHC because one is probing smaller $x$ 
in the wave function. Measurements of saturation scales for nuclei at the EIC 
will independently corroborate equations such as Eq.~\ref{eq:41} and therefore 
the picture of heavy ion collisions outlined above. Further, a systematic study 
of energy loss in cold matter will help constrain extrapolations of pQCD~\cite{Wang} 
used to study jet quenching in hot matter.

\subsubsection{Proton/Deuteron-nucleus versus electron-nucleus collisions
as probes of high parton densities}

Both p/D-A and e-A collisions probe the small $x$ region at high energies. Both are important to ascertain truly universal aspects of  novel 
physics. e-A collisions, owing to the independent "lever" arm in $x$ and $Q^2$, as well as the simpler 
lepton-quark vertex, are better equipped for precision measurements. For example, in e-A collisions, information about gluon distributions can be extracted from scaling violations and from photon-gluon fusion processes. 
In both cases, high precision measurements are feasible. In p-A collisions, one can extract gluon distributions from 
scaling violations in Drell-Yan and gluon-gluon and quark-gluon fusion channels such as open charm 
and direct photon measurements respectively. However, for both scaling violations and fusion processes, one has more convolutions 
and kinematic constraints in p-A than in e-A. These limit both the precision and range of measurements. In Drell-Yan, in 
contrast to $F_2$, clear scaling violations in the data are very hard to see and data are limited to $M^2 > 16$ GeV$^2$, 
above the $J/\psi$ and $\psi^\prime$ thresholds. 

A clear difference between p/D-A and e-A collisions is in hard diffractive final states. 
At HERA, these constituted approximately 10$\%$ of the total cross section. At eRHIC, 
these may constitute 30-40$\%$ of 
the cross section~\cite{Strikman,Levin}. Also, factorization theorems 
derived for diffractive 
parton distributions only apply to lepton-hadron processes~\cite{Collins}. Spectator 
interactions in p/D-A collisions will destroy rapidity gaps. 
A comparative study of p/D-A and e-A collisions thus has
great potential for unravelling universal aspects of event structures in high energy
QCD.

\section{Electron-Ion Collider: Accelerator Issues}

With the scientific interest in a high luminosity lepton-ion collider 
gathering momentum during 
the last several years, there has been a substantial  effort in parallel 
to develop a preliminary technical design for such a machine. A team of 
physicists from BNL, MIT-Bates, DESY and the Budker Institute have developed 
a realistic design~\cite{eRHIC_ZDR} 
for a machine using RHIC, which would attain 
an e-p collision luminosity of $0.4 \times 10^{33} \,{\rm cm}^{-2} {\rm s}^{-1}$ 
and could with minimal R\&D start construction as soon as funding becomes available. 
Other more ambitious lepton-ion collider concepts which would use a high intensity 
electron linac to attain higher luminosity are under active 
consideration~\cite{eRHIC_ZDR,merm}.  This section gives an overview of the 
activities currently underway related to the accelerator design. 

The physics program described above sets clear requirements and goals for the 
lepton-ion collider to be a successful and efficient tool. These goals include: 
a sufficiently high  luminosity; a significant range of beam collision energies; 
and polarized beam (both lepton and nucleon) capability.  On the other hand, to be 
realistic, the goals should be based on the present understanding of the existing 
RHIC machine and limitations which arise from the machine itself. 
Realistic machine upgrades should be considered to overcome existing limitations and to 
achieve advanced machine parameters, but those upgrades should be cost-effective.

The intent to minimize required upgrades in the existing RHIC rings affects the 
choice of parameters and the set of goals. For example, the design assumed 
simultaneous collisions of both ion-ion and lepton-ion beams. In the main design line, 
collisions in two ion-ion interaction regions, at the ``6'' and ``8 o'clock'' 
locations, have to be allowed in parallel with electron-ion collisions.

Taking these considerations into account, the following goals were defined for the 
accelerator design:
\begin{itemize}
\item The machine should be able to provide beams in the following energy ranges:
for the electron accelerator, 5-10 GeV polarized electrons, 10 GeV polarized positrons;
for the ion accelerator, 50-250 GeV polarized protons, 100 GeV/u Gold ions.

\item Luminosity: in the $10^{32} - 10^{33}\, {\rm cm}^{-2} {\rm s}^{-1}$ range for 
e-p collisions; in the $10^{30} - 10^{31} \, {\rm cm}^{-2} {\rm s}^{-1}$ range for 
e-Au collisions

\item 70\% polarization for both lepton and proton beams
\item Longitudinal polarization in the collision point for both lepton and proton beams
\end{itemize}

An additional design goal was to include the possibility of accelerating polarized ions, 
especially polarized  $^3$He ions.

\subsection{eRHIC: ring-ring design}

The primary eRHIC design centers on a 10 GeV lepton storage ring which intersects with 
one of the RHIC ion beams at one of the interaction regions (IRs), 
not used by any of the ion-ion collision experiments.  RHIC uses superconducting dipole 
and quadrupole magnets to maintain ion beams circulating in two rings on a 3834 meter 
circumference.  The ion energy range covers 10.8 to 100 GeV/u for gold ions and 25 to 
250 GeV for protons.  There are in total 6 intersection points where two ion rings, 
Blue and Yellow, cross each other.  Four of these intersections points are currently 
in use by physics experiments.

\begin{figure}[!h]
\hspace*{32mm}
\epsfig{file=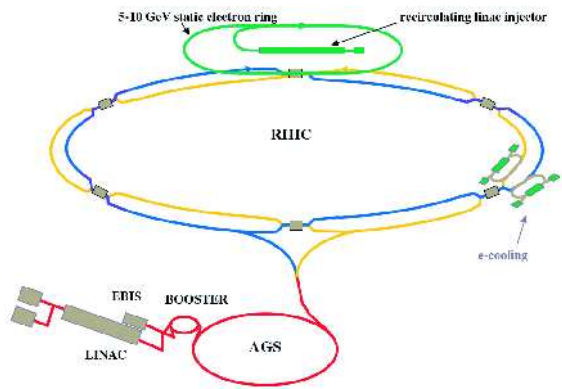,width=8cm,angle=0}
\caption{Schematic layout of the ring-ring eRHIC collider.
\label{fig:ring-ring}}
\end{figure}

 A general layout of the ring-ring eRHIC collider is shown in Figure \ref{fig:ring-ring} 
with the lepton-ion collisions occurring in the ``12 o'clock'' interaction region.   
Plans have been made for a new detector, developed and optimized for electron-ion 
collision studies, to be constructed in that interaction region.

The electron beam in this design is produced by a polarized electron source and accelerated in a linac injector to energies of 5 to 10 GeV. To reduce the injector size and cost, the injector design includes recirculation arcs, so that the electron beam passes through the same accelerating linac sections multiple times.  Two possible linac designs, superconducting and normal conducting, have been considered. The beam is accelerated by the linac to the required collision energy and injected into the storage ring. The electron storage ring is designed to be capable of electron beam storage in the energy range of 5 to 10 GeV with appropriate beam emittance values. It does not provide any additional acceleration for the beam. The electron ring should minimize depolarization effects in order to keep the electron beam polarization lifetime longer than 
the typical storage time of several hours.  

The injector system also includes the conversion system for positron production. After production the positrons are accelerated to 10 GeV energy and injected into the storage ring similarly to the electrons. Obviously the field polarities of all ring magnets should be reversed
in the positron operation mode. Unlike electrons, the positrons are produced unpolarized and have to be polarized using radiative self-polarization in the ring.  Therefore, the design of the ring should allow for a sufficiently small self polarization time. The current ring design provides a self polarization time of about 20 min at 10 GeV. But with polarization time increasing sharply as beam energy goes down the use of a
polarized positron beam in the present design is limited to 10 GeV energy.

The design of the eRHIC interaction region involves both accelerator and detector considerations.  Figure \ref{fig:ring-ring} shows the electron accelerator located at the 
``12 o'clock'' region.  Another possible location for the electron accelerator and for 
electron-ion collisions might be the ``4 o'clock'' region. For collisions with electrons 
the ion beam in the RHIC Blue ring will be used, because the 
Blue ring can operate alone, even with  the other ion ring, Yellow, being down. The interaction region design provides for fast beam separation for electron and Blue ring ion bunches as well as for strong focusing at the collision point. In this design,  
the other (Yellow) ion ring makes a 3m vertical excursion around the collision region, avoiding collisions both with electrons and the Blue ion beam. The eRHIC interaction region includes spin rotators, in both the electron and the Blue ion rings, to produce longitudinally 
polarized beams of leptons and protons at the collision point. 

The electron cooling system in RHIC~\cite{park,benz} is one of the essential upgrades required for eRHIC. The cooling is necessary to reach the luminosity goals for lepton collisions with Gold ions and low (below 150 GeV) energy protons. Electron cooling is considered an essential upgrade of RHIC to attain higher luminosity in ion-ion collisions.

In addition, the present eRHIC design assumes a total ion beam current higher than that being used at present in RHIC operation. This is attained by operating RHIC with 360 bunches. 

The eRHIC collision luminosity is limited mainly by the maximum achievable beam-beam parameters and by the interaction region magnet aperture limitations. To understand this, it is most
convenient to use a luminosity expression in terms of beam-beam parameters 
($\xi_e\xi_i$) and rms angular spread in the interaction point ($\sigma^\prime_{xi},
\sigma^\prime_{ye}$):
\begin{eqnarray*}
L= f_c {{\pi\gamma_i\gamma_e}\over{r_ir_e}} \xi_{xi} \xi_{ye} \sigma^\prime_{xi} \sigma^\prime_{ye} {{(1 + K)^2}\over{K}}
\end{eqnarray*}

The $f_c = 28.15$ MHz is a collision frequency, assuming 360 bunches in the ion ring and 120 bunches in the electron ring. The parameter $K=\sigma_y/\sigma_x$ presents the ratio of beam sizes in the interaction point. One of the basic conditions which defines the choice of  beam parameters is a requirement on equal beam sizes of ion and electron beams at the interaction point: $\sigma_{xe} = \sigma_{xi}$ and $\sigma_{ye} = \sigma_{ye}$. The requirement is based on the operational experience at the  HERA collider and on the reasonable intention to minimize the amount of one beam passing through the strongly nonlinear field in the outside area of the counter-rotating beam.

According to the above expression, the luminosity reaches a limiting value at the maximum values of beam-beam parameters, or at the beam-beam parameter limits. For protons (and ions) the total beam-beam parameter limit was assumed to be 0.02 , following the experience and observation from other proton machines as well as initial experience from RHIC operation.  With three beam-beam interaction points, two for proton-proton and one for electron-proton collisions, the beam-beam parameter per interaction point should not exceed 0.007.  

For the electron (or positron) beam a limiting value of the beam-beam parameter has been put at 0.08 for 10 GeV beam energy, following the results of beam-beam simulations, as well as from the experience at electron machines of similar energy range. Because 
the beam-beam limit decreases proportionally with the beam energy, the limiting 
value for 5 GeV is reduced to 0.04.

The available magnet apertures in the interaction region also put a limit on the achievable luminosity. The work on the interaction region design revealed considerable difficulties to provide an acceptable design for collisions of round beams. The IR has been designed to provide low beta focusing and efficient separation of elliptical beams, with beam size ratio $K=1/2$. 
The main aperture limitation comes from the septum magnet, which leads to the 
limiting values of  $\sigma^\prime_{xp} = 93\mu$rad.

Another limitation which must be taken into account is a minimum acceptable value of the beta-function at the interaction point ($\beta^*$). With the proton rms bunch length of 20~cm, decreasing $\beta^*$ well below this number results in a luminosity degradation due to the hour-glass effect.  The limiting value $\beta^*= 19$~cm has been used for the design, which results in a luminosity reduction of only about 12\%. A bunch length of 20~cm for Au ions 
would be achieved with  electron cooling.

Tables \ref{tab1} and \ref{tab2} show design luminosities and beam parameters. The positron beam intensity is assumed to be identical to the electron beam intensity, hence the luminosities for collisions involving a positron beam are equivalent to electron-ion collision luminosities.  To achieve the high luminosity in the low energy tune in Table \ref{tab1}, 
the electron cooling has to be used to reduce the normalized transverse emittance of the lower energy proton beam to $5\pi$mm$\cdot$mrad. 
Also, in that case the proton beam should have collisions only with the electron beam. Proton-proton collisions in the other two interaction points have to be avoided  to allow for a
higher proton beam-beam parameter. The maximum luminosity achieved in the present design is $4.4\times 10^{32} \,{\rm cm}^{-2} {\rm s}^{-1}$ 
in the high energy collision mode (10 GeV leptons on 250 GeV protons).  Possible paths to luminosities as high as $10^{33} \,{\rm cm}^{-2} {\rm s}^{-1}$ are being explored, with studies planned to investigate the feasibility of higher electron beam intensity operation.  To achieve and maintain the Au normalized transverse beam emittances shown in Table \ref{tab2}, electron cooling of the Au beam will be used. For the lower energy tune of electron-Gold 
collisions, the intensity of the Gold beam is considerably reduced because of the 
reduced value of the beam-beam parameter limit for the electron beam.

\begin{table}[h!]
\begin{center}
\begin{tabular}{|l|c|c|c|c|}
\hline
High energy tune & p & e & p & e  \\
\hline
Energy, GeV & 250 & 10 & 50 & 5 \\
Bunch intensity, 10$^{11}$ & 1 & 1 & 1 & 1 \\
Ion normalized & & & & \\ 
emittance,  & 15/15 & & 5/5  & \\
$\pi$ mm $\cdot$ mrad, $x/y$ & & & & \\
rms emittance, nm, $x/y$ & 9.5/9.5 & 53/9.5 & 16.1/16.1 & 85/38 \\
$\beta^*$, cm, $x/y$ & 108/27 & 19/27 & 186/46 & 35/20 \\
Beam-beam parameters, & 0.0065/0.0 & 0.03/0.08 & 0.019/0.00 & 0.036/0.04 \\
$x/y$ & 03 & & 95  & \\
$\kappa=\xi_y/\xi_x$ & 1 & 0.18 & 1 & 0.45 \\
\hline\hline
Luminosity, $1.0 \times 10^{32}$ & & & & \\
cm$^{-2}s^{-1}$ & \multicolumn{2}{c|}{4.4} & \multicolumn{2}{c|}{1.5} \\
\hline
\end{tabular}
\caption{Luminosities and main beam parameters for e$^{\pm}$-p collisions}
\label{tab1}
\end{center}
\end{table}

\begin{table}[h!]
\begin{center}
\begin{tabular}{|l|c|c|c|c|}
\hline

High energy tune

& Au & e & Au & e  \\
\hline
Energy, GeV/u & 100 & 10 & 100 & 5 \\

Bunch intensity, 10$^{11}$ & 0.01 & 1 & 0.0045 & 1 \\

Ion normalized & & & & \\ 

emittance,  & 6/6 & & 6/6  & \\

$\pi$ mm $\cdot$ mrad, $x/y$ & & & & \\

rms emittance, nm, $x/y$ & 9.5/9.5 & 54/7.5 & 9.5/9.5 & 54/13.5 \\

$\beta^*$, cm, $x/y$ & 108/27 & 19/34 & 108/27 & 19/19 \\

Beam-beam parameters, & 0.0065/0.0 & 0.0224/0.0 & 0.0065/0.0 & 0.02/0.04 \\

$x/y$ & 03 & 8 & 03  & \\

$\kappa=\xi_y/\xi_x$ & 1 & 0.14 & 1 & 0.25 \\
\hline\hline

Luminosity, $1.0 \times 10^{32}$ & & & & \\

cm$^{-2}s^{-1}$ & \multicolumn{2}{c|}{4.4} & \multicolumn{2}{c|}{2.0} \\
\hline
\end{tabular}
\caption{Luminosities and main beam parameters for e$^{\pm}$-Au collisions}
\label{tab2}
\end{center}
\end{table}

\subsection{eRHIC: linac-ring design}

A linac-ring design for eRHIC is also under active consideration.  This configuration uses a fresh electron beam bunch for each collision and so the tune shift limit on the electron beam is removed.  This provides the important possibility to attain significantly higher luminosity (up to $10^{34} \,{\rm cm}^{-2} {\rm s}^{-1}$) than the ring-ring design.  A second advantage of the linac-ring design is the ability to reverse the electron spin polarization on each bunch.  A disadvantage of the linac-ring design is the inability to deliver polarized positrons.  The realization of the linac beam is technically challenging and the polarized electron source requirements are well beyond  present capabilities~\cite{fark2}.  

\begin{figure}[!h]
\vspace*{3mm}
\hspace*{32mm}
\epsfig{file=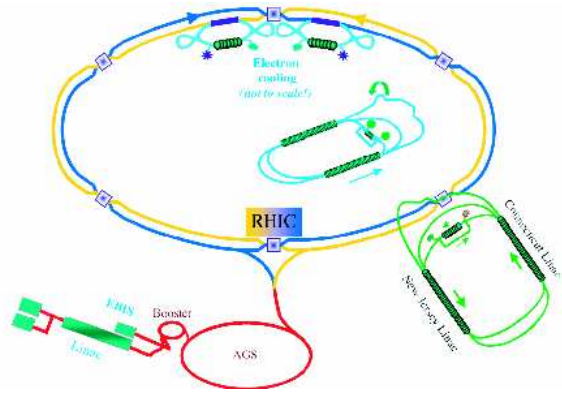,width=8cm,angle=0}
\caption{Schematic layout of a possible linac-ring eRHIC design. 
\label{fig:linac-ring}}
\end{figure}

Figure \ref{fig:linac-ring} shows a schematic layout of a possible linac-ring eRHIC 
design.  A 450 mA polarized electron beam is accelerated in an Energy Recovery Linac 
(ERL). After colliding with the RHIC beam in as many as four interaction points, 
the electron beam is decelerated to an energy of a few MeV and dumped.  The energy thus 
recovered is used for accelerating subsequent bunches to the energy of the experiment.

\subsection{Other lepton-ion collider designs: ELIC}

A very ambitious electron-ion collider design seeking to attain luminosities up to $10^{35} 
\, {\rm cm}^{-2} {\rm s}^{-1}$ is underway at Jefferson Laboratory~\cite{merm}. This Electron Light Ion Collider (ELIC) design is based on use of polarized 5 to 7 GeV electrons in a superconducting ERL upgrade of the present CEBAF accelerator and a 30 to 150 GeV ion storage ring (polarized p, d, $^3$He, Li and unpolarized nuclei up to Ar, all totally stripped).  The ultra-high luminosity is envisioned to be achievable with short ion bunches and crab-crossing at 1.5 GHz bunch collision rate in up to four interaction regions.  The ELIC design also includes a recirculating electron ring that would help to reduce the linac and polarized source requirements compared to the linac-ring eRHIC design of section 4.2.

\begin{figure}[!h]
\hspace*{32mm}
\epsfig{file=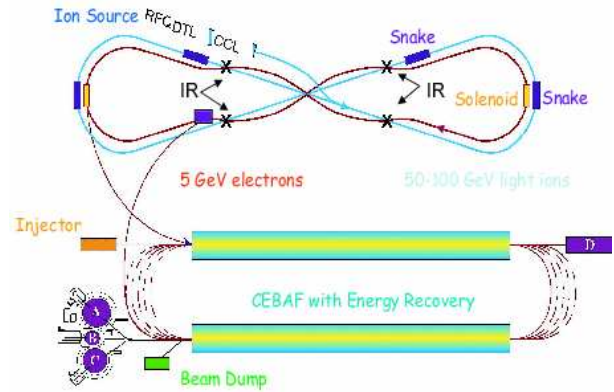,width=8cm,angle=0}
\caption{Schematic layout of a possible electron light ion collider at Jefferson Laboratory.
\label{fig:43}}
\end{figure}

The ELIC proposal is at an early stage of development.  A number of technical challenges must be resolved, and several R\&D projects have been started.  These include development of a high average current polarized electron source with a high bunch charge, electron cooling of protons/ions, energy recovery at high current and high energy, and the design of an interaction region that supports the combination of high luminosity and high detector acceptance and resolution.

\section{Detector Ideas for the EIC \label{detecsec}}
The experience gained at HERA with the H1 and ZEUS detectors~\cite{hera_physics} 
provides useful guidance for the conceptual design of a detector that will measure 
a complete event ($4\pi$-coverage) produced in collisions of energetic electrons with 
protons and ions, at different beam energies and polarizations. The H1 and ZEUS detectors 
are general-purpose magnetic detectors with nearly hermetic calorimetric coverage. The 
differences between them are based 
on their approach to calorimetry. The H1 detector collaboration emphasized the electron 
identification and energy resolution, whereas the ZEUS collaboration puts more emphasis 
on optimizing hadronic calorimetry. The differences in
their physics philosophy were reflected in their overall design: H1 had a 
liquid Argon calorimeter inside the large diameter magnet, whereas 
ZEUS chose to build a Uranium scintillator sampling calorimeter with equal response to 
electrons and hadrons. They put their tracking detectors inside a 
superconducting solenoid surrounded by calorimeters and muon chambers. H1 placed
their tracking chambers inside the calorimeter surrounded by their magnet. In addition, 
both collaborations placed their luminosity and electron detectors downstream in the 
direction of the proton and electron directions, respectively. Both 
collaborations added low angle forward proton spectrometers and neutron 
detectors in the proton beam direction.

Both detectors have good angular coverage (approximately, $3^{\circ} 
< \theta < 175^{\circ}$, 
where the angle is measured with respect to the incoming proton beam direction) for 
electromagnetic (EM) calorimetry, with energy resolutions of 1-3\% and for electromagnetic 
showers $\sigma /E  \sim  15\%/\sqrt{E~{\rm (GeV)}} + 1\%$. Hadronic energy 
scale uncertainties of 3\% were achieved for both, with some differences in the
$\sigma/E$ for hadronic showers, which were $\sim  R/\sqrt{E~{\rm (GeV)}}
+2\%$, where $R=35\%$ and $50\%$ 
for ZEUS and H1, respectively. With the central tracking fields 
$\approx 1.5$~T covering a region similar to the calorimetric angular acceptance, 
momentum resolution $\sigma /p_{T} < 0.01 \, p_{T} {\rm (GeV)}$ was generally achieved 
for almost all acceptances, except for the forward and backward directions.
These directions were regarded at the beginning as being less interesting. 
However, the unexpected physics of low $x$ and low $Q^{2}$ (including diffraction in 
e-p scattering) came from this rather poorly instrumented region. And since 
the luminosity upgrade program, the low $\beta^{*}$ magnets installed close to the 
interaction point to enhance the luminosity of e-p collisions (HERA-II) have further 
deteriorated the acceptance of detectors in these specific geometric regions.

The EIC detector design ideas are already being guided by the lessons learned from the 
triumphs and tribulations of the HERA experience. 
All advantages of the HERA detectors such as the almost 
4$\pi$ coverage and the functionality with respect to spatial 
orientation will be preserved. The EIC detector will have enhanced capability
in the very forward and backward directions to measure continuously the low $x$ and 
low $Q^2$ regions that are not comprehensively accessible
at HERA. The detector design directly impacts the interaction region 
design and hence the accelerator parameters for the two beam elements:
the effective interaction luminosity and the effective polarization of the 
two beams at the interaction point. Close interaction between the detector design
and the IR design is hence needed in the very early stage of the project, which 
has already started \cite{eRHIC_ZDR}. It is expected that the detector design and 
the IR design will evolve over the next few years. The e-p and e-A collisions at EIC will 
produce very asymmetric event topologies, not unlike HERA events. These asymmetries, 
properly exploited, allow precise measurements of energy and 
color flow in collisions of large and small-$x$ partons. They also allow observation of 
interactions of electrons with photons that are coherently emitted by the relativistic 
heavy ions. The detector for EIC must detect: the scattered electrons, the quark 
fragmentation products and the centrally produced hadrons. It will be the 
first collider detector to measure the fragmentation region of the proton or 
the nucleus, a domain not covered effectively at HERA. The detector design, in 
addition, should pose no difficulties for important measurements such as precision 
beam polarization (electron as well as hadron beam) and collision luminosity.

The EIC detector design will allow measurements of partons from hard processes in the 
region around $90^{\circ}$ scattering angle with respect to the beam pipes. This 
central region could have a jet tracker with an EM calorimeter backed by an 
instrumented iron yoke.
Electrons from DIS are also emitted into this region and will utilize the tracking and 
the EM calorimetry. Electrons from photo-production and from DIS at intermediate and low 
momentum transfer will have to be detected by specialized backward detectors. With these 
guiding ideas, one could imagine that the EIC barrel might have a time projection chamber 
(TPC) backed by an EM Calorimeter inside a superconducting coil. One could use Spaghetti 
Calorimetry (SPACAL) for endcaps and GEM-type micro-vertex detector to complement the 
tracking capacity of the TPC in the central as well as forward/backward (endcap) regions. 
This type of central  and end-cap detector geometry is now fairly standard. Details of 
the design could be finalized in the next few years using the state of the art technology 
and experience from more recent detectors such as BaBar at SLAC (USA) and Belle at KEK 
(Japan). To accommodate tracking and particle ID requirements for
the different center-of-mass energy running ($\sqrt{s}= 30-100$~GeV) resulting at
different beam energies, the central spectrometer magnet will 
have multiple field strength operation capabilities, including radial dependence 
of field strengths. Possible spectrometry based on dipole and toroidal fields is also 
being considered at this time.

The forward and backward regions (hadron and electron beam directions) in e-p collisions 
were instrumented at HERA up to a pseudo-rapidity of $\eta \approx 3$. A specialized 
detector added later extended this range with difficulty to $\eta \approx 4$. Although 
acceptance enhancement in the regions beyond $\eta =4$ is possible with conventional 
ideas such as forward calorimetry and tracking using beam elements and silicon strip
based Roman Pot Detectors~\cite{Roman_pots}, it is imperative for the EIC that 
this region be well instrumented. A recent detector design for eRHIC 
developed by the experimental group at the Max-Planck Institute, 
Munich, accomplishes just this~\cite{Caldwell} by allowing continuous access to physics 
up to $\eta \approx 6$. The main difference with respect to a conventional collider 
detector is a dipole field, rather than a solenoid, that separates the low energy 
scattered electron from the beam. 
High precision silicon tracking stations capable of achieving 
$\Delta p/p \sim 2\%$, EM calorimetry with energy resolution better than $20\%/\sqrt{E}$, 
an excellent $e/ \pi$ separation over a large $Q^{2}$ range, all in the backward region 
(in the electron beam direction) are attainable. 
In the forward region, the dipole field allows excellent 
tracking and a combination of EM and hadronic calorimetry with $20\%/\sqrt{E {\rm (GeV)}}$ 
and $50\%/\sqrt{E {\rm (GeV)}}$ energy resolution, respectively. This allows access 
to very high $x \sim 0.9$ with excellent accuracy. This region of high $x$ 
is largely unexplored both in polarized and in unpolarized DIS. A significant 
distance away from the EIC central detector and IR, there may be Roman Pots, high rigidity 
spectrometers including EM calorimetry and forward electron taggers, all placed to improve 
the measurement of low angle scattering at high energy.

Although significant effort will be made to avoid design conflicts, the 
conventional detector using a solenoid magnet and the one described above  
may not coexist in certain scenarios being considered 
for the accelerator designs of the EIC at BNL. The main design line, presently 
the ring-ring design, may be particularly difficult with only one IR.
Options such as time sharing between two detectors at the same IR with the two 
detectors residing on parallel rails may be considered. In the case of the linac-ring 
scenario, several other options are available. Because the physics of low $x$ and 
low $Q^2$ does not require a large luminosity, nor is presently the beam polarization 
a crucial requirement for the physics~\cite{Caldwell}, an interaction point 
with sufficient beam luminosity would be possible with innovative layouts of 
the accelerator complex. These and other details will be worked out
in the next several years. Depending on the interest shown by the experimental 
community, accelerator designs that incorporate up to four collision points 
(while still allowing two hadron-hadron collision points at RHIC) will be considered 
and developed.

\section*{Acknowledgments}
This review draws generously on the whitepaper for the Electron Ion 
Collider (BNL-68933-02/07-REV), and we thank all our co-authors 
on this publication for their efforts. We are especially grateful to
Marco Stratmann and Mark Strikman for valuable advice, discussions and 
comments. W.V.\  and A.D.\ are  grateful to RIKEN and Brookhaven National 
Laboratory. R.M. is supported by the Department of
Energy Cooperative Agreement DE-FC02-94ER40818.
W.V.\ and R.V.\  were supported  by Department of Energy 
(contract number DE-AC02-98CH10886). R.V.\ 's research was also supported 
in part by a research award from the A. Von Humboldt Foundation.

\end{document}